\documentclass[preprintnumbers,amsmath,amssymb,amsthm,nofootinbib,superscriptaddress,floatfix,onecolumn,reprint,
groupedaddress,
eqsecnum,
amsfonts,aps]{revtex4}

\pdfoutput=1
\usepackage[normalem]{ulem}
\usepackage{graphicx}
\usepackage[colorlinks,citecolor=blue,urlcolor=blue,linkcolor=blue]{hyperref}
\usepackage[T1]{fontenc} 
\usepackage{subfigure}
\usepackage{enumitem}
\usepackage[normalem]{ulem}
\usepackage{graphicx}
\usepackage{subeqnarray}
\usepackage{cases}

\usepackage{subfigure}
\usepackage{latexsym}
\usepackage{mathrsfs}
\usepackage{color}
\usepackage{dcolumn}
\usepackage{bm}

\newcommand{\red}{\textcolor[rgb]{1.00,0.00,0.00}}

\definecolor{dyellow}{rgb}{1.,0.8,.0}
\definecolor{myblue}{rgb}{.1,.1,.7}
\definecolor{dcyan}{rgb}{.0,.6,.6}
\definecolor{dmagenta}{rgb}{0.6,0.0,0.6}
\definecolor{brown}{rgb}{0.6,0.2,0.}
\definecolor{darkblue}{rgb}{.0,.0,0.5}
\definecolor{darkred}{rgb}{0.75,0.0,0.0}
\definecolor{orange}{rgb}{1.,.6,.0}
\definecolor{dorange}{rgb}{0.8,.4,.0}
\definecolor{darkgreen}{rgb}{0.0,0.6,0.0}
\definecolor{purple}{rgb}{.4,.0,.4}
\definecolor{grey}{rgb}{0.5,0.5,0.5}
\def\black{\color{black}}
\def\red{\color{red}}

\begin{document}
\hyphenpenalty=1000
\title{A signature invariant geometric algebra framework for spacetime physics and its applications in relativistic dynamics of a massive particle and gyroscopic precession}

\author{Bofeng Wu}
\affiliation{Department of Physics, College of Sciences, Northeastern University, Shenyang 110819, China}


\begin{abstract}
\centerline{\textbf{ABSTRACT}}
\bigskip
A signature invariant geometric algebra framework for spacetime physics is formulated. By following the original idea of David Hestenes in the spacetime algebra of signature $(+,-,-,-)$, the techniques related to relative vector and spacetime split are built up in the spacetime algebra of signature $(-,+,+,+)$. The even subalgebras of the spacetime algebras of signatures $(\pm,\mp,\mp,\mp)$ share the same operation rules, so that they could be treated as one algebraic formalism, in which spacetime physics is described in a signature invariant form. Based on the two spacetime algebras and their ``common'' even subalgebra, rotor techniques on Lorentz transformation and relativistic dynamics of a massive particle in curved spacetime are constructed. A signature invariant treatment of the general Lorentz boost with velocity in an arbitrary direction and the general spatial rotation in an arbitrary plane is presented. For a massive particle, the spacetime splits of the velocity, acceleration, momentum, and force four-vectors with the normalized four-velocity of the fiducial observer, at rest in the coordinate system of the spacetime metric, are given, where the proper time of the fiducial observer is identified, and the contribution of the bivector connection is considered, and with these results, a three-dimensional analogue of Newton's second law for this particle in curved spacetime is achieved. Finally, as a comprehensive application of the techniques constructed in this paper, a geometric algebra approach to gyroscopic precession is provided, where for a gyroscope moving in the Lense-Thirring spacetime, the precessional angular velocity of its spin is derived in a signature invariant manner.
\end{abstract}
\maketitle
\section{Introduction}
\label{sec:first}
William Kingdon Clifford introduced geometric algebra (GA) based on the earlier work of Hamilton and Grassmann~\cite{Clifford1882}, and then, David Hestenes developed it by inventing geometric calculus and formulating spacetime algebra (STA)~\cite{Hestenes1966}. GA is a unified language for mathematics and physics~\cite{Hestenes1984}, and has important applications in theoretical physics~\cite{Hestenes1986,Application.electrodynamics,Application.gravity,Application.quantum}. STA, as the GA for spacetime, provides a synthetic framework for spacetime physics~\cite{Doran2003}. One of the remarkable advantages of STA is that Lorentz boost and spatial rotation can be handled with rotor techniques in an elegant and highly condensed manner~\cite{Doran2003,Rotortechnology,Sabbata2006}. Therefore, for those topics involving a knowledge of Lorentz boost and spatial rotation, such as gyroscopic precession~\cite{MTW1973,Ignazio1995}, it could be expected that a more efficient approach to dealing with them will be found in the language of STA.

STA can be generated by an orthonormal frame with respect to the Minkowski metric. Since the signature $(+,-,-,-)$ is widely used in STA~\cite{Doran2003} whereas the opposite signature $(-,+,+,+)$ is often adopted in literatures on relativity~\cite{Eric2010}, when STA is applied to relativistic physics the change of signature from one to another will cause inconvenience. In fact, the STA of signature $(-,+,+,+)$ is also used~\cite{Oppositesignature}, and however, because the techniques related to relative vector and spacetime split have not been developed in this algebraic formalism, its applications are quite limited. One of the purposes of this paper is to build up these techniques in the STA of signature $(-,+,+,+)$ by following the original idea of David Hestenes in the STA of signature $(+,-,-,-)$, which will definitely facilitate the study of relativistic physics in the language of GA.

Throughout the paper, the following notation and rules are adopted unless stated otherwise:
\begin{itemize}[itemsep=0pt,topsep=4pt,parsep=0pt]
\item For two multivectors $A$ and $B$ in spacetime, their geometric product, inner product, outer product, and commutator product are represented by $AB,A\cdot B,A\wedge B$, and $A\times B$, respectively;
\item For a multivector $M$ in spacetime, $\tilde{M}$ and $\langle M\rangle_{p}\ (p=0,1,2,3,4)$ denote its reverse and $p$-vector part, respectively, where $\langle M\rangle_{0}$ is abbreviated as $\langle M\rangle$;
\item The Greek letters, denoting the spacetime indices, range from 0 to 3, whereas the Latin letters, denoting the space indices, range from 1 to 3;
\item The sum should be taken over, when repeated indices appear within a term;
\item The international system of units is used.
\end{itemize}
Let $\{\gamma_{\alpha}^{+}\}$ and $\{\gamma_{\alpha}^{-}\}$ be orthonormal frames with respect to the Minkowski metrics in the signatures $(+,-,-,-)$ and $(-,+,+,+)$, respectively, and
the STAs of the two signatures can be generated by them. In these two STAs, we find the following important conclusions:
\begin{itemize}[itemsep=0pt,topsep=4pt,parsep=0pt]
\item Denote $\{\gamma^{\alpha}_{\pm}\}$ as the reciprocal frames of $\{\gamma_{\alpha}^{\pm}\}$, and frames of relative vectors are constructed by $\{\boldsymbol{\sigma}_{k}^{\pm}:=\gamma_{k}^{\pm}\gamma^{0}_{\pm}\}$, where both $\{\boldsymbol{\sigma}_{k}^{+}\}$ and $\{\boldsymbol{\sigma}_{k}^{-}\}$, spanning the relative spaces orthogonal to the timelike vectors $\gamma_{0}^{+}$ and $\gamma_{0}^{-}$, respectively, provide representation-free versions of the Pauli matrices;
\item The two relative spaces are both the Euclidean spaces of dimension 3 with $\{\boldsymbol{\sigma}_{k}^{\pm}\}$ as right-handed orthonormal bases, where the inner product and the cross product in these two spaces can be defined as their conventional ones, respectively;
\item The even subalgebras of the STAs of signatures $(\pm,\mp,\mp,\mp)$ are generated by $\{\boldsymbol{\sigma}_{k}^{\pm}\}$, and they share the same operation rules;
\item For vectors $b^{\pm}=b^{\alpha}_{\pm}\gamma_{\alpha}^{\pm}$, their spacetime splits with $\gamma_{0}^{\pm}$ are $b^{\pm}\gamma^{0}_{\pm}=b^{0}_{\pm}+\boldsymbol{b}^{\pm}$, where $\boldsymbol{b}^{\pm}=b^{i}_{\pm}\boldsymbol{\sigma}_{i}^{\pm}$, as bivectors in spacetime, are called the relative vectors of $b^{\pm}$;
\item For operators $\partial^{\pm}:=\gamma^{\alpha}_{\pm}\partial_{\alpha}$, their spacetime splits with $\gamma_{0}^{\pm}$ are $\gamma_{0}^{\pm}\partial^{\pm}=\partial_{0}+\boldsymbol{\nabla}^{\pm}$, where $\partial_{\mu}:=\partial/\partial x^{\mu}$ and $\boldsymbol{\nabla}^{\pm}:=\boldsymbol{\sigma}^{k}_{\pm}\partial_{k}$ with $x^{\mu}$ and $\{\boldsymbol{\sigma}^{k}_{\pm}:=\gamma_{0}^{\pm}\gamma^{k}_{\pm}\}$ as
    coordinates in spacetime and the reciprocal frames of $\{\boldsymbol{\sigma}_{k}^{\pm}\}$ in the relative spaces, respectively.
\end{itemize}
Since the even subalgebras of the two STAs share the same operation rules, we will no longer distinguish them strictly and treat them as one algebraic formalism hereafter. In Appendix B of
this paper, a detailed presentation of this algebraic formalism is given. It will be shown that the ``common'' even subalgebra of the STAs of signatures $(\pm,\mp,\mp,\mp)$ actually provides a signature invariant GA framework for spacetime physics. In order to give an application paradigm of the two STAs and their ``common'' even subalgebra, we need to make use of them to study some specific problems in spacetime physics, and gyroscopic precession is such a typical topic.

According to the prediction of General Relativity, the spin of a gyroscope precesses relative to the asymptotic
inertial frames as it moves around a rotating spherical source~\cite{Ignazio1995}. The conventional method to describe gyroscopic precession under the weak-field and slow-motion (WFSM) approximation in tensor language is presented in Refs.~\cite{MTW1973,Ignazio1995}.
For a uniformly rotating spherical source, the external gravitational field is stationary, and only the leading pole moments need to be considered, so that the spacetime geometry is described by the Lense-Thirring metric~\cite{Wu:2021uws}. As a result, the corresponding spacetime is known as the Lense-Thirring spacetime. When a torque-free gyroscope is moving in this spacetime, there exist three types of precession for its spin, namely, the de Sitter precession, the Lense-Thirring precession, and the Thomas precession, where these phenomena are, respectively, resulted from gyroscopic motion through the spacetime curved by the mass of the source, rotation of the source, and gyroscopic non-geodesic motion~\cite{Everitt:2011hp}.

In the traditional description for gyroscopic precession based on tensor language, one always needs to work with the components of some tensor in a chosen coordinate frame, which often leads to many equations with a low degree of clarity. The language of STA could provide a physically clear approach to dealing with this topic, since one just involves geometric objects during calculation~\cite{Lasenby:2016lfl}. As a preliminary attempt, another purpose of the present paper is to handle gyroscopic precession by applying the STAs of signatures $(\pm,\mp,\mp,\mp)$ and their ``common'' even subalgebra, so that for a gyroscope moving in the Lense-Thirring spacetime, a signature invariant derivation of the precessional angular velocity of its spin could be achieved. \emph{For brevity, in later applications, the signs ``$\pm$'' associated with multivectors and operators will be suppressed, and for equalities like $A=F(\pm B)$ and $C=G(\mp D)$, the signs ``$+$'' and ``$-$'' in the former equation correspond to the cases in the signatures $(+,-,-,-)$ and $(-,+,+,+)$, respectively, and the situation in the latter equation is reverse.}

Before analyzing gyroscopic precession, rotor techniques on Lorentz transformation and relativistic dynamics of a massive particle in curved spacetime need to be addressed in the two STAs. Rotor techniques are available in the STA of signature $(+,-,-,-)$~\cite{Doran2003,Rotortechnology,Sabbata2006}, and however, since the STA of signature $(-,+,+,+)$ is rarely employed, these techniques have not been fully developed in this algebraic formalism, where in particular the expressions of the rotors inducing Lorentz boost and spatial rotation should be clearly established. Being the third purpose of this paper, by virtue of the rotors constructed in the ``common'' even subalgebra of the two STAs, the general Lorentz boost with velocity in an arbitrary direction and the general spatial rotation in an arbitrary plane are handled in a signature invariant manner. How to study physics in curved spacetime based on STA is a fundamental problem. By following GA techniques for General Relativity formulated in Ref.~\cite{Francis:2003xi}, the treatment of gyroscopic precession in this paper is able to be put on a solid theoretical footing. To generate the STAs of signatures $(\pm,\mp,\mp,\mp)$ in a curved spacetime, one just needs to define a local orthonormal tetrad $\{\gamma_{\alpha}\}$ by the orthonormalization of a coordinate frame (in either signature), and then, by applying these two STAs and their ``common'' even subalgebra, the relevant topics in spacetime physics can be dealt with.

Relativistic dynamics of a massive particle in curved spacetime should be studied so as to describe the motion of a gyroscope moving around a gravitating source~\cite{Weinberg2014}. We assume that a collection of fiducial observers is distributed over space, and each fiducial observer is at rest in the coordinate system of the spacetime metric. For a massive particle, the spacetime splits of the velocity, acceleration, momentum, and force four-vectors with the normalized four-velocity $\gamma_{0}$ of the fiducial observer need to be derived, which is easy when spacetime is flat. However, in curved spacetime, some subtleties appear and ought to be seriously analyzed. For instance, the proper time of fiducial observers should be identified, and the contribution of the bivector connection $\omega(u)$ associated with $\{\gamma_{\alpha}\}$ (cf.~Ref.~\cite{Francis:2003xi}) should also be considered. In this paper, after overcoming these difficulties, the results are given, and with them, a three-dimensional analogue of Newton's second law for the particle in curved spacetime is achieved, which is the fourth purpose of the present paper. Besides, the Fermi-Walker derivatives presented in tensor language are recast in the STAs of signatures $(\pm,\mp,\mp,\mp)$ so that the motion of the spin of a gyroscope can be depicted in these two STAs~\cite{MTW1973}.

With the aid of the GA techniques constructed before, an efficient treatment of gyroscopic precession could be provided in the two STAs. Considering a gyroscope moving in the Lense-Thirring spacetime, some significant results like the three-dimensional generalized equation of motion for the gyroscope are first given on the basis of relativistic dynamics of a massive particle. Then, the rotor techniques are employed to handle the spin of the gyroscope, and the direct result shows that a bivector field $\varOmega(\tau)$ along its worldline completely determines the motion of its spin, where $\tau$ is the proper time. The bivector field $\varOmega(\tau)$ is dependent on the rotor $\hat{L}$ generating the pure Lorentz boost from the gyroscope's four-velocity $u$ to the fiducial observer's four-velocity $c\gamma_{0}$ and the bivector connection $\omega(u)$ associated with $\{\gamma_{\alpha}\}$, where $c$ is the velocity of light in vacuum. Just like the Faraday bivector, namely the electromagnetic field strength, the bivector field $\varOmega(\tau)$ can also be decomposed into the electric part $\varOmega^{(E)}(\tau)$ and the magnetic part $\varOmega^{(B)}(\tau)$. Let $\{\gamma^{\beta}\}$ be the reciprocal tetrad of $\{\gamma_{\alpha}\}$, and technically, if the condition $\tilde{\hat{L}}a\hat{L}\gamma^{0}=c\varOmega^{(E)}(\tau)$ is fulfilled, the spin of the gyroscope always precesses relative to its comoving frame, determined by the pure Lorentz boost generated by the rotor $\hat{L}$, with $\varOmega^{(B)}(\tau)$ as the precessional angular velocity.

The key point is to write down signature invariant expression of the bivector field $\varOmega(\tau)$ and the spacetime split of the gyroscope's four-acceleration $a$ with the normalized four-velocity $\gamma_{0}$ of the fiducial observer based on the ``common'' even subalgebra of the two STAs. According to Refs.~\cite{Francis:2003xi,Snygg1997}, the bivector connection $\omega(u)$ associated with $\{\gamma_{\alpha}\}$ can be directly derived, and then, by recasting it in terms of the relative vectors $\{\boldsymbol{\sigma}_{k}\}$, its  signature invariant expression and those of its electric part $\omega^{(E)}(u)$ and magnetic part $\omega^{(B)}(u)$ are obtained. Moreover, by applying the rotor techniques, the pure Lorentz boost $\hat{L}$ from $u$ to $c\gamma_{0}$ can also be derived. Thus, as noted before, the signature invariant expression of $\varOmega(\tau)$ and those of $\varOmega^{(E)}(\tau)$ and $\varOmega^{(B)}(\tau)$ are completely determined. As to $a$, its spacetime split with $\gamma_{0}$ could be directly obtained from the relevant conclusion in relativistic dynamics of a massive particle. Thus, with $a$, $\hat{L}$, and $\varOmega^{(E)}(\tau)$, one is capable of verifying that the condition $\tilde{\hat{L}}a\hat{L}\gamma^{0}=c\varOmega^{(E)}(\tau)$ holds by means of various operations in the ``common'' even subalgebra of the two STAs, and hence, the spin of the gyroscope indeed precesses in the comoving frame with $\varOmega^{(B)}(\tau)$ as the precessional angular velocity. After expanding $\varOmega^{(B)}(\tau)$ up to $1/c^3$ order with $1/c$ as the WFSM parameter~\cite{Blanchet:2013haa}, the gyroscope spin's angular velocities of the de Sitter precession, the Lense-Thirring precession, and the Thomas precession are able to be read out, and their expressions, in the form of geometric objects, are equivalent to their conventional ones in component form, respectively.

The whole derivation implies that the ``common'' even subalgebra of the STAs of signatures $(\pm,\mp,\mp,\mp)$ does provide a signature invariant GA framework for spacetime physics, and the rotors, presented in a signature invariant form, can be used to generate Lorentz transformations in these two STAs. The treatment of relativistic dynamics of a massive particle and gyroscopic precession intuitively displays the basic method of dealing with specific topics in curved spacetime within the signature invariant GA framework, which suggests that the GA techniques established in this paper are efficient and reliable. No doubt, if these techniques are directly applied to gyroscopic precession in alternate theories of gravity, such as $f(R)$ gravity~\cite{fRtheory,Wu:2021uws}, $f(R,\mathcal{G})$ gravity~\cite{fRGtheory}, and $f(X,Y,Z)$ gravity~\cite{Stabile:2010mz}, they will definitely facilitate the relevant studies, where $\mathcal{G}$ is the Gauss-Bonnet invariant, $X:=R$ is the Ricci scalar, $Y:=R_{\mu\nu}R^{\mu\nu}$ is the quadratic contraction of two Ricci tensors, and $Z:=R_{\mu\nu\sigma\rho}R^{\mu\nu\sigma\rho}$ is the quadratic contraction of two Riemann tensors. Furthermore, by developing other types of techniques, the method in this paper could also be applied to more fields, and in fact, some topics in classical mechanics and electrodynamics have been described in such a manner. The applications of this method will be expected to be extended to a wider range in the future, so that the study of spacetime physics in the language of GA could be greatly promoted.

This paper is organized as follows. In Sec.~\ref{Sec:second}, the STAs of signatures $(\pm,\mp,\mp,\mp)$ and their ``common'' even subalgebra are formulated. In Sec.~\ref{Sec:third}, rotor techniques on Lorentz transformation and relativistic dynamics of a massive particle in curved spacetime are constructed. In Sec.~\ref{Sec:fourth}, a GA approach to gyroscopic precession in the Lense-Thirring spacetime is given. In Sec.~\ref{Sec:fifth}, some concluding remarks will be made. In Appendix A, operation rules of blades in the STAs of signatures $(\pm,\mp,\mp,\mp)$ are summarized. In Appendix B, the ``common'' even subalgebra of the STAs of signatures $(\pm,\mp,\mp,\mp)$ is introduced in detail. In Appendix C, a local orthonormal tetrad $\{\gamma_{\alpha}\}$ and the bivector connection $\omega(u)$ associated with it in the Lense-Thirring spacetime are derived.
\section{STAs of signatures $(\pm,\mp,\mp,\mp)$ and their ``common'' even subalgebra~\label{Sec:second}}
STA, introduced in the classical literature \emph{Space-Time Algebra} by David Hestenes (1966), can provide a synthetic framework for relativistic physics~\cite{Doran2003}, so it has attracted widespread attention in the physical community. Since the establishment of STA, the signature $(+,-,-,-)$ has been widely used, and however, in relativistic physics, one of the main application fields of STA, the opposite signature $(-,+,+,+)$ is often adopted~\cite{Doran2003,Eric2010}. Thus, when one intends to apply STA to relativistic physics, the change of signature from one to another will cause inconvenience even though these two signatures differ only by a minus sign. In fact, the STA of signature $(-,+,+,+)$ was also used~\cite{Oppositesignature}, but a lack of long-term attention to it results in that the techniques related to relative vector and spacetime split have not been developed in this algebraic formalism so that its applications are quite limited. In this section, by following the original idea of David Hestenes, we will build up these techniques in the STA of signature $(-,+,+,+)$ so that a more convenient approach to relativistic physics could be given in the language of GA. For the ease of writing, we will directly formulate the STAs of signatures $(\pm,\mp,\mp,\mp)$, and analyze the operation rules of multivectors.

In spacetime, the STAs of signatures $(\pm,\mp,\mp,\mp)$ can be generated by corresponding orthogonal vectors $\{\gamma_{\alpha}^{\pm}\}$ satisfying
\begin{equation}\label{equ2.1}
\gamma_{\alpha}^{\pm}\cdot\gamma_{\beta}^{\pm}=\eta_{\alpha\beta}^{\pm}=\text{diag}(\pm,\mp,\mp,\mp),
\end{equation}
respectively, where $\eta_{\alpha\beta}^{\pm}$ are the Minkowski metrics in the two signatures. With these vector generators $\{\gamma_{\alpha}^{\pm}\}$, explicit bases for both the STAs are defined, namely
\begin{equation}\label{equ2.2}
\left\{1,\ \ \gamma_{\alpha}^{\pm},\ \ \gamma_{\mu}^{\pm}\wedge\gamma_{\nu}^{\pm}\ \left(\mu<\nu\right),\ \ \gamma_{\rho}^{\pm}\wedge\gamma_{\sigma}^{\pm}\wedge\gamma_{\lambda}^{\pm}\ \left(\rho<\sigma<\lambda\right),\ \ \gamma_{0}^{\pm}\wedge\gamma_{1}^{\pm}\wedge\gamma_{2}^{\pm}\wedge\gamma_{3}^{\pm}\right\},
\end{equation}
where, in either signature, one scalar, four vectors, six bivectors, four trivectors, and one pseudoscalar are contained.
One can perform operations between any two multivectors in spacetime by expanding them in a basis, once operation rules of blades of different grades are given, where the term ``blade'' here denotes a multivector written as the outer product of a set of vectors (cf.~Ref.~\cite{Doran2003}). In Appendix A of this paper, a detail list of operation rules of blades in the two STAs is presented, and based on these rules, the ``common'' even subalgebra of these two STAs will be constructed in the following.

According to Eqs.~(\ref{equA1}) and (\ref{equA7}), the orthogonality between the vector generators $\{\gamma_{\alpha}^{\pm}\}$ implies that the bases (\ref{equ2.2}) can be rewritten as
\begin{equation}\label{equ2.3}
\left\{1,\ \ \gamma_{\alpha}^{\pm},\quad \gamma_{\mu}^{\pm}\gamma_{\nu}^{\pm}\ \left(\mu<\nu\right),\ \ \gamma_{\rho}^{\pm}\gamma_{\sigma}^{\pm}\gamma_{\lambda}^{\pm}\ \left(\rho<\sigma<\lambda\right),\ \ I^{\pm}:=\gamma_{0}^{\pm}\gamma_{1}^{\pm}\gamma_{2}^{\pm}\gamma_{3}^{\pm}\right\},
\end{equation}
where the geometric products of $\{\gamma_{\alpha}^{\pm}\}$ are obviously anticommutative,
\begin{equation}\label{equ2.4}
\gamma_{\mu}^{\pm}\gamma_{\nu}^{\pm}=-\gamma_{\nu}^{\pm}\gamma_{\mu}^{\pm},\quad \left(\mu\neq\nu\right).
\end{equation}
By making use of the anticommutation of $\{\gamma_{\alpha}^{\pm}\}$, the pseudoscalars $I^{\pm}$ also have the expressions,
\begin{equation}\label{equ2.5}
I^{\pm}=\frac{1}{3!}\epsilon_{ijk}\gamma_{0}^{\pm}\gamma_{i}^{\pm}\gamma_{j}^{\pm}\gamma_{k}^{\pm}
\end{equation}
with $\epsilon_{ijk}$ as the three-dimensional Levi-Civit\`{a} symbol. Among the basis blades, those of even grade,
\begin{equation}\label{equ2.6}
\big\{1,\ \ \gamma_{0}^{\pm}\gamma_{k}^{\pm},\ \ \gamma_{i}^{\pm}\gamma_{j}^{\pm}\ \left(i<j\right),\ \ I^{\pm}\big\},
\end{equation}
form bases for the even subalgebras of the two STAs. Now, we will first discuss some properties of the bivectors $\{\gamma_{0}^{\pm}\gamma_{k}^{\pm}\}$.
With Eqs.~(\ref{equ2.1}), (\ref{equ2.4}), and (\ref{equA14}), one can directly derive the following equalities,
\begin{eqnarray}
\label{equ2.7}\left(\gamma_{0}^{\pm}\gamma_{i}^{\pm}\right)\cdot\big(\gamma_{0}^{\pm}\gamma_{j}^{\pm}\big)&=&\delta_{ij},\\
\label{equ2.8}\left(\gamma_{0}^{\pm}\gamma_{i}^{\pm}\right)\times\big(\gamma_{0}^{\pm}\gamma_{j}^{\pm}\big)&=&\mp\gamma_{i}^{\pm}\wedge\gamma_{j}^{\pm}=\mp\epsilon_{ijk}\left(\gamma_{0}^{\pm}\gamma_{k}^{\pm}\right)I^{\pm},\\
\label{equ2.9}\left(\gamma_{0}^{\pm}\gamma_{1}^{\pm}\right)\left(\gamma_{0}^{\pm}\gamma_{2}^{\pm}\right)\left(\gamma_{0}^{\pm}\gamma_{3}^{\pm}\right)&=&\mp I^{\pm},
\end{eqnarray}
where $\delta_{ij}$ is the Kronecker symbol, and in the second step of~(\ref{equ2.8}), Eqs.~(\ref{equ2.5}), (\ref{equA5}), and (\ref{equA10}) have been used. These equalities show that relative vectors, spanning the relative spaces orthogonal to the timelike vectors $\gamma_{0}^{\pm}$, could be defined as $\{\boldsymbol{\sigma}_{k}^{\pm}=\mp\gamma_{0}^{\pm}\gamma_{k}^{\pm}=\gamma_{k}^{\pm}\gamma^{0}_{\pm}\}$ with $\{\gamma^{\alpha}_{\pm}\}$ as the reciprocal frames of $\{\gamma_{\alpha}^{\pm}\}$, so that they have the similar algebraic properties to the Pauli matrices,
\begin{eqnarray}
\label{equ2.10}\boldsymbol{\sigma}_{i}^{\pm}\cdot\boldsymbol{\sigma}_{j}^{\pm}&=&\delta_{ij},\\
\label{equ2.11}\boldsymbol{\sigma}_{i}^{\pm}\times\boldsymbol{\sigma}_{j}^{\pm}&=&\epsilon_{ijk}\boldsymbol{\sigma}_{k}^{\pm}I^{\pm},\\
\label{equ2.12}\boldsymbol{\sigma}_{1}^{\pm}\boldsymbol{\sigma}_{2}^{\pm}\boldsymbol{\sigma}_{3}^{\pm}&=&I^{\pm}.
\end{eqnarray}
Clearly, $\{\boldsymbol{\sigma}_{k}^{+}=\gamma_{k}^{+}\gamma^{0}_{+}=\gamma_{k}^{+}\gamma_{0}^{+}\}$ is the frame of relative vectors introduced in the STA of signature $(+,-,-,-)$~\cite{Hestenes1966,Doran2003,Lasenby:2016lfl}, whereas $\{\boldsymbol{\sigma}_{k}^{-}=\gamma_{k}^{-}\gamma^{0}_{-}=-\gamma_{k}^{-}\gamma_{0}^{-}\}$ is the one in the STA of signature $(-,+,+,+)$. Further properties of $\{\boldsymbol{\sigma}_{k}^{\pm}\}$ can also be obtained. Eqs.~(\ref{equA19}) and (\ref{equA20}) yield
\begin{eqnarray}
\label{equ2.13}\gamma_{i}^{\pm}\cdot\gamma_{j}^{\pm}&=&\frac{\gamma_{i}^{\pm}\gamma_{j}^{\pm}+\gamma_{j}^{\pm}\gamma_{i}^{\pm}}{2},\\
\label{equ2.14}\gamma_{i}^{\pm}\wedge\gamma_{j}^{\pm}&=&\frac{\gamma_{i}^{\pm}\gamma_{j}^{\pm}-\gamma_{j}^{\pm}\gamma_{i}^{\pm}}{2},
\end{eqnarray}
and then, by inserting Eqs.~(\ref{equ2.13}) and (\ref{equ2.14}) into Eqs.~(\ref{equ2.7}) and (\ref{equ2.8}), respectively, we get
\begin{eqnarray}
\label{equ2.15}&&\boldsymbol{\sigma}_{i}^{\pm}\boldsymbol{\sigma}_{j}^{\pm}+\boldsymbol{\sigma}_{j}^{\pm}\boldsymbol{\sigma}_{i}^{\pm}=2\delta_{ij},\\
\label{equ2.16}&&\boldsymbol{\sigma}_{i}^{\pm}\boldsymbol{\sigma}_{j}^{\pm}-\boldsymbol{\sigma}_{j}^{\pm}\boldsymbol{\sigma}_{i}^{\pm}=2\epsilon_{ijk}\boldsymbol{\sigma}_{k}^{\pm}I^{\pm},\\
\label{equ2.17}&&\boldsymbol{\sigma}_{i}^{\pm}\boldsymbol{\sigma}_{j}^{\pm}=\delta_{ij}+\epsilon_{ijk}\boldsymbol{\sigma}_{k}^{\pm}I^{\pm},
\end{eqnarray}
which prove once again that the algebraic properties of $\{\boldsymbol{\sigma}_{k}^{\pm}\}$ are similar to those of the Pauli matrices. In fact, as mentioned in Ref.~\cite{Lasenby:2016lfl}, $\{\boldsymbol{\sigma}_{k}^{+}\}$ or $\{\boldsymbol{\sigma}_{k}^{-}\}$ provide a representation-free version of the Pauli matrices.

Eqs.~(\ref{equ2.10}) and (\ref{equ2.12}) show that the relative spaces orthogonal to $\gamma_{0}^{\pm}$ are both the Euclidean spaces of dimension 3 with $\{\boldsymbol{\sigma}_{k}^{\pm}\}$
and $I^{\pm}$ as orthonormal bases and pseudoscalars, respectively. In relative space, a relative vector, although being a bivector in STA, is actually treated as a multivector of grade 1, and thus, in this sense, the inner product and the cross product between two relative vectors can be defined.
Let $\boldsymbol{a}^{\pm}=a_{i}^{\pm}\boldsymbol{\sigma}_{i}^{\pm}$ and $\boldsymbol{b}^{\pm}=b_{j}^{\pm}\boldsymbol{\sigma}_{j}^{\pm}$ be relative vectors, and then, with the help of Eqs.~(\ref{equ2.10}) and (\ref{equ2.11}), the inner products and the cross products between $\boldsymbol{a}^{\pm}$ and $\boldsymbol{b}^{\pm}$ are defined as
\begin{eqnarray}
\label{equ2.18}\boldsymbol{a}^{\pm}\cdot\boldsymbol{b}^{\pm}&=&\left\langle\boldsymbol{a}^{\pm}\boldsymbol{b}^{\pm}\right\rangle=a_{k}^{\pm}b_{k}^{\pm},\\
\label{equ2.19}\boldsymbol{a}^{\pm}\times_{3}\boldsymbol{b}^{\pm}&=&-I^{\pm}\left(\boldsymbol{a}^{\pm}\times\boldsymbol{b}^{\pm}\right)=\epsilon_{ijk}a_{i}^{\pm}b_{j}^{\pm}\boldsymbol{\sigma}_{k}^{\pm},
\end{eqnarray}
where the commutator products between $\boldsymbol{a}^{\pm}$ and $\boldsymbol{b}^{\pm}$,
\begin{eqnarray}
\label{equ2.20}&&\boldsymbol{a}^{\pm}\times\boldsymbol{b}^{\pm}=\left\langle\boldsymbol{a}^{\pm}\boldsymbol{b}^{\pm}\right\rangle_{2}=a_{i}^{\pm}b_{j}^{\pm}\epsilon_{ijk}\boldsymbol{\sigma}_{k}^{\pm}I^{\pm}
\end{eqnarray}
and
\begin{eqnarray}
\label{equ2.21}\boldsymbol{a}^{\pm}I^{\pm}=I^{\pm}\boldsymbol{a}^{\pm},\quad \left(I^{\pm}\right)^2=I^{\pm}I^{\pm}=-1
\end{eqnarray}
have been used. Obviously, the above definitions of inner product and cross product are identical to their conventional ones, respectively. The cross products defined in Eqs.~(\ref{equ2.19}) determine the handedness of $\{\boldsymbol{\sigma}_{k}^{\pm}\}$, and by applying them, one easily gets
\begin{eqnarray}
\label{equ2.22}\boldsymbol{\sigma}_{i}^{\pm}\times_{3}\boldsymbol{\sigma}_{j}^{\pm}&=&\epsilon_{ijk}\boldsymbol{\sigma}_{k}^{\pm},
\end{eqnarray}
which clearly suggest that $\{\boldsymbol{\sigma}_{k}^{\pm}\}$ are both right-handed bases. Next, we will employ relative vectors to reconstruct bases of the even subalgebras of the STAs of signatures $(\pm,\mp,\mp,\mp)$. The definitions of $\{\boldsymbol{\sigma}_{k}^{\pm}\}$ provide
\begin{eqnarray}
\label{equ2.23}\gamma_{0}^{\pm}\gamma_{k}^{\pm}=\mp\boldsymbol{\sigma}_{k}^{\pm},
\end{eqnarray}
and then, by further using Eqs.~(\ref{equ2.1}) and (\ref{equ2.4}), there are
\begin{eqnarray}
\label{equ2.24}\gamma_{i}^{\pm}\gamma_{j}^{\pm}&=&\mp\boldsymbol{\sigma}_{i}^{\pm}\boldsymbol{\sigma}_{j}^{\pm}\quad (i<j).
\end{eqnarray}
After inserting Eqs.~(\ref{equ2.23}), (\ref{equ2.24}), and (\ref{equ2.12}) into~(\ref{equ2.6}), we know that bases of the even subalgebras of the two STAs can be reconstructed as
\begin{equation}\label{equ2.25}
\big\{1,\ \ \boldsymbol{\sigma}_{k}^{\pm},\ \ \boldsymbol{\sigma}_{i}^{\pm}\boldsymbol{\sigma}_{j}^{\pm}\ (i<j),\quad \boldsymbol{\sigma}_{1}^{\pm}\boldsymbol{\sigma}_{2}^{\pm}\boldsymbol{\sigma}_{3}^{\pm}\big\},
\end{equation}
which indicates that $\{\boldsymbol{\sigma}_{k}^{\pm}\}$ are actually the vector generators of the two subalgebras. Eqs.~(\ref{equ2.11}) and (\ref{equ2.17}) imply that equalities
\begin{eqnarray}
\label{equ2.26}\boldsymbol{\sigma}_{i}^{\pm}\boldsymbol{\sigma}_{j}^{\pm}=\boldsymbol{\sigma}_{i}^{\pm}\times\boldsymbol{\sigma}_{j}^{\pm}\quad (i\neq j)
\end{eqnarray}
hold, and thus, the anticommutation of $\{\boldsymbol{\sigma}_{k}^{\pm}\}$,
\begin{eqnarray}
\label{equ2.27}\boldsymbol{\sigma}_{i}^{\pm}\boldsymbol{\sigma}_{j}^{\pm}=-\boldsymbol{\sigma}_{j}^{\pm}\boldsymbol{\sigma}_{i}^{\pm}\quad (i\neq j),
\end{eqnarray}
is explicitly obtained. As a consequence, there exist three types of basic homogeneous multivectors (cf.~Ref.~\cite{Doran2003}) in the even subalgebras of the two STAs, namely,
\begin{eqnarray}
\label{equ2.28}\boldsymbol{a}^{\pm}&=&a_{i}^{\pm}\boldsymbol{\sigma}_{i}^{\pm},\\
\label{equ2.29}\boldsymbol{a}^{\pm}\times\boldsymbol{b}^{\pm}&=&a_{i}^{\pm}b_{j}^{\pm}\boldsymbol{\sigma}_{i}^{\pm}\times\boldsymbol{\sigma}_{j}^{\pm}=\sum_{i<j}\left(a_{i}^{\pm}b_{j}^{\pm}-a_{j}^{\pm}b_{i}^{\pm}\right)\boldsymbol{\sigma}_{i}^{\pm}\boldsymbol{\sigma}_{j}^{\pm},\\
\label{equ2.30}\left(\boldsymbol{a}^{\pm}\times\boldsymbol{b}^{\pm}\right)\wedge \boldsymbol{c}^{\pm}&=&\text{det}\left(\begin{array}{ccc}
a_{1}^{\pm},&\ b_{1}^{\pm},&\ c_{1}^{\pm}\\
a_{2}^{\pm},&\ b_{2}^{\pm},&\ c_{2}^{\pm}\\
a_{3}^{\pm},&\ b_{3}^{\pm},&\ c_{3}^{\pm}
\end{array}\right)\boldsymbol{\sigma}_{1}^{\pm}\boldsymbol{\sigma}_{2}^{\pm}\boldsymbol{\sigma}_{3}^{\pm}\quad \text{with}\quad \boldsymbol{c}^{\pm}=c_{k}^{\pm}\boldsymbol{\sigma}_{k}^{\pm}.
\end{eqnarray}
In view of~(\ref{equ2.12}), $\left(\boldsymbol{a}^{\pm}\times\boldsymbol{b}^{\pm}\right)\wedge \boldsymbol{c}^{\pm}$ in~(\ref{equ2.30}) are able to be written in the form of multiplications of the pseudoscalars $I^{\pm}$ by real numbers, and in fact, from the bases~(\ref{equ2.25}), all multivectors of grade 4 could be expressed in such a form. Thus, Eq.~(\ref{equA5}) states that the geometric product between any multivector and a pseudoscalar is equivalent to their inner product. Keep this conclusion in mind, and then, with the help of the following formulas,
\begin{eqnarray}
\label{equ2.31}\left(\boldsymbol{\sigma}_{i}^{\pm}\times\boldsymbol{\sigma}_{j}^{\pm}\right)I^{\pm}=-\epsilon_{ijk}\boldsymbol{\sigma}_{k}^{\pm}\quad\Leftrightarrow\quad\boldsymbol{\sigma}_{k}^{\pm}I^{\pm}=\frac{1}{2}\epsilon_{kij}\left(\boldsymbol{\sigma}_{i}^{\pm}\times\boldsymbol{\sigma}_{j}^{\pm}\right),
\end{eqnarray}
one gets a convenient way to carry out operations involving multivectors of grade 4, where in the derivation of~(\ref{equ2.31}), Eqs.~(\ref{equ2.20}) and (\ref{equ2.21}) have been used. Eqs.~(\ref{equ2.23}) and (\ref{equ2.8}) show that both $\boldsymbol{\sigma}_{k}^{\pm}$ and $\boldsymbol{\sigma}_{i}^{\pm}\times\boldsymbol{\sigma}_{j}^{\pm}\ (i\neq j)$ are bivectors in the two STAs, where the former contain timelike components, whereas the latter do not. The geometric products of them also need to be derived, where according to Eq.~(\ref{equA2}), we have
\begin{eqnarray}
\label{equ2.32}\boldsymbol{\sigma}_{i}^{\pm}\boldsymbol{\sigma}_{j}^{\pm}&=&\boldsymbol{\sigma}_{i}^{\pm}\cdot\boldsymbol{\sigma}_{j}^{\pm}+\boldsymbol{\sigma}_{i}^{\pm}\times\boldsymbol{\sigma}_{j}^{\pm},\\
\label{equ2.33}\boldsymbol{\sigma}_{k}^{\pm}\left(\boldsymbol{\sigma}_{i}^{\pm}\times\boldsymbol{\sigma}_{j}^{\pm}\right)&=&\boldsymbol{\sigma}_{k}^{\pm}\times\left(\boldsymbol{\sigma}_{i}^{\pm}\times\boldsymbol{\sigma}_{j}^{\pm}\right)+\boldsymbol{\sigma}_{k}^{\pm}\wedge\left(\boldsymbol{\sigma}_{i}^{\pm}\times\boldsymbol{\sigma}_{j}^{\pm}\right),\\
\label{equ2.34}\left(\boldsymbol{\sigma}_{i}^{\pm}\times\boldsymbol{\sigma}_{j}^{\pm}\right)\boldsymbol{\sigma}_{k}^{\pm}&=&\left(\boldsymbol{\sigma}_{i}^{\pm}\times\boldsymbol{\sigma}_{j}^{\pm}\right)\times\boldsymbol{\sigma}_{k}^{\pm}+\left(\boldsymbol{\sigma}_{i}^{\pm}\times\boldsymbol{\sigma}_{j}^{\pm}\right)\wedge\boldsymbol{\sigma}_{k}^{\pm},\\
\label{equ2.35}\big(\boldsymbol{\sigma}_{i}^{\pm}\times\boldsymbol{\sigma}_{j}^{\pm}\big)\big(\boldsymbol{\sigma}_{p}^{\pm}\times\boldsymbol{\sigma}_{q}^{\pm}\big)&=&\big(\boldsymbol{\sigma}_{i}^{\pm}\times\boldsymbol{\sigma}_{j}^{\pm}\big)\cdot\big(\boldsymbol{\sigma}_{p}^{\pm}\times\boldsymbol{\sigma}_{q}^{\pm}\big)+\big(\boldsymbol{\sigma}_{i}^{\pm}\times\boldsymbol{\sigma}_{j}^{\pm}\big)\times\big(\boldsymbol{\sigma}_{p}^{\pm}\times\boldsymbol{\sigma}_{q}^{\pm}\big).
\end{eqnarray}
By further using Eqs.~(\ref{equ2.1}), (\ref{equ2.4}), (\ref{equA8}), and (\ref{equA15}), the terms on the right-hand sides of Eqs.~(\ref{equ2.33}), (\ref{equ2.34}), and (\ref{equ2.35}) are achieved,
\begin{eqnarray}
\label{equ2.36}\boldsymbol{\sigma}_{k}^{\pm}\times\big(\boldsymbol{\sigma}_{i}^{\pm}\times\boldsymbol{\sigma}_{j}^{\pm}\big)&=&-\big(\boldsymbol{\sigma}_{i}^{\pm}\times\boldsymbol{\sigma}_{j}^{\pm}\big)\times\boldsymbol{\sigma}_{k}^{\pm}=\big(\boldsymbol{\sigma}_{k}^{\pm}\cdot\boldsymbol{\sigma}_{i}^{\pm}\big)\boldsymbol{\sigma}_{j}^{\pm}-\big(\boldsymbol{\sigma}_{k}^{\pm}\cdot\boldsymbol{\sigma}_{j}^{\pm}\big)\boldsymbol{\sigma}_{i}^{\pm},\\
\label{equ2.37}\boldsymbol{\sigma}_{k}^{\pm}\wedge\left(\boldsymbol{\sigma}_{i}^{\pm}\times\boldsymbol{\sigma}_{j}^{\pm}\right)&=&\left(\boldsymbol{\sigma}_{i}^{\pm}\times\boldsymbol{\sigma}_{j}^{\pm}\right)\wedge\boldsymbol{\sigma}_{k}^{\pm}=\boldsymbol{\sigma}_{i}^{\pm}\wedge\left(\boldsymbol{\sigma}_{j}^{\pm}\times\boldsymbol{\sigma}_{k}^{\pm}\right)\nonumber\\
&=&\boldsymbol{\sigma}_{j}^{\pm}\wedge\left(\boldsymbol{\sigma}_{k}^{\pm}\times\boldsymbol{\sigma}_{i}^{\pm}\right),\\
\label{equ2.38}\left(\boldsymbol{\sigma}_{i}^{\pm}\times\boldsymbol{\sigma}_{j}^{\pm}\right)\cdot\big(\boldsymbol{\sigma}_{p}^{\pm}\times\boldsymbol{\sigma}_{q}^{\pm}\big)&=&\big(\boldsymbol{\sigma}_{j}^{\pm}\cdot\boldsymbol{\sigma}_{p}^{\pm}\big)\big(\boldsymbol{\sigma}_{i}^{\pm}\cdot\boldsymbol{\sigma}_{q}^{\pm}\big)-\big(\boldsymbol{\sigma}_{i}^{\pm}\cdot\boldsymbol{\sigma}_{p}^{\pm}\big)\big(\boldsymbol{\sigma}_{j}^{\pm}\cdot\boldsymbol{\sigma}_{q}^{\pm}\big),\\
\label{equ2.39}\left(\boldsymbol{\sigma}_{i}^{\pm}\times\boldsymbol{\sigma}_{j}^{\pm}\right)\times\big(\boldsymbol{\sigma}_{p}^{\pm}\times\boldsymbol{\sigma}_{q}^{\pm}\big)&=&\big(\boldsymbol{\sigma}_{j}^{\pm}\cdot\boldsymbol{\sigma}_{p}^{\pm}\big)\big(\boldsymbol{\sigma}_{i}^{\pm}\times\boldsymbol{\sigma}_{q}^{\pm}\big)+\big(\boldsymbol{\sigma}_{i}^{\pm}\cdot\boldsymbol{\sigma}_{q}^{\pm}\big)\big(\boldsymbol{\sigma}_{j}^{\pm}\times\boldsymbol{\sigma}_{p}^{\pm}\big)\nonumber\\
&-&\big(\boldsymbol{\sigma}_{i}^{\pm}\cdot\boldsymbol{\sigma}_{p}^{\pm}\big)\big(\boldsymbol{\sigma}_{j}^{\pm}\times\boldsymbol{\sigma}_{q}^{\pm}\big)-\big(\boldsymbol{\sigma}_{j}^{\pm}\cdot\boldsymbol{\sigma}_{q}^{\pm}\big)\big(\boldsymbol{\sigma}_{i}^{\pm}\times\boldsymbol{\sigma}_{p}^{\pm}\big).
\end{eqnarray}
With the aid of the above operation rules of the basic homogeneous multivectors, namely Eqs.~(\ref{equ2.31})---(\ref{equ2.39}), one can carry out operations of any two multivectors in the even subalgebras of the STAs of signatures $(\pm,\mp,\mp,\mp)$. Evidently, as shown in these formulas, the two even subalgebras share the same operation rules, and thus, when dealing with specific problems, such as relativistic dynamics of a massive particle and gyroscopic precession in the next two sections, we will no longer distinguish them strictly and treat them as one algebraic formalism. In Appendix B of the present paper, a detailed presentation of this ``common'' even subalgebra of the two STAs is given. It will be shown that this algebraic formalism provides a signature invariant GA framework for spacetime physics.

When STA is used to describe relativistic physics, the techniques on spacetime split are also of significance, where in the STA of signature $(+,-,-,-)$, these techniques provide an extremely efficient tool for comparing physical effects in different frames~\cite{Hestenes1966,Doran2003}. Of course, these techniques can also be constructed in the STA of signature $(-,+,+,+)$. Let $b^{\pm}=b^{\alpha}_{\pm}\gamma_{\alpha}^{\pm}$ be vectors in spacetime, and the spacetime splits of $b^{\pm}$ with $\gamma_{0}^{\pm}$ are defined as
\begin{eqnarray}
\label{equ2.40}b^{\pm}\gamma^{0}_{\pm}=b^{0}_{\pm}+\boldsymbol{b}^{\pm},
\end{eqnarray}
where $\boldsymbol{b}^{\pm}=b^{i}_{\pm}\boldsymbol{\sigma}_{i}^{\pm}$ are called the relative vectors of $b^{\pm}$. Besides, as for operators $\partial^{\pm}:=\gamma^{\alpha}_{\pm}\partial_{\alpha}$, their spacetime splits with $\gamma_{0}^{\pm}$ are given by
\begin{eqnarray}
\label{equ2.41}\gamma_{0}^{\pm}\partial^{\pm}=\partial_{0}+\boldsymbol{\nabla}^{\pm},
\end{eqnarray}
where $\partial_{\mu}:=\partial/\partial x^{\mu}$ and $\boldsymbol{\nabla}^{\pm}:=\boldsymbol{\sigma}^{k}_{\pm}\partial_{k}$ with $x^{\mu}$ and $\{\boldsymbol{\sigma}^{k}_{\pm}:=\gamma_{0}^{\pm}\gamma^{k}_{\pm}\}$ as coordinates in spacetime and the reciprocal frames of $\{\boldsymbol{\sigma}_{k}^{\pm}\}$ in the relative spaces, respectively. As clearly shown, the spacetime splits of $b^{+}$ and $\partial^{+}$ are indeed the same as those introduced in the STA of signature $(+,-,-,-)$~\cite{Hestenes1966,Doran2003}, and the spacetime splits of $b^{-}$ and $\partial^{-}$ are those defined in the STA of signature $(-,+,+,+)$.

The timelike vectors $c\gamma_{0}^{\pm}$ could be recognized as the four-velocities of some observer, so the spacetime split introduced above is observer dependent, and consequently, one of the most powerful applications of the techniques on spacetime split is that they can greatly simplify the study of effects involving different observers~\cite{Hestenes1966,Doran2003}. Technically, spacetime split actually encodes the crucial geometric relationship between STA and its even subalgebra~\cite{Hestenes1966}, where with these techniques, many calculations between vectors in spacetime are able to be transformed into those in the even subalgebra of STA. As a result, based on various operations in this algebraic formalism, a large number of specific problems could be solved efficiently. Moreover, since the even subalgebras of the STAs of signatures $(\pm,\mp,\mp,\mp)$ share the same operation rules, by resorting to the techniques on spacetime split, one is capable of managing to acquire a signature invariant approach to these problems. We will see that the above advantages of spacetime split play a key role in the following treatment of relevant topics.
\section{Rotor techniques on Lorentz transformation and relativistic dynamics of a massive particle in curved spacetime~\label{Sec:third}}
It is well known that one of the remarkable advantages of STA is that Lorentz boost and spatial rotation can be handled with rotor techniques in an elegant and highly condensed manner~\cite{Doran2003,Rotortechnology,Sabbata2006}. As shown in classical literatures~\cite{MTW1973,Ignazio1995}, a knowledge of Lorentz boost and spatial rotation is heavily involved in the description of gyroscopic precession, and hence, it could be expected that a more efficient approach to dealing with this topic will be found in the language of STA. Besides, in Sec.~\ref{Sec:second}, it is claimed that the ``common'' even subalgebra of the STAs of signatures $(\pm,\mp,\mp,\mp)$ provides a signature invariant GA framework for spacetime physics, and thus, when this framework is applied to gyroscopic precession, a signature invariant GA derivation of the precessional angular velocity of the gyroscope spin could be achieved. Therefore, as a preliminary attempt, making use of the two STAs and their ``common'' even subalgebra to study gyroscopic precession is one objective of the present paper, which, if successful, will definitely become an application paradigm of STA. In view that many relevant techniques need to be constructed in this section, the detailed treatment of gyroscopic precession will be left to the next section.

In the analyse of gyroscopic precession, rotor techniques on Lorentz boost and spatial rotation are widely used, and therefore, these techniques need to be specifically addressed in the two STAs. Rotor techniques are available in the STA of signature $(+,-,-,-)$~\cite{Doran2003,Rotortechnology,Sabbata2006}, and however, since the STA of signature $(-,+,+,+)$ is rarely employed, these techniques have not been fully developed in this algebraic formalism, where in particular the expressions of the rotors inducing Lorentz boost and spatial rotation should be clearly established. In this section, by constructing the rotors on the basis of the exponential function defined on the ``common'' even subalgebra of the two STAs, the general Lorentz boost with velocity in an arbitrary direction and the general spatial rotation in an arbitrary plane are handled in a signature invariant manner. In addition, relativistic dynamics of a massive particle in curved spacetime ought be studied so as to describe the motion of a gyroscope moving around a gravitating source~\cite{Weinberg2014}. To this end, for a massive particle, the spacetime splits of the velocity, acceleration, momentum, and force four-vectors with the normalized four-velocity
of the fiducial observer, at rest in the coordinate system of the spacetime metric, are first derived, and then with these results, a three-dimensional analogue of Newton's second law for this particle in curved spacetime is achieved. Furthermore, in order to describe the motion of the spin of a gyroscope, the Fermi-Walker derivative in the STA of signature $(-,+,+,+)$ is also constructed by following the way in the $(+,-,-,-)$ signature.

In Appendix B of this paper, the signs ``$\pm$'' associated with multivectors have been omitted in the ``common'' even subalgebra of the STAs of signatures $(\pm,\mp,\mp,\mp)$, so that all the formulas in this algebraic formalism are presented in a neat form. Inspired by this, when formulas in the two STAs are involved hereafter, the following convention will be adopted for brevity: \emph{The signs ``$\pm$'' associated with multivectors and operators are suppressed, and for equalities like $A=F(\pm B)$ and $C=G(\mp D)$, the signs ``$+$'' and ``$-$'' in the former equation correspond to the cases in the signatures $(+,-,-,-)$ and $(-,+,+,+)$, respectively, and the situation in the latter equation is reverse.}
\subsection{Rotor techniques on Lorentz boost and spatial rotation~\label{Sec:3.1}}
In GA, a rotor $R$ is defined as an even multivector satisfying $R\tilde{R}=1$ and the property that the map defined by $b\mapsto Rb\tilde{R}$ transforms any vector into another one~\cite{Doran2003}. Rotors encode an important geometric object and can provide a more elegant scheme for performing orthogonal transformations in spaces of arbitrary signature, where mathematically, rotor group, formed by the set of rotors, provides a double-cover representation of the connected subgroup of the special orthogonal group. In the present paper, we are only interested in rotors in spacetime, and in such a case, the rotor group in spacetime is a representation of the group of proper orthochronous Lorentz transformations~\cite{Doran2003}.

In the STA of signature $(+,-,-,-)$, rotor techniques on Lorentz boost and spatial rotation have been established~\cite{Doran2003,Rotortechnology,Sabbata2006}, which greatly promotes the application of STA in spacetime physics. Of course, in order to complete the necessary discussion on gyroscopic precession in a signature invariant manner, these techniques also need to be explicitly constructed in the STA of signature $(-,+,+,+)$. To facilitate the writing, as in Sec.~\ref{Sec:second}, we will directly build up rotor techniques in the STAs of signatures $(\pm,\mp,\mp,\mp)$.

In Appendix B of this paper, a simple method to construct rotor is presented, and it has been shown that for a real number $\alpha$ and a unit 2-blade $B$, $\text{e}^{\alpha B}$
is a rotor. Here, we will make use of $\text{e}^{\alpha B}$ to handle Lorentz boost and spatial rotation in the two STAs. From Eqs.~(\ref{equ2.8}) and (\ref{equ2.23}), $\boldsymbol{\sigma}_{k}$ and $\boldsymbol{\sigma}_{i}\times\boldsymbol{\sigma}_{j}$ have the forms
\begin{subequations}
\begin{eqnarray}
\label{equ3.1a}\boldsymbol{\sigma}_{k}&=&\mp\gamma_{0}\wedge\gamma_{k},\\
\label{equ3.1b}\boldsymbol{\sigma}_{i}\times\boldsymbol{\sigma}_{j}&=&\mp\gamma_{i}\wedge\gamma_{j},
\end{eqnarray}
\end{subequations}
and their squares are deduced by applying Eqs.~(\ref{equB3}) and (\ref{equB17}),
\begin{subequations}
\begin{eqnarray}
\label{equ3.2a}\left(\boldsymbol{\sigma}_{k}\right)^2&=&1,\\
\label{equ3.2b}\left(\boldsymbol{\sigma}_{i}\times\boldsymbol{\sigma}_{j}\right)^2&=&-1\quad (i\neq j).
\end{eqnarray}
\end{subequations}
Clearly, both $\boldsymbol{\sigma}_{k}$ and $\boldsymbol{\sigma}_{i}\times\boldsymbol{\sigma}_{j}\ (i\neq j)$ are unit 2-blades, and the signs of
their squares are different, which suggests that there are two types of unit 2-blades in spacetime. It is based on the exponential functions of these two types of unit 2-blades that
the rotors inducing Lorentz boost and spatial rotation can be constructed. Let $\boldsymbol{v}=v^{k}\boldsymbol{\sigma}_{k}, \boldsymbol{m}=m^{i}\boldsymbol{\sigma}_{i}$, and $\boldsymbol{n}=n^{j}\boldsymbol{\sigma}_{j}$ be three arbitrary relative vectors. Consider the bivectors $\boldsymbol{v}$ and $\boldsymbol{m}\times\boldsymbol{n}$, and the following results can be easily given by means of Eqs.~(\ref{equ3.1a})---(\ref{equ3.2b}):
\begin{subequations}
\begin{eqnarray}
\label{equ3.3a}\boldsymbol{v}&=&\mp \gamma_{0}\wedge\big(v^{k}\gamma_{k}\big),\\
\label{equ3.3b}\boldsymbol{m}\times\boldsymbol{n}&=&\mp\big(m^{i}\gamma_{i}\big)\wedge\big(n^{j}\gamma_{j}\big)
\end{eqnarray}
\end{subequations}
and
\begin{subequations}
\begin{eqnarray}
\label{equ3.4a}\boldsymbol{v}^2&=&v^{k}v^{k},\\
\label{equ3.4b}(\boldsymbol{m}\times\boldsymbol{n})^2&=&-\sum_{i<j}\left(m^{i}n^{j}-m^{j}n^{i}\right)^2.
\end{eqnarray}
\end{subequations}
The former two equations indicate that both $\boldsymbol{v}$ and $\boldsymbol{m}\times\boldsymbol{n}$ are 2-blades, and thus, with the latter two equations, two unit 2-blades are derived,
\begin{subequations}
\begin{eqnarray}
\label{equ3.5a}\boldsymbol{e}_{v}&:=&\frac{\boldsymbol{v}}{\sqrt{\boldsymbol{v}^2}},\\
\label{equ3.5b}\boldsymbol{I}_{2}&:=&\frac{\boldsymbol{m}\times\boldsymbol{n}}{\sqrt{-(\boldsymbol{m}\times\boldsymbol{n})^2}},
\end{eqnarray}
\end{subequations}
where a direct calculation verifies that
\begin{subequations}
\begin{eqnarray}
\label{equ3.6a}\left(\boldsymbol{e}_{v}\right)^2&=&1,\\
\label{equ3.6b}\left(\boldsymbol{I}_{2}\right)^2&=&-1.
\end{eqnarray}
\end{subequations}

According to Ref.~\cite{Doran2003}, a proper orthochronous Lorentz transformation can be generated by a rotor $R$ in spactime, and under this transformation, a general multivector $M$ will be transformed double-sidedly as $M\mapsto R^{-1}MR$. Let $\theta$ and $\varphi$ be two real numbers, and the corresponding rotors associated with $\boldsymbol{e}_{v}$ and $\boldsymbol{I}_{2}$ are constructed as
$\text{e}^{\frac{\theta}{2}\boldsymbol{e}_{v}}$ and $\text{e}^{\frac{\varphi}{2}\boldsymbol{I}_{2}}$, respectively. When they act on vectors $x$ and $y$, two new vectors $x'$ and $y'$ are obtained,
\begin{subequations}
\begin{eqnarray}
\label{equ3.7a}x'&=&\text{e}^{-\frac{\theta}{2}\boldsymbol{e}_{v}}x\,\text{e}^{\frac{\theta}{2}\boldsymbol{e}_{v}},\\ \label{equ3.7b}y'&=&\text{e}^{-\frac{\varphi}{2}\boldsymbol{I}_{2}}y\,\text{e}^{\frac{\varphi}{2}\boldsymbol{I}_{2}}.
\end{eqnarray}
\end{subequations}
In order to analyze the generated Lorentz transformations in the ``common'' even subalgebra of the two STAs, the techniques on spacetime split need to be applied. From Eqs.~(\ref{equ3.3a})---(\ref{equ3.5b}) and (\ref{equA1}), the orthogonality and anticommutation of $\{\gamma_{\alpha}\}$ imply
\begin{subequations}
\begin{eqnarray}
\label{equ3.8a}\gamma_{0}\boldsymbol{e}_{v}&=&-\boldsymbol{e}_{v}\gamma_{0},\\
\label{equ3.8b}\gamma_{0}\boldsymbol{I}_{2}&=&\boldsymbol{I}_{2}\gamma_{0},
\end{eqnarray}
\end{subequations}
and then, with the help of Eq.~(\ref{equB39}), one gets
\begin{subequations}
\begin{eqnarray}
\label{equ3.9a}x'\gamma^{0}&=&\text{e}^{-\frac{\theta}{2}\boldsymbol{e}_{v}}x\gamma^{0}\text{e}^{-\frac{\theta}{2}\boldsymbol{e}_{v}},\\
\label{equ3.9b}y'\gamma^{0}&=&\text{e}^{-\frac{\varphi}{2}\boldsymbol{I}_{2}}y\gamma^{0}\text{e}^{\frac{\varphi}{2}\boldsymbol{I}_{2}}.
\end{eqnarray}
\end{subequations}
The spacetime splits of $x,y,x',$ and $y'$ with $\gamma_{0}$ are provided by applying Eq.~(\ref{equ2.40}),
\begin{subequations}
\begin{eqnarray}
\label{equ3.10a}x\gamma^{0}&=&x^{0}+\boldsymbol{x},\\
\label{equ3.10b}x'\gamma^{0}&=&x'^{0}+\boldsymbol{x}',\\
\label{equ3.10c}y\gamma^{0}&=&y^{0}+\boldsymbol{y},\\
\label{equ3.10d}y'\gamma^{0}&=&y'^{0}+\boldsymbol{y}',
\end{eqnarray}
\end{subequations}
and substituting them in Eqs.~(\ref{equ3.9a}) and (\ref{equ3.9b}), Eqs.~(\ref{equ3.7a}) and (\ref{equ3.7b}) are recast in a signature invariant form,
\begin{subequations}
\begin{eqnarray}
\label{equ3.11a}x'^{0}+\boldsymbol{x}'&=&\text{e}^{-\frac{\theta}{2}\boldsymbol{e}_{v}}\left(x^{0}+\boldsymbol{x}\right)\text{e}^{-\frac{\theta}{2}\boldsymbol{e}_{v}},\\
\label{equ3.11b}y'^{0}+\boldsymbol{y}'&=&\text{e}^{-\frac{\varphi}{2}\boldsymbol{I}_{2}}\left(y^{0}+\boldsymbol{y}\right)\text{e}^{\frac{\varphi}{2}\boldsymbol{I}_{2}}.
\end{eqnarray}
\end{subequations}
For the relative vectors $\boldsymbol{x},\boldsymbol{x}',\boldsymbol{y}$, and $\boldsymbol{y}'$, Eqs.~(\ref{equ3.5b}), (\ref{equB8}), and (\ref{equB9}) yield the decompositions,
\begin{subequations}
\begin{eqnarray}
\label{equ3.12a}\boldsymbol{x}&=&\big(\boldsymbol{x}\cdot\boldsymbol{e}_{v}\big)\boldsymbol{e}_{v}+\big(\boldsymbol{x}\times\boldsymbol{e}_{v}\big)\boldsymbol{e}_{v},\\
\label{equ3.12b}\boldsymbol{x}'&=&\big(\boldsymbol{x}'\cdot\boldsymbol{e}_{v}\big)\boldsymbol{e}_{v}+\big(\boldsymbol{x}'\times\boldsymbol{e}_{v}\big)\boldsymbol{e}_{v},\\
\label{equ3.12c}\boldsymbol{y}&=&\big(\boldsymbol{y}\times\boldsymbol{I}_{2}\big)\boldsymbol{I}_{2}^{-1}+\big(\boldsymbol{y}\wedge\boldsymbol{I}_{2}\big)\boldsymbol{I}_{2}^{-1},\\
\label{equ3.12d}\boldsymbol{y}'&=&\big(\boldsymbol{y}'\times\boldsymbol{I}_{2}\big)\boldsymbol{I}_{2}^{-1}+\big(\boldsymbol{y}'\wedge\boldsymbol{I}_{2}\big)\boldsymbol{I}_{2}^{-1}.
\end{eqnarray}
\end{subequations}
Since one can directly check that
\begin{subequations}
\begin{eqnarray}
\label{equ3.13a}\big(\boldsymbol{x}\times\boldsymbol{e}_{v}\big)\wedge\boldsymbol{e}_{v}&=&0,\\
\label{equ3.13b}\big(\boldsymbol{x}'\times\boldsymbol{e}_{v}\big)\wedge\boldsymbol{e}_{v}&=&0,\\
\label{equ3.13c}\big(\boldsymbol{y}\times\boldsymbol{I}_{2}\big)\wedge\boldsymbol{I}_{2}^{-1}&=&0,\\
\label{equ3.13d}\big(\boldsymbol{y}'\times\boldsymbol{I}_{2}\big)\wedge\boldsymbol{I}_{2}^{-1}&=&0
\end{eqnarray}
\end{subequations}
by virtue of Eqs.~(\ref{equ3.5b}), (\ref{equ3.6b}), (\ref{equB15}), and (\ref{equB16}), the following relative vectors can be defined with Eqs.~(\ref{equB8})---(\ref{equB10}) and (\ref{equB13}):
\begin{subequations}
\begin{eqnarray}
\label{equ3.14a}\boldsymbol{x}_{\parallel}:=&\big(\boldsymbol{x}\cdot\boldsymbol{e}_{v}\big)\boldsymbol{e}_{v},\qquad\qquad\qquad\qquad\qquad\ \ \boldsymbol{x}_{\perp}&:=\big(\boldsymbol{x}\times\boldsymbol{e}_{v}\big)\times\boldsymbol{e}_{v}=\big(\boldsymbol{x}\times\boldsymbol{e}_{v}\big)\boldsymbol{e}_{v},\\
\label{equ3.14b}\boldsymbol{x}'_{\parallel}:=&\big(\boldsymbol{x}'\cdot\boldsymbol{e}_{v}\big)\boldsymbol{e}_{v},\qquad\qquad\qquad\qquad\qquad\ \boldsymbol{x}'_{\perp}&:=\big(\boldsymbol{x}'\times\boldsymbol{e}_{v}\big)\times\boldsymbol{e}_{v}=\big(\boldsymbol{x}'\times\boldsymbol{e}_{v}\big)\boldsymbol{e}_{v},\\
\label{equ3.14c}\boldsymbol{y}_{\parallel}:=&\big(\boldsymbol{y}\times\boldsymbol{I}_{2}\big)\times\boldsymbol{I}_{2}^{-1}=\big(\boldsymbol{y}\times\boldsymbol{I}_{2}\big)\boldsymbol{I}_{2}^{-1},\quad\ \ \, \boldsymbol{y}_{\perp}&:=\big(\boldsymbol{y}\wedge\boldsymbol{I}_{2}\big)\cdot\boldsymbol{I}_{2}^{-1}=\big(\boldsymbol{y}\wedge\boldsymbol{I}_{2}\big)\boldsymbol{I}_{2}^{-1},\\
\label{equ3.14d}\boldsymbol{y}'_{\parallel}:=&\big(\boldsymbol{y}'\times\boldsymbol{I}_{2}\big)\times\boldsymbol{I}_{2}^{-1}=\big(\boldsymbol{y}'\times\boldsymbol{I}_{2}\big)\boldsymbol{I}_{2}^{-1},\quad \ \boldsymbol{y}'_{\perp}&:=\big(\boldsymbol{y}'\wedge\boldsymbol{I}_{2}\big)\cdot\boldsymbol{I}_{2}^{-1}=\big(\boldsymbol{y}'\wedge\boldsymbol{I}_{2}\big)\boldsymbol{I}_{2}^{-1}
\end{eqnarray}
\end{subequations}
with
\begin{subequations}
\begin{eqnarray}
\label{equ3.15a}\boldsymbol{x}&=&\boldsymbol{x}_{\parallel}+\boldsymbol{x}_{\perp},\\
\label{equ3.15b}\boldsymbol{x}'&=&\boldsymbol{x}'_{\parallel}+\boldsymbol{x}'_{\perp},\\
\label{equ3.15c}\boldsymbol{y}&=&\boldsymbol{y}_{\parallel}+\boldsymbol{y}_{\perp},\\
\label{equ3.15d}\boldsymbol{y}'&=&\boldsymbol{y}'_{\parallel}+\boldsymbol{y}'_{\perp}
\end{eqnarray}
\end{subequations}
and
\begin{subequations}
\begin{eqnarray}
\label{equ3.16a}\boldsymbol{x}_{\parallel}\times\boldsymbol{e}_{v}=&\big\langle\big(\boldsymbol{x}\cdot\boldsymbol{e}_{v}\big)\boldsymbol{e}_{v}\boldsymbol{e}_{v}\big\rangle_{2}=0,\qquad\quad\
\boldsymbol{x}_{\perp}\cdot\boldsymbol{e}_{v}&=\big\langle\big(\boldsymbol{x}\times\boldsymbol{e}_{v}\big)\boldsymbol{e}_{v}\boldsymbol{e}_{v}\big\rangle_{0}=0,\\
\label{equ3.16b}\boldsymbol{x}'_{\parallel}\times\boldsymbol{e}_{v}=&\big\langle\big(\boldsymbol{x}'\cdot\boldsymbol{e}_{v}\big)\boldsymbol{e}_{v}\boldsymbol{e}_{v}\big\rangle_{2}=0,\qquad\quad
\boldsymbol{x}'_{\perp}\cdot\boldsymbol{e}_{v}&=\big\langle\big(\boldsymbol{x}'\times\boldsymbol{e}_{v}\big)\boldsymbol{e}_{v}\boldsymbol{e}_{v}\big\rangle_{0}=0,\\
\label{equ3.16c}\boldsymbol{y}_{\parallel}\wedge\boldsymbol{I}_{2}=&\big\langle\big(\boldsymbol{y}\times\boldsymbol{I}_{2}\big)\boldsymbol{I}_{2}^{-1}\boldsymbol{I}_{2}\big\rangle_{4}=0,\qquad\
\boldsymbol{y}_{\perp}\times\boldsymbol{I}_{2}&=\big\langle\big(\boldsymbol{y}\wedge\boldsymbol{I}_{2}\big)\boldsymbol{I}_{2}^{-1}\boldsymbol{I}_{2}\big\rangle_{2}=0,\\
\label{equ3.16d}\boldsymbol{y}'_{\parallel}\wedge\boldsymbol{I}_{2}=&\big\langle\big(\boldsymbol{y}'\times\boldsymbol{I}_{2}\big)\boldsymbol{I}_{2}^{-1}\boldsymbol{I}_{2}\big\rangle_{4}=0,\qquad
\boldsymbol{y}'_{\perp}\times\boldsymbol{I}_{2}&=\big\langle\big(\boldsymbol{y}'\wedge\boldsymbol{I}_{2}\big)\boldsymbol{I}_{2}^{-1}\boldsymbol{I}_{2}\big\rangle_{2}=0.
\end{eqnarray}
\end{subequations}

As stated in Sec.~\ref{Sec:second}, the relative space is an Euclidean space of dimension 3, and a relative vector, although being a spacetime bivector, could be
treated as a multivector of grade 1, which implies that in terms of the three-dimensional geometric meaning, a relative vector is just a vector~\cite{Doran2003}. Similarly, the commutator product of two relative vectors, referred to as the relative bivector, also has three-dimensional geometric meaning. After comparing  Eq.~(\ref{equB8}) with Eq.~(\ref{equA1}), one is able to find that the commutator product of two relative vectors serves as the role of the wedge product of two vectors in general finite dimensional GA, and thus, in the three-dimensional relative space, it encodes an oriented plane~\cite{Hestenes1984,Doran2003}. In such a sense, Eqs.~(\ref{equ3.16a}) and (\ref{equ3.16b}) indicate that $\boldsymbol{x}_{\parallel}\,(\boldsymbol{x}'_{\parallel})$ and $\boldsymbol{x}_{\perp}\,(\boldsymbol{x}'_{\perp})$ are, respectively, the components of $\boldsymbol{x}\,(\boldsymbol{x}')$ parallel and perpendicular to $\boldsymbol{e}_{v}$.

Of course, $\boldsymbol{I}_{2}$ also defines an oriented plane in the relative space. Let
\begin{eqnarray}
\label{equ3.17}\boldsymbol{l}&:=&\frac{\boldsymbol{m}\times_{3}\boldsymbol{n}}{\sqrt{(\boldsymbol{m}\times_{3}\boldsymbol{n})^2}},
\end{eqnarray}
and then, together with Eqs.~(\ref{equ3.5b}), (\ref{equB14}), (\ref{equB20}), and (\ref{equB23}), one obtains
\begin{eqnarray}
\label{equ3.18}\boldsymbol{l}&=&-\boldsymbol{I}_{2}I\Leftrightarrow \boldsymbol{I}_{2}=\boldsymbol{l}I.
\end{eqnarray}
By further applying Eq.~(\ref{equB26}), one can verify that
\begin{subequations}
\begin{eqnarray}
\label{equ3.19a}\boldsymbol{l}^{2}&=&1,\\
\label{equ3.19b}\boldsymbol{m}\cdot\boldsymbol{l}&=&0,\\
\label{equ3.19c}\boldsymbol{n}\cdot\boldsymbol{l}&=&0,
\end{eqnarray}
\end{subequations}
which mean that the relative vector $\boldsymbol{l}$ is a unit normal vector to the plane encoded by $\boldsymbol{I}_{2}$. Thus, for any relative vector $\boldsymbol{a}$,
\begin{subequations}
\begin{eqnarray}
\label{equ3.20a}\boldsymbol{a}\times\boldsymbol{I}_{2}&=&\langle\boldsymbol{a}\boldsymbol{l}I\rangle_{2}=(\boldsymbol{a}\times\boldsymbol{l})I,\\
\label{equ3.20b}\boldsymbol{a}\wedge\boldsymbol{I}_{2}&=&\langle\boldsymbol{a}\boldsymbol{l}I\rangle_{4}=(\boldsymbol{a}\cdot\boldsymbol{l})I
\end{eqnarray}
\end{subequations}
hold, where Eqs.~(\ref{equ3.18}) and (\ref{equB13}) are used. With the aid of these results, Eqs.~(\ref{equ3.16c}) and (\ref{equ3.16d}) can be transformed into
\begin{subequations}
\begin{eqnarray}
\label{equ3.21a}\boldsymbol{y}_{\parallel}\cdot\boldsymbol{l}=&0,\qquad\
\boldsymbol{y}_{\perp}\times\boldsymbol{l}&=0,\\
\label{equ3.21b}\boldsymbol{y}'_{\parallel}\cdot\boldsymbol{l}=&0,\qquad
\boldsymbol{y}'_{\perp}\times\boldsymbol{l}&=0,
\end{eqnarray}
\end{subequations}
which explicitly show that $\boldsymbol{y}_{\parallel}\,(\boldsymbol{y}'_{\parallel})$ and $\boldsymbol{y}_{\perp}\,(\boldsymbol{y}'_{\perp})$ are, respectively, the components of $\boldsymbol{y}\,(\boldsymbol{y}')$ parallel and perpendicular to the plane defined by $\boldsymbol{I}_{2}$.

When the relative vectors $\boldsymbol{x},\boldsymbol{x}',\boldsymbol{y}$, and $\boldsymbol{y}'$ in Eqs.~(\ref{equ3.11a}) and (\ref{equ3.11b}) are replaced by their decompositions, namely Eqs.~(\ref{equ3.15a})---(\ref{equ3.15d}), it will be seen that a clear physical explanation of the Lorentz transformations induced by $\boldsymbol{e}_{v}$ and $\boldsymbol{I}_{2}$ in Eqs.~(\ref{equ3.7a}) and (\ref{equ3.7b}) is able to be achieved. To this end, the following properties of the components of $\boldsymbol{x},\boldsymbol{x}',\boldsymbol{y}$, and $\boldsymbol{y}'$ need to be first derived by means of the combination of Eqs.~(\ref{equ3.5b}), (\ref{equ3.16a}), (\ref{equ3.16c}), and the relevant formulas in Appendix B,
\begin{subequations}
\begin{eqnarray}
\label{equ3.22a}\boldsymbol{x}_{\parallel}\boldsymbol{e}_{v}=&\boldsymbol{e}_{v}\boldsymbol{x}_{\parallel},\qquad\quad \boldsymbol{x}_{\perp}\boldsymbol{e}_{v}&=-\boldsymbol{e}_{v}\boldsymbol{x}_{\perp},\\
\label{equ3.22b}\boldsymbol{y}_{\parallel}\boldsymbol{I}_{2}=&-\boldsymbol{I}_{2}\boldsymbol{y}_{\parallel},\qquad\
\boldsymbol{y}_{\perp}\boldsymbol{I}_{2}&=\boldsymbol{I}_{2}\boldsymbol{y}_{\perp}.
\end{eqnarray}
\end{subequations}
With these equalities and Eqs.~(\ref{equB39})---(\ref{equB41}), after substituting Eqs.~(\ref{equ3.15a})---(\ref{equ3.15d}) in Eqs.~(\ref{equ3.11a}) and (\ref{equ3.11b}), important intermediate results are obtained,
\begin{subequations}
\begin{eqnarray}
\label{equ3.23a}x'^{0}+\boldsymbol{x}'_{\parallel}+\boldsymbol{x}'_{\perp}&=&\text{e}^{-\theta\boldsymbol{e}_{v}}\left(x^{0}+\boldsymbol{x}_{\parallel}\right)+\boldsymbol{x}_{\perp},\\
\label{equ3.23b}y'^{0}+\boldsymbol{y}'_{\parallel}+\boldsymbol{y}'_{\perp}&=&y^{0}+\text{e}^{-\varphi\boldsymbol{I}_{2}}\boldsymbol{y}_{\parallel}+\boldsymbol{y}_{\perp}.
\end{eqnarray}
\end{subequations}
In order to handle these two equations, $\text{e}^{-\theta\boldsymbol{e}_{v}}$ and $\text{e}^{-\varphi\boldsymbol{I}_{2}}$ should be rewritten as
\begin{subequations}
\begin{eqnarray}
\label{equ3.24a}\text{e}^{-\theta\boldsymbol{e}_{v}}&=&\cosh{\theta}-\boldsymbol{e}_{v}\sinh{\theta}=:\gamma\left(1-\boldsymbol{\beta}\right),\\
\label{equ3.24b}\text{e}^{-\varphi\boldsymbol{I}_{2}}&=&\cos{\varphi}-\boldsymbol{I}_{2}\sin{\varphi}
\end{eqnarray}
\end{subequations}
with
\begin{subequations}
\begin{eqnarray}
\label{equ3.25a}\beta&:=&\tanh{\theta},\\
\label{equ3.25b}\boldsymbol{\beta}&:=&\beta\boldsymbol{e}_{v},\\
\label{equ3.25c}\gamma&:=&\cosh{\theta}=\frac{1}{\sqrt{1-\boldsymbol{\beta}^2}}.
\end{eqnarray}
\end{subequations}
Plug Eqs.~(\ref{equ3.24a}) and (\ref{equ3.24b}) into (\ref{equ3.23a}) and (\ref{equ3.23b}), respectively, and then, by using the grade operator $\langle\cdots\rangle$ and the orthogonal projection operator successively, we finally arrive at
\begin{subequations}
\begin{eqnarray}
\label{equ3.26a}&&\left\{\begin{array}{l}
\displaystyle x'^{0}=\gamma\big(x^{0}-\boldsymbol{\beta}\cdot\boldsymbol{x}_{\parallel}\big),\smallskip\\
\displaystyle \boldsymbol{x}'_{\parallel}=\gamma\big(\boldsymbol{x}_{\parallel}-x^{0}\boldsymbol{\beta}\big),\smallskip\\
\displaystyle \boldsymbol{x}'_{\perp}=\boldsymbol{x}_{\perp},
\end{array}\right.\\
\label{equ3.26b}&&\left\{\begin{array}{l}
\displaystyle y'^{0}=y^{0},\smallskip\\
\displaystyle \boldsymbol{y}'_{\parallel}=\text{e}^{-\varphi\boldsymbol{I}_{2}}\boldsymbol{y}_{\parallel}=\cos{\varphi}\boldsymbol{y}_{\parallel}-\sin{\varphi}\boldsymbol{I}_{2}\boldsymbol{y}_{\parallel},\smallskip\\
\displaystyle \boldsymbol{y}'_{\perp}=\boldsymbol{y}_{\perp}.
\end{array}\right.
\end{eqnarray}
\end{subequations}
In the above derivation, Eqs.~(\ref{equ3.16a}) and (\ref{equB8}) have been employed, and besides, one also needs to note that in view of Eqs.~(\ref{equ3.5b}), (\ref{equ3.16c}), (\ref{equ3.19b}), (\ref{equ3.19c}), (\ref{equB10}), (\ref{equB15}), and (\ref{equB17}),
\begin{subequations}
\begin{eqnarray}
\label{equ3.27a}\boldsymbol{I}_{2}\boldsymbol{y}_{\parallel}&=&\boldsymbol{I}_{2}\times\boldsymbol{y}_{\parallel},\\
\label{equ3.27b}\left(\boldsymbol{I}_{2}\boldsymbol{y}_{\parallel}\right)\cdot \boldsymbol{l}&=&\langle\boldsymbol{I}_{2}\boldsymbol{y}_{\parallel}\boldsymbol{l}\rangle=\boldsymbol{I}_{2}\cdot(\boldsymbol{y}_{\parallel}\times\boldsymbol{l})=0
\end{eqnarray}
\end{subequations}
hold, and hence, $\boldsymbol{I}_{2}\boldsymbol{y}_{\parallel}$ is indeed a relative vector parallel to the plane defined by $\boldsymbol{I}_{2}$.
Remember that $\theta$ is a free parameter, and if one defines
$$
\label{equ3.28a}\tanh{\theta}=\frac{\sqrt{\boldsymbol{v}^2}}{c},\eqno{(3.28\text{a})}
$$
because of Eqs.~(\ref{equ3.5a}), (\ref{equ3.25a}), and (\ref{equ3.25b}),
$$
\label{equ3.28b}\boldsymbol{\beta}=\frac{\boldsymbol{v}}{c}.\eqno{(3.28\text{b})}
$$
Thus, the equivalent expression of Eq.~(\ref{equ3.26a}) is given by making use of Eqs.~(\ref{equ3.14a}), (\ref{equ3.15a}), (\ref{equ3.15b}), and (\ref{equ3.16a}),
\setcounter{equation}{28}
\begin{eqnarray}
\label{equ3.29}&&\left\{\begin{array}{l}
\displaystyle x'^{0}=\gamma\left(x^{0}-\frac{\boldsymbol{v}\cdot\boldsymbol{x}}{c}\right),\smallskip\\
\displaystyle \boldsymbol{x}'=\boldsymbol{x}+\boldsymbol{v}\left((\gamma-1)\frac{\boldsymbol{x}\cdot\boldsymbol{v}}{\boldsymbol{v}^2}-\gamma\frac{x^{0}}{c}\right).
\end{array}\right.
\end{eqnarray}
As for Eq.~(\ref{equ3.26b}), by virtue of Eqs.~(\ref{equ3.22b}), (\ref{equB8}), (\ref{equB39}), and (\ref{equB41}), one can achieve
\begin{subequations}
\begin{eqnarray}
\label{equ3.30a}\boldsymbol{y}'_{\parallel}\cdot\boldsymbol{y}'_{\parallel}&=&\left\langle\boldsymbol{y}_{\parallel}\text{e}^{\varphi\boldsymbol{I}_{2}}\text{e}^{-\varphi\boldsymbol{I}_{2}}\boldsymbol{y}_{\parallel}\right\rangle=\boldsymbol{y}_{\parallel}\cdot\boldsymbol{y}_{\parallel},\\
\label{equ3.30b}\boldsymbol{y}_{\parallel}\boldsymbol{y}'_{\parallel}&=&\left(\boldsymbol{y}_{\parallel}\cdot\boldsymbol{y}_{\parallel}\right)\text{e}^{\varphi\boldsymbol{I}_{2}}\Rightarrow\left\{\begin{array}{l}
\displaystyle \boldsymbol{y}_{\parallel}\cdot\boldsymbol{y}'_{\parallel}=\left(\boldsymbol{y}_{\parallel}\cdot\boldsymbol{y}_{\parallel}\right)\cos{\varphi},\smallskip\\
\displaystyle \boldsymbol{y}_{\parallel}\times\boldsymbol{y}'_{\parallel}=\left(\boldsymbol{y}_{\parallel}\cdot\boldsymbol{y}_{\parallel}\right)\sin{\varphi}\boldsymbol{I}_{2}.
\end{array}\right.
\end{eqnarray}
\end{subequations}
Evidently, these results suggest that the Lorentz transformations induced by $\boldsymbol{e}_{v}$ and $\boldsymbol{I}_{2}$ in Eqs.~(\ref{equ3.7a}) and (\ref{equ3.7b}) are, respectively, a Lorentz boost with the velocity $\boldsymbol{v}$~\cite{Application.electrodynamics,Jackson1998} and a spatial rotation through an angle $\varphi$ in the plane encoded by $\boldsymbol{I}_{2}$. Here, in order to reasonably interpret relevant equations obtained in this subsection, the active view for Lorentz transformation needs to be adopted~\cite{Application.electrodynamics}. Moreover, it also needs to be stressed that for the spatial rotation, Eq.~(\ref{equ3.30b}) shows that if $\varphi>0$, the relative bivector $\boldsymbol{y}_{\parallel}\times\boldsymbol{y}'_{\parallel}$ has the same orientation as $\boldsymbol{I}_{2}$ in the three-dimensional geometry. Let us recall that the relative vectors $\boldsymbol{v}, \boldsymbol{m}$, and $\boldsymbol{n}$ were chosen arbitrarily in the beginning, and therefore, with the rotors $\text{e}^{\frac{\theta}{2}\boldsymbol{e}_{v}}$ and $\text{e}^{\frac{\varphi}{2}\boldsymbol{I}_{2}}$, the general Lorentz boost with velocity in an arbitrary direction and the general spatial rotation in an arbitrary plane can be handled. Furthermore, considering that Eqs.~(\ref{equ3.26a}), (\ref{equ3.26b}), (\ref{equ3.29}), (\ref{equ3.30a}), and (\ref{equ3.30b}) are derived in the ``common'' even subalgebra of the STAs of signatures $(\pm,\mp,\mp,\mp)$, all of these equations are presented in a signature invariant form.

According to the previous discussion, the Lorentz boost and the spatial rotation have been first generated in Eqs.~(\ref{equ3.7a}) and (\ref{equ3.7b}), and however, until these two equations were transformed into those in the ``common'' even subalgebra of the two STAs, their physical explanations were achieved in the three-dimensional geometry. In this process,
the techniques on spacetime split have been employed, which implies that the intuitive pictures formed in the relative space are observer dependent. In addition, one may also have found that it is since the ``common'' even subalgebra of the two STAs are independent of the signatures that the original equation (\ref{equ3.7a}) or (\ref{equ3.7b}) has the same three-dimensional meaning in the two signatures, and thus, a signature invariant method for handling Lorentz boost and spatial rotation is gained. In fact, many topics in spacetime physics can be dealt with in such a manner, and inspired by this, we will apply this method to studying gyroscopic precession in the next section, so that a signature invariant GA derivation of the precessional angular velocity of the gyroscope spin could be found.

As the final task of this subsection, the pure Lorentz boost (cf.~Ref.~\cite{Doran2003}) between two vectors of the same magnitude will be discussed based on the previous results.
Assuming that $x'=c\gamma_{0}$, Eqs.~(\ref{equ3.10a}), (\ref{equ3.10b}), (\ref{equ3.25c}), and (\ref{equ3.26a}) yield $x\gamma^{0}=\gamma(c+\boldsymbol{v})$, and then, with Eq.~(\ref{equA1}) and $\boldsymbol{v}=c\boldsymbol{\beta}$,
one obtains
\begin{subequations}
\begin{eqnarray}
\label{equ3.31a}\pm x\cdot x'&=&\gamma c^2,\\
\label{equ3.31b}\pm x\wedge x'&=&\gamma c^2\boldsymbol{\beta}
\end{eqnarray}
\end{subequations}
and
\begin{subequations}
\begin{eqnarray}
\label{equ3.32a}x^2&=&x'^2=\pm c^2,\\
\label{equ3.32b}\boldsymbol{e}_{v}&=&\pm\frac{x\wedge x'}{\sqrt{\left(x\wedge x'\right)^2}},
\end{eqnarray}
\end{subequations}
where Eq.~(\ref{equ3.32a}) implies that the vectors $x$ and $x'$ could be thought of as the four-velocities of observers.
In such a case, by means of Eqs.~(\ref{equB39}) and (\ref{equ3.25a})---(\ref{equ3.25c}), the rotor $\text{e}^{\frac{\theta}{2}\boldsymbol{e}_{v}}$ can be expressed as
\begin{eqnarray}
\label{equ3.33}\text{e}^{\frac{\theta}{2}\boldsymbol{e}_{v}}&=&\displaystyle\frac{1+\cosh{\theta}+\boldsymbol{e}_{v}\sinh{\theta}}{\sqrt{2\left(1+\cosh{\theta}\right)}}
=\displaystyle\frac{1+\gamma+\gamma\boldsymbol{\beta}}{\sqrt{2\left(1+\gamma\right)}}=\displaystyle\frac{c^2\pm xx'}{\sqrt{2c^2\left(c^2\pm x\cdot x'\right)}}=\text{e}^{\pm\frac{\theta}{2}\frac{x\wedge x'}{\sqrt{\left(x\wedge x'\right)^2}}},
\end{eqnarray}
and thus, Eq.~(\ref{equ3.7a}) states that under the Lorentz boost generated by the rotor
\begin{eqnarray}
\label{equ3.34}\hat{L}:=\frac{c^2\pm xx'}{\sqrt{2c^2\left(c^2\pm x\cdot x'\right)}}=\text{e}^{\pm\frac{\theta}{2}\frac{x\wedge x'}{\sqrt{\left(x\wedge x'\right)^2}}},
\end{eqnarray}
$x$ is mapped to $x'$ by
\begin{eqnarray}
\label{equ3.35}x'=\tilde{\hat{L}}x\hat{L}.
\end{eqnarray}
According to Ref.~\cite{Doran2003}, the above $\hat{L}$ in the $(+,-,-,-)$ signature is exactly the rotor that determines the pure Lorentz boost between $x$ and $x'$, and motivated by this, we claim that the above $\hat{L}$ in the $(-,+,+,+)$ signature also plays the same role. It should be noted that the validity of Eq.~(\ref{equ3.35}) is able to be directly verified only by Eqs.~(\ref{equ3.32a}) and (\ref{equ3.34}), which does not depend on the selection of the frame $\{\gamma_{\alpha}\}$. In the treatment of gyroscopic precession in the next section, Eqs.~(\ref{equ3.34}) and (\ref{equ3.35}) will be used to generate the pure Lorentz boost between a comoving orthonormal frame of the gyroscope and a local orthonormal tetrad at rest in the coordinate system of the spacetime metric, which greatly improves the computational efficiency.
\subsection{Relativistic dynamics of a massive particle in curved spacetime~\label{Sec:3.2}}
As mentioned previously, the description of the motion of a gyroscope requires that relativistic dynamics of a massive particle in curved spacetime should be studied~\cite{Weinberg2014}, and to this end, a brief introduction to relevant GA techniques for General Relativity formulated in Ref.~\cite{Francis:2003xi} needs to be given, so that the treatment of gyroscopic precession in the following can be put on a solid theoretical footing. In order to develop a GA description of curved spacetime, one should define a local orthonormal tetrad $\{\gamma_{\alpha}\}$ by the orthonormalization of a coordinate frame and then generate the corresponding STA. Let $x^{\mu}$ and $\{g_{\mu}\}$ be local coordinates in a curved spacetime and the associated coordinate frame, respectively. Assume that a collection of fiducial observers is distributed over space, and each fiducial observer is at rest in the coordinate system. Then, the components of the metric with respect to the coordinate frame $\{g_{\mu}\}$,
\begin{eqnarray}
\label{equ3.36}g_{\mu\nu}&:=&g_{\mu}\cdot g_{\nu},
\end{eqnarray}
satisfy the conditions~\cite{Landau1971}
\begin{subequations}
\begin{eqnarray}
\label{equ3.37a}\pm g_{0}\cdot g_{0}&=&\pm g_{00}>0,\\
\label{equ3.37b}-(g_{1}\wedge g_{0})\cdot(g_{0}\wedge g_{1})&=&-\det\left(\begin{array}{cc}
g_{00},&\ g_{01}\\
g_{10},&\ g_{11}
\end{array}\right)>0,\\
\label{equ3.37c}\pm(g_{2}\wedge g_{1}\wedge g_{0})\cdot(g_{0}\wedge g_{1}\wedge g_{2})&=&\pm\det\left(\begin{array}{ccc}
g_{00},&\ g_{01},&\ g_{02}\\
g_{10},&\ g_{11},&\ g_{12}\\
g_{20},&\ g_{21},&\ g_{22}
\end{array}\right)>0,\\
\label{equ3.37d}-(g_{3}\wedge g_{2}\wedge g_{1}\wedge g_{0})\cdot(g_{0}\wedge g_{1}\wedge g_{2}\wedge g_{3})&=&-\det\left(\begin{array}{cccc}
g_{00},&\ g_{01},&\ g_{02},&\ g_{03}\\
g_{10},&\ g_{11},&\ g_{12},&\ g_{13}\\
g_{20},&\ g_{21},&\ g_{22},&\ g_{23}\\
g_{30},&\ g_{31},&\ g_{32},&\ g_{33}
\end{array}\right)>0,
\end{eqnarray}
\end{subequations}
where in the last three equations, Eqs.~(\ref{equA8}), (\ref{equA11}), and (\ref{equA13}) are used. By means of the GA technique on the Gram-Schmidt orthogonalization procedure provided in Ref.~\cite{Hestenes1984}, the coordinate frame $\{g_{\mu}\}$ is able to be orthonormalized conveniently,
\begin{eqnarray}
\label{equ3.38}&&\left\{\begin{array}{l}
\displaystyle \gamma_{0}=\frac{g_{0}}{\sqrt{\pm g_{0}\cdot g_{0}\phantom{)}}},\smallskip\\
\displaystyle \gamma_{1}=\pm\frac{g_{0}(g_{0}\wedge g_{1})}{\sqrt{\pm g_{0}\cdot g_{0}\phantom{)}}\sqrt{-(g_{1}\wedge g_{0})\cdot(g_{0}\wedge g_{1})}},\smallskip\\
\displaystyle \gamma_{2}=-\frac{(g_{1}\wedge g_{0})(g_{0}\wedge g_{1}\wedge g_{2})}{\sqrt{-(g_{1}\wedge g_{0})\cdot(g_{0}\wedge g_{1})}\sqrt{\pm(g_{2}\wedge g_{1}\wedge g_{0})\cdot(g_{0}\wedge g_{1}\wedge g_{2})}},\smallskip\\
\displaystyle \gamma_{3}=\pm\frac{(g_{2}\wedge g_{1}\wedge g_{0})(g_{0}\wedge g_{1}\wedge g_{2}\wedge g_{3})}{\sqrt{\pm(g_{2}\wedge g_{1}\wedge g_{0})\cdot(g_{0}\wedge g_{1}\wedge g_{2})}\sqrt{-(g_{3}\wedge g_{2}\wedge g_{1}\wedge g_{0})\cdot(g_{0}\wedge g_{1}\wedge g_{2}\wedge g_{3})}}.
\end{array}\right.
\end{eqnarray}
With the relevant formulas in Appendix A, one can immediately verify that $\{\gamma_{\alpha}\}$, as a local orthonormal tetrad, satisfies
\begin{eqnarray}
\label{equ3.39}\gamma_{\alpha}\cdot\gamma_{\beta}&=&\eta_{\alpha\beta}=\text{diag}(\pm,\mp,\mp,\mp)
\end{eqnarray}
and
\begin{subequations}
\begin{eqnarray}
\label{equ3.40a}\gamma_{0}\wedge \gamma_{1}&=&\frac{g_{0}\wedge g_{1}}{\sqrt{-(g_{1}\wedge g_{0})\cdot(g_{0}\wedge g_{1})}},\\
\label{equ3.40b}\gamma_{0}\wedge \gamma_{1}\wedge\gamma_{2}&=&\frac{g_{0}\wedge g_{1}\wedge g_{2}}{\sqrt{\pm(g_{2}\wedge g_{1}\wedge g_{0})\cdot(g_{0}\wedge g_{1}\wedge g_{2})}},\\
\label{equ3.40c}\gamma_{0}\wedge \gamma_{1}\wedge\gamma_{2}\wedge\gamma_{3}&=&\frac{g_{0}\wedge g_{1}\wedge g_{2}\wedge g_{3}}{\sqrt{-(g_{3}\wedge g_{2}\wedge g_{1}\wedge g_{0})\cdot(g_{0}\wedge g_{1}\wedge g_{2}\wedge g_{3})}}.
\end{eqnarray}
\end{subequations}

Within the framework of General Relativity, the covariant derivative $\nabla$ on the spacetime manifold can be defined in the standard way~\cite{Wald1984}, where one of its important properties is that it will reduce to $\partial$ when acting on scalar functions. Suppose that $\nabla$ is the unique torsion-free and metric-compatible derivative operator. Then, according to Ref.~\cite{Francis:2003xi}, the covariant derivative of a multivector $A$ along a vector $b$ is evaluated by the formula
\begin{eqnarray}
\label{equ3.41}b\cdot\nabla A=b\cdot\partial A+\omega(b)\times A.
\end{eqnarray}
Here, the operator $b\cdot\partial$ satisfies
\begin{subequations}
\begin{eqnarray}
\label{equ3.42a}b\cdot\partial \gamma_{\alpha}&=&b\cdot\partial \gamma^{\beta}=0,\\
\label{equ3.42b}b\cdot\partial \phi&=&b\cdot\nabla\phi
\end{eqnarray}
\end{subequations}
with $\{\gamma^{\beta}\}$ and $\phi$ as the reciprocal tetrad of $\{\gamma_{\alpha}\}$ and a scalar field in spacetime, respectively. $\omega(b)$, being the bivector connection associated with $\{\gamma_{\alpha}\}$, is defined by
\begin{eqnarray}
\label{equ3.43}b\cdot\nabla \gamma_{\alpha}=\omega(b)\times\gamma_{\alpha},
\end{eqnarray}
where if $b=b^{\mu}g_{\mu}$, the expression of $\omega(b)$ is given by~\cite{Francis:2003xi,Snygg1997}
\begin{eqnarray}
\label{equ3.44}\omega(b)=b^{\mu}\omega(g_{\mu})
\end{eqnarray}
with
\begin{eqnarray}
\label{equ3.45}\omega(g_{\mu})=\frac{1}{2}g^{\rho}\wedge g^{\sigma}\big(g_{\sigma}\cdot\partial{g_{\mu\rho}}\big)+\frac{1}{2}g^{\rho}\wedge\big(g_{\mu}\cdot\partial g_{\rho}\big).
\end{eqnarray}
With the aid of the corresponding GA technique~\cite{Hestenes1984}, $\{g^{\nu}\}$, as the reciprocal frame of $\{g_{\mu}\}$, is constructed as
\begin{eqnarray}
\label{equ3.46}&&\left\{\begin{array}{l}
\displaystyle g^{0}=\left(g_{1}\wedge g_{2}\wedge g_{3}\right)\left(g_{0}\wedge g_{1}\wedge g_{2}\wedge g_{3}\right)^{-1},\smallskip\\
\displaystyle g^{1}=-\left(g_{0}\wedge g_{2}\wedge g_{3}\right)\left(g_{0}\wedge g_{1}\wedge g_{2}\wedge g_{3}\right)^{-1},\smallskip\\
\displaystyle g^{2}=\left(g_{0}\wedge g_{1}\wedge g_{3}\right)\left(g_{0}\wedge g_{1}\wedge g_{2}\wedge g_{3}\right)^{-1},\smallskip\\
\displaystyle g^{3}=-\left(g_{0}\wedge g_{1}\wedge g_{2}\right)\left(g_{0}\wedge g_{1}\wedge g_{2}\wedge g_{3}\right)^{-1},
\end{array}\right.
\end{eqnarray}
where from Eq.~(\ref{equ3.38}), the coordinate frame $\{g_{\mu}\}$ can be expanded in the local orthonormal tetrad $\{\gamma_{\alpha}\}$,
\begin{eqnarray}
\label{equ3.47}&&\left\{\begin{array}{l}
\displaystyle g_{0}=\sqrt{\pm g_{0}\cdot g_{0}\phantom{)}}\;\gamma_{0},\smallskip\\
\displaystyle g_{1}=\pm\frac{g_{01}}{\sqrt{\pm g_{0}\cdot g_{0}\phantom{)}}}\gamma_{0}+\frac{\sqrt{-(g_{1}\wedge g_{0})\cdot(g_{0}\wedge g_{1})}}{\sqrt{\pm g_{0}\cdot g_{0}\phantom{)}}}\gamma_{1},\smallskip\\
\displaystyle g_{2}=\pm\frac{g_{02}}{\sqrt{\pm g_{0}\cdot g_{0}\phantom{)}}}\gamma_{0}-\frac{(g_{2}\wedge g_{0})\cdot(g_{0}\wedge g_{1})}{\sqrt{\pm g_{0}\cdot g_{0}\phantom{)}} \sqrt{-(g_{1}\wedge g_{0})\cdot(g_{0}\wedge g_{1})}}\gamma_{1}\smallskip\\
\displaystyle\phantom{\displaystyle g_{2}=}+\frac{\sqrt{\pm(g_{2}\wedge g_{1}\wedge g_{0})\cdot(g_{0}\wedge g_{1}\wedge g_{2})}}{\sqrt{-(g_{1}\wedge g_{0})\cdot(g_{0}\wedge g_{1})}}\gamma_{2},\smallskip\\
\displaystyle g_{3}=\pm\frac{g_{03}}{\sqrt{\pm g_{0}\cdot g_{0}\phantom{)}}}\gamma_{0}-\frac{(g_{3}\wedge g_{0})\cdot(g_{0}\wedge g_{1})}{\sqrt{\pm g_{0}\cdot g_{0}\phantom{)}} \sqrt{-(g_{1}\wedge g_{0})\cdot(g_{0}\wedge g_{1})}}\gamma_{1}\smallskip\\
\displaystyle\phantom{g_{3}=}\pm\frac{(g_{3}\wedge g_{1}\wedge g_{0})\cdot(g_{0}\wedge g_{1}\wedge g_{2})}{\sqrt{-(g_{1}\wedge g_{0})\cdot(g_{0}\wedge g_{1})}\sqrt{\pm(g_{2}\wedge g_{1}\wedge g_{0})\cdot(g_{0}\wedge g_{1}\wedge g_{2})}}\gamma_{2}\smallskip\\
\displaystyle\phantom{g_{3}=}+\frac{\sqrt{-(g_{3}\wedge g_{2}\wedge g_{1}\wedge g_{0})\cdot(g_{0}\wedge g_{1}\wedge g_{2}\wedge g_{3})}}{\sqrt{\pm(g_{2}\wedge g_{1}\wedge g_{0})\cdot(g_{0}\wedge g_{1}\wedge g_{2})}}\gamma_{3}.
\end{array}\right.
\end{eqnarray}
Because only the knowledge of covariant derivative and bivector connection will be involved in the discussion of gyroscopic precession, other GA techniques for General Relativity
will not be covered here, and the reader wishing to go into more details may consult Ref.~\cite{Francis:2003xi}.

Next, for a massive particle, the spacetime splits of the velocity, acceleration, momentum, and force four-vectors with the normalized four-velocity of the fiducial observer will be discussed, so that relativistic dynamics of this particle in curved spacetime can be studied. Let us first identify the proper time of fiducial observers. As indicated earlier, fiducial observers are at rest in the coordinate system $x^{\mu}$, which means that their worldlines are the coordinate curves with $x^{i}=\text{const.}\ (i=1,2,3)$, namely, $t:=x^{0}/c$ coordinate curves. As a consequence, if we let $t_{0}$ denote the proper time of each fiducial observer,
$\pm c^2(dt_{0})^2=g_{00}c^2(dt)^2$ hold along his worldline, and then,
\begin{eqnarray}
\label{equ3.48}\frac{dt_{0}}{dt}=\sqrt{\pm g_{00}}.
\end{eqnarray}
Assuming that $x^{\mu}(\tau)$ is the worldline of a massive particle with $\tau$ as the proper time, the four-velocity of the particle can be rewritten as~\cite{Ignazio1995,Wu:2021uws}
\begin{eqnarray}
\label{equ3.49}u&=&\gamma_{u}\left(c\gamma_{0}+u^{i}\gamma_{i}\right).
\end{eqnarray}
We will prove that
\begin{eqnarray}
\label{equ3.50}\gamma_{u}&=&\frac{dt_{0}}{d\tau}.
\end{eqnarray}
Consider an event $P$ on the particle's worldline. The $t$ coordinate curve with $x^{i}=x^{i}(P)\ (i=1,2,3)$ passes through $P$ and is the worldline of a fiducial observer.
Based on the orthonormal tetrad $\{\gamma_{\mu}|_{x^{i}=x^{i}(P)}\}$ carried by this fiducial observer, his proper reference frame can be defined, and thus, a local coordinate system $\left(y^{0}=:ct_{0},y^{1},y^{2},y^{3}\right)$ covering a finite domain near his worldline can also be defined.
In this coordinate system, if the worldline of the particle is $y^{\mu}(\tau)$, its four-velocity at the event $P$ is
\begin{eqnarray}
\label{equ3.52}u|_{P}=\frac{dt_{0}}{d\tau}\bigg|_{P}\left(c\gamma_{0}|_{P}+\frac{dy^{i}}{dt_{0}}\bigg|_{P}\red\gamma_{i}\black|_{P}\right).
\end{eqnarray}
Comparing Eq.~(\ref{equ3.52}) with Eq.~(\ref{equ3.49}), we get
$$
\gamma_{u}|_{P}=\frac{dt_{0}}{d\tau}\bigg|_{P}.
$$
$P$ is an arbitrary event on the particle's worldline, and due to Eq.~(\ref{equ3.48}),
\begin{eqnarray}
\label{equ3.53}\frac{dt_{0}}{d\tau}=\frac{dt_{0}}{dt}\frac{dt}{d\tau}=\sqrt{\pm g_{00}}\frac{dt}{d\tau}
\end{eqnarray}
does not depend on the selection of the coordinate system $y^{\mu}$, so Eq.~(\ref{equ3.50}) holds. By applying Eq.~(\ref{equ2.40}), the spacetime split of the four-velocity of the particle with $\gamma_{0}$ yields
\begin{eqnarray}
\label{equ3.54}u\gamma^{0}=\gamma_{u}\left(c+\boldsymbol{u}\right)\quad \text{with}\quad \boldsymbol{u}:=u^{i}\boldsymbol{\sigma}_{i},
\end{eqnarray}
where because of $(\widetilde{u\gamma^{0}})\cdot(u\gamma^{0})=\pm u^2=c^2$, one is able to achieve
\begin{eqnarray}
\label{equ3.55}\gamma_{u}=\frac{1}{\sqrt{1-\displaystyle\frac{\boldsymbol{u}^2}{c^2}}}.
\end{eqnarray}
Since $c\gamma_{0}$ could be identified as the four-velocity of some fiducial observer, $\boldsymbol{u}$ is actually the relative velocity measured in his orthonormal tetrad, which is also able to be inferred from Eq.~(\ref{equ3.52}).

After clarifying many concepts, we are in a position to derive the spacetime split of the four-acceleration of the particle with $\gamma_{0}$, which is an essential ingredient in the formalism of relativistic dynamics. The four-acceleration of the particle, $a=Du/d\tau=u\cdot\nabla u$, is immediately gained from Eq.~(\ref{equ3.41}), and then, by employing Eq.~(\ref{equ2.40}), its spacetime split with $\gamma_{0}$ is provided,
\begin{eqnarray}
\label{equ3.56}a\gamma^{0}=\left(u\cdot\partial u\right)\gamma^{0}+\left(\omega(u)\times u\right)\gamma^{0}.
\end{eqnarray}
The first term is
\begin{eqnarray}
\left(u\cdot\partial u\right)\gamma^{0}&=&\big(u\cdot\partial{(c\gamma_{u})}\big)+\left(u\cdot\partial{(\gamma_{u}u^{i})}\right)\boldsymbol{\sigma}_{i}\nonumber\\
&=&\big(u\cdot\nabla{(c\gamma_{u})}\big)+\left(u\cdot\nabla{(\gamma_{u}u^{i})}\right)\boldsymbol{\sigma}_{i}\nonumber\\
\label{equ3.57}&=&c\frac{d\gamma_{u}}{d\tau}+\frac{d\gamma_{u}}{d\tau}\boldsymbol{u}+\gamma_{u}\frac{d\boldsymbol{u}}{d\tau}\nonumber\\
&=&\gamma_{u}^4\frac{\boldsymbol{u}\cdot\boldsymbol{a}}{c}+\gamma_{u}^4\frac{\boldsymbol{u}\cdot\boldsymbol{a}}{c^2}\boldsymbol{u}+\gamma_{u}^2\boldsymbol{a},
\end{eqnarray}
in which, Eqs.~(\ref{equ3.42a}), (\ref{equ3.42b}), (\ref{equ3.49}), (\ref{equ3.50}), (\ref{equ3.54}), $\boldsymbol{\sigma}_{i}=\gamma_{i}\gamma^{0}$, and
\begin{subequations}
\begin{eqnarray}
\label{equ3.58a}\boldsymbol{a}:&=&\frac{d\boldsymbol{u}}{dt_{0}},\\
\label{equ3.58b}\frac{d\gamma_{u}}{dt_{0}}&=&\gamma_{u}^3\frac{\boldsymbol{u}\cdot\boldsymbol{a}}{c^2}
\end{eqnarray}
\end{subequations}
have been used. Explicitly, the above $\boldsymbol{a}$ is the relative acceleration measured by the fiducial observer.
By virtue of Eqs.~(\ref{equA1}), (\ref{equA6}), (\ref{equA14}), and (\ref{equ3.54}), the second term of Eq.~(\ref{equ3.56}) is
\begin{eqnarray}
\label{equ3.59}\left(\omega(u)\times u\right)\gamma^{0}&=&\left(\omega(u)\cdot u\right)\cdot\gamma^{0}+\left(\omega(u)\cdot u\right)\wedge\gamma^{0}\nonumber\\
&=&\omega(u)\cdot\left(u\wedge\gamma^{0}\right)+\gamma^{0}\wedge\left(u\cdot\omega(u)\right)\nonumber\\
&=&\gamma_{u}\boldsymbol{u}\cdot\omega^{(E)}(u)+\gamma_{u}\boldsymbol{u}\cdot\omega^{(B)}(u)+\gamma^{0}\wedge\left(u\cdot\omega^{(E)}(u)\right)\nonumber\\
&&+\gamma^{0}\wedge\left(u\cdot\omega^{(B)}(u)\right).
\end{eqnarray}
Here, just like the Faraday bivector, namely the electromagnetic field strength, the bivector connection $\omega(u)$ has been decomposed into the electric part $\omega^{(E)}(u)$ and the magnetic part $\omega^{(B)}(u)$,
\begin{subequations}
\begin{eqnarray}
\label{equ3.60a}\omega^{(E)}(u):&=&\Big(\omega(u)\cdot\big(\gamma^{k}\wedge\gamma^{0}\big)\Big)\gamma_{0}\wedge\gamma_{k},\\
\label{equ3.60b}\omega^{(B)}(u):&=&\sum_{i<j}\Big(\omega(u)\cdot\big(\gamma^{j}\wedge\gamma^{i}\big)\Big)\gamma_{i}\wedge\gamma_{j},\\
\label{equ3.60c}\omega(u)&=&\omega^{(E)}(u)+\omega^{(B)}(u),
\end{eqnarray}
\end{subequations}
and by making use of Eq.~(\ref{equA1}) and the anticommutation of $\{\gamma_{\alpha}\}$, the important equalities are obtained,
\begin{subequations}
\begin{eqnarray}
\label{equ3.61a}\gamma_{0}\omega(u)\gamma^{0}&=&-\omega^{(E)}(u)+\omega^{(B)}(u),\\
\label{equ3.61b}\omega^{(E)}(u)&=&\frac{1}{2}\left(\omega(u)-\gamma_{0}\omega(u)\gamma^{0}\right),\\
\label{equ3.61c}\omega^{(B)}(u)&=&\frac{1}{2}\left(\omega(u)+\gamma_{0}\omega(u)\gamma^{0}\right).
\end{eqnarray}
\end{subequations}
Finally, let us deal with the last three terms in Eq.~(\ref{equ3.59}) with Eqs.~(\ref{equ3.49}), (\ref{equ3.54}), (\ref{equ3.60a}), and (\ref{equ3.60b}),
\begin{subequations}
\begin{eqnarray}
\label{equ3.62a}\gamma_{u}\boldsymbol{u}\cdot\omega^{(B)}(u)&=&\gamma_{u}u^{i}\big(\gamma_{i}\wedge\gamma^{0}\big)\cdot\omega^{(B)}(u)=0,\\
\label{equ3.62b}\gamma^{0}\wedge\left(u\cdot\omega^{(E)}(u)\right)&=&\gamma^{0}\wedge\left(c\gamma_{u}\gamma_{0}\cdot\omega^{(E)}(u)\right)+\gamma^{0}\wedge\left(\gamma_{u}u^{i}\gamma_{i}\cdot\omega^{(E)}(u)\right)\nonumber\\
&=&c\gamma_{u}\gamma^{0}\wedge\left(\gamma_{0}\cdot\omega^{(E)}(u)\right)=c\gamma_{u}\omega^{(E)}(u),\\
\label{equ3.62c}\gamma^{0}\wedge\left(u\cdot\omega^{(B)}(u)\right)&=&\gamma^{0}\wedge\left(c\gamma_{u}\gamma_{0}\cdot\omega^{(B)}(u)\right)+\gamma^{0}\wedge\left(\gamma_{u}u^{i}\gamma_{i}\cdot\omega^{(B)}(u)\right)\nonumber\\
&=&\gamma^{0}\wedge\left(\gamma_{u}u^{i}\gamma_{i}\cdot\omega^{(B)}(u)\right)+\gamma^{0}\cdot\left(\gamma_{u}u^{i}\gamma_{i}\wedge\omega^{(B)}(u)\right)\nonumber\\
&=&\gamma_{u}\left\langle\gamma^{0}u^{i}\gamma_{i}\omega^{(B)}(u)\right\rangle_{2}=\gamma_{u}\omega^{(B)}(u)\times\boldsymbol{u},
\end{eqnarray}
\end{subequations}
in which, Eqs.~(\ref{equA7}), (\ref{equA8}), and $\boldsymbol{\sigma}_{i}=\gamma_{i}\gamma^{0}$ are used.
Substituting them in Eq.~(\ref{equ3.59}) and together with Eqs.~(\ref{equ3.56}) and (\ref{equ3.57}), one finally arrives at
\begin{eqnarray}
\label{equ3.63}a\gamma^{0}&=&\gamma_{u}^4\frac{\boldsymbol{u}\cdot\boldsymbol{a}}{c}+\gamma_{u}^4\frac{\boldsymbol{u}\cdot\boldsymbol{a}}{c^2}\boldsymbol{u}+\gamma_{u}^2\boldsymbol{a}\nonumber\\
&+&\gamma_{u}\boldsymbol{u}\cdot\omega^{(E)}(u)+c\gamma_{u}\omega^{(E)}(u)-\gamma_{u}\boldsymbol{u}\times\omega^{(B)}(u).
\end{eqnarray}

Let $m$ be the rest mass of the particle. The spacetime splits of its four-momentum $p=mu$ and the four-force $f=Dp/d\tau=u\cdot\nabla p$ acting on it also need to be evaluated so that a three-dimensional analogue of Newton's second law in curved spacetime will be achieved. Starting from Eq.~(\ref{equ3.54}), the spacetime split of the particle's four-momentum $p$ with $\gamma_{0}$ is
\begin{eqnarray}
\label{equ3.64}p\gamma^{0}=\frac{E}{c}+\boldsymbol{p},
\end{eqnarray}
where
\begin{subequations}
\begin{eqnarray}
\label{equ3.65a}E&:=&\gamma_{u}mc^{2}=cp\cdot\gamma^{0},\\
\label{equ3.65b}\boldsymbol{p}&:=&\gamma_{u}m\boldsymbol{u}=p\wedge\gamma^{0}
\end{eqnarray}
\end{subequations}
are the energy and the relative momentum of the particle measured by the fiducial observer (cf.~Ref.~\cite{Doran2003}), respectively. The relationship between $E$ and $\boldsymbol{p}$ can be directly obtained from
$(\widetilde{p\gamma^{0}})\cdot(p\gamma^{0})=m^2c^2$,
\begin{eqnarray}
\label{equ3.66}E^2=\boldsymbol{p}^2c^2+m^2c^4,
\end{eqnarray}
which is exactly the same as that in Special Relativity. Assuming that the particle's rest mass remains unchanged as it moves, namely $dm/d\tau=0$, the four-force $f$ acting on it is able be expressed as
\begin{eqnarray}
\label{equ3.67}f&=&ma.
\end{eqnarray}
When the spacetime is flat and $x^{\mu}$ are coordinates in an inertial frame of reference with $g_{\mu\nu}=\eta_{\mu\nu}$, by definition, fiducial observers reduce to inertial observers. In such a case, Eq.~(\ref{equ3.48}) suggests that $dt_{0}=dt$, and the relative force $\boldsymbol{f}=f^{i}\boldsymbol{\sigma}_{i}$ acting on the particle should be given by $f^{i}=dp^{i}/dt$~\cite{Michael2019}. Thus, using $\boldsymbol{\sigma}_{i}=\gamma_{i}\gamma^{0}$, one is capable of recasting $\boldsymbol{f}$ as
\begin{eqnarray*}
\boldsymbol{f}&=&\frac{dp^{i}}{dt_{0}}\boldsymbol{\sigma}_{i}=\frac{dp^{\lambda}}{dt_{0}}\gamma_{\lambda}\wedge\gamma^{0}=\frac{dp}{dt_{0}}\wedge\gamma^{0}.
\end{eqnarray*}
In curved spacetime, we claim that the corresponding relative force $\boldsymbol{f}$ measured by the fiducial observer is related to $Dp/dt_{0}$ in the same way,
\begin{eqnarray}
\label{equ3.68}\boldsymbol{f}&=&\frac{Dp}{dt_{0}}\wedge\gamma^{0}=\frac{d\tau}{dt_{0}}\frac{Dp}{d\tau}\wedge\gamma^{0}=\frac{1}{\gamma_{u}}ma\wedge\gamma^{0}\nonumber\\
&=&m\bigg(\gamma_{u}^3\frac{\boldsymbol{u}\cdot\boldsymbol{a}}{c^2}\boldsymbol{u}+\gamma_{u}\boldsymbol{a}+c\,\omega^{(E)}(u)-\boldsymbol{u}\times\omega^{(B)}(u)\bigg),
\end{eqnarray}
in which, Eqs.~(\ref{equ3.50}), (\ref{equ3.63}), and (\ref{equ3.67}) have been used. Furthermore, by employing Eq.~(\ref{equ3.55}), the power delivered by the relative force $\boldsymbol{f}$ is evaluated as
\begin{eqnarray}
\label{equ3.69}\boldsymbol{f}\cdot\boldsymbol{u}&=&m\left(\gamma_{u}\boldsymbol{u}\cdot\boldsymbol{a}\left(\frac{\gamma_{u}^2}{c^2}\boldsymbol{u}^2+1\right)+c\omega^{(E)}(u)\cdot\boldsymbol{u}-\left(\boldsymbol{u}\times\omega^{(B)}(u)\right)\cdot\boldsymbol{u}\right)\nonumber\\
&=&m\left(\gamma_{u}^{3}\boldsymbol{u}\cdot\boldsymbol{a}+c\boldsymbol{u}\cdot\omega^{(E)}(u)\right)=c\frac{Dp}{dt_{0}}\cdot\gamma^{0}.
\end{eqnarray}
Thus, with Eqs.~(\ref{equ3.63}) and (\ref{equ3.67})---(\ref{equ3.69}), one can verify that
\begin{eqnarray}
\label{equ3.70}f\gamma^{0}&=&\gamma_{u}\left(\frac{\boldsymbol{f}\cdot\boldsymbol{u}}{c}+\boldsymbol{f}\right).
\end{eqnarray}

Eq.~(\ref{equ3.68}) is a three-dimensional analogue of Newton's second law in curved spacetime, which constitutes the core content of relativistic dynamics of a massive particle. In the above discussion, the key point is that the relative velocity, relative acceleration, relative momentum, and relative force for the particle could be reasonably defined in the orthonormal tetrad carried by the fiducial observer. Evidently, in terms of the three-dimensional geometric meaning in the relative space, these relative vectors ought to be interpreted as their corresponding three-vectors in tensor language. When the spacetime is flat, the bivector connection $\omega(u)$ and its electric part $\omega^{(E)}(u)$ and magnetic part $\omega^{(B)}(u)$ vanish. In this case, via considering the components of these relative vectors in the rest frame of the fiducial observer, namely $\{\boldsymbol{\sigma}_{k}\}$, one is able to verify that all the above results reduce to those in Special Relativity. Therefore, the formalism of relativistic dynamics of a massive particle constructed in this subsection is an elegant generalization of the classical one in flat spacetime.

In the tetrad formalism of General Relativity~\cite{Yepez:2011bw}, the covariant derivative of a vector $b=b^{\alpha}\gamma_{\alpha}$ along the coordinate frame vector $g_{\mu}$ is given by
\begin{eqnarray}
\label{equ3.71}g_{\mu}\cdot\nabla b=\Big(D_{\mu}b^{\alpha}\Big)\gamma_{\alpha}= \left(\partial_{\mu}b^{\alpha}+\omega_{\mu\phantom{\alpha}\beta}^{\phantom{\mu}\alpha} b^{\beta}\right)\gamma_{\alpha},
\end{eqnarray}
where
\begin{eqnarray}
\label{equ3.72}\omega_{\mu\phantom{\alpha}\beta}^{\phantom{\mu}\alpha}:=\left(g_{\mu}\cdot\nabla\gamma_{\beta}\right)\cdot\gamma^{\alpha}
\end{eqnarray}
are the spin connection coefficients, and due to the metric compatibility condition,
they satisfy~\cite{Maurizio2017}
\begin{eqnarray}
\label{equ3.73}\omega_{\mu\alpha\beta}=-\omega_{\mu\beta\alpha}\quad\text{with}\quad \omega_{\mu\alpha\beta}=\omega_{\mu\phantom{\delta}\beta}^{\phantom{\mu}\delta}\eta_{\delta\alpha}.
\end{eqnarray}
Using Eqs.~(\ref{equ3.43}) and (\ref{equA14}), one obtains
\begin{eqnarray}
\label{equ3.74}\omega_{\mu\alpha\beta}=\left(g_{\mu}\cdot\nabla\gamma_{\beta}\right)\cdot\gamma_{\alpha}=\left(\omega(g_{\mu})\cdot\gamma_{\beta}\right)\cdot\gamma_{\alpha}=\omega(g_{\mu})\cdot\left(\gamma_{\beta}\wedge\gamma_{\alpha}\right),
\end{eqnarray}
which means that the bivector connection $\omega(g_{\mu})$ can be expressed as
\begin{eqnarray}
\label{equ3.75}\omega(g_{\mu})=\frac{1}{2}\omega_{\mu\alpha\beta}\gamma^{\alpha}\wedge\gamma^{\beta}.
\end{eqnarray}
The above discussion suggests that it could be expected that when the relative vectors in Eq.~(\ref{equ3.68}) are expanded in the frame $\{\boldsymbol{\sigma}_{k}\}$, the corresponding generalization of Newton's second law in the tetrad formalism will also be acquired. Compared with those results in the tetrad formalism, the results in this paper are presented in the form of geometric objects, so they are endowed with a higher degree of clarity. Besides, as highlighted before, since the operations in the ``common'' even subalgebra of the STAs of signatures $(\pm,\mp,\mp,\mp)$ are independent of the signatures, the relevant results like Eqs.~(\ref{equ3.55}), (\ref{equ3.66}), (\ref{equ3.68}), and (\ref{equ3.69}) are able to be handled in a signature invariant manner. As a primary application of the signature invariant GA framework provided by the ``common'' even subalgebra of the two STAs, the treatment of  relativistic dynamics of a massive particle in this subsection provides a paradigm on how to achieve a signature invariant approach to spacetime physics in curved spacetime.

In order to depict the motion of the spin of a gyroscope, the behaviors of vector fields along the worldline of the particle also need to be studied, and here, we only focus our attention on the Fermi-Walker derivatives in the $(\pm,\mp,\mp,\mp)$ signatures. In fact, their classical forms written in tensor language have been available in Refs.~\cite{Peter2016,Hawking1973}, and recasting them in the STAs of the two signatures is a straightforward task. Hence, the results are directly provided as follows: The Fermi-Walker derivatives of a vector field $p(\tau)$ along the particle's worldline in the STAs of signatures $(\pm,\mp,\mp,\mp)$ are
\begin{eqnarray}
\label{equ3.76}\frac{D_{F}p(\tau)}{d\tau}=\frac{Dp(\tau)}{d\tau}\pm\frac{1}{c^2}(u\wedge a)\cdot p(\tau),
\end{eqnarray}
where if $D_{F}p(\tau)/d\tau=0$, the vector field $p(\tau)$ is said to be Fermi-Walker transported along the particle's worldline. For a torque-free gyroscope moving in
spacetime, any nongravitational forces acting on it are applied at its center of mass, and in this case, the spin of the gyroscope experiences the Fermi-Walker transport along its worldline~\cite{MTW1973}. In the next section, we will regard the transport equation satisfied by the gyroscope spin as the starting point for the discussion of gyroscopic precession. Interestingly, by means of the Leibniz rule and the formula~\cite{Hestenes1984}
\begin{eqnarray}
\label{equ3.77}B\times(C\wedge D)=(B\times C)\wedge D+C\wedge(B\times D)
\end{eqnarray}
with $B$ as a bivector in spacetime, the above forms of Fermi-Walker derivative can readily be extended to a multivector field $A(\tau)$ along the worldline of the particle, namely,
\begin{eqnarray}
\label{equ3.78}\frac{D_{F}A(\tau)}{d\tau}=\frac{DA(\tau)}{d\tau}\pm\frac{1}{c^2}(u\wedge a)\times A(\tau),
\end{eqnarray}
and readers who are interested in this conclusion could attempt to prove it.
\section{A GA approach to gyroscopic precession in the Lense-Thirring spacetime~\label{Sec:fourth}}
According to the prediction of General Relativity, the spin of a gyroscope precesses relative to the asymptotic
inertial frames as it moves around a rotating spherical source~\cite{Ignazio1995}. Conventionally, by following the standard method in tensor language~\cite{MTW1973,Ignazio1995}, the precessional angular velocity of the gyroscope spin is able to be evaluated under the WFSM approximation. In General Relativity, the time-dependent metric, presented in the form of multipole expansion, for the external gravitational field of a spatially compact supported source is derived under the WFSM approximation in Ref.~\cite{Wu:2021uws}. Since we are only interested in uniformly rotating spherical sources like the Earth in this paper, the spacetime is stationary, and only the leading pole moments of the source need to be considered. Consequently, in such a case, the metric reduces to the Lense-Thirring metric~\cite{Wu:2021uws}, and the spacetime is accordingly known as the Lense-Thirring spacetime. When a torque-free gyroscope is moving in this spacetime, there exist three types of precession for its spin, namely, the de Sitter precession, the Lense-Thirring precession, and the Thomas precession, where these phenomena are, respectively, resulted from gyroscopic motion through the spacetime curved by the mass of the source, rotation of the source, and gyroscopic non-geodesic motion~\cite{Everitt:2011hp}.  Today, the type of experiments designed according to these effects of gyroscopic precession have become an important method to test gravitational theories.

In the traditional description for gyroscopic precession based on tensor language, since one always needs to work with the components of some tensor in a chosen coordinate frame, many equations are given a low degree of clarity. In the language of STA, it could be expected that a physically clear approach to handling this topic will be found, since one just involves geometric objects during calculation~\cite{Lasenby:2016lfl}. In this section, as a comprehensive application of the STAs of signatures $(\pm,\mp,\mp,\mp)$ formulated in Sec.~\ref{Sec:second} and the GA techniques constructed in Sec.~\ref{Sec:third}, a GA approach to gyroscopic precession will be provided, where for a gyroscope moving in the Lense-Thirring spacetime, the precessional angular velocity of its spin will be derived in a signature invariant manner. The GA description of curved spacetime and the relevant GA techniques for General Relativity introduced at the beginning of Sec.~\ref{Sec:3.2} will still be adopted, and here, we let $x^{\mu}$ and $\{g_{\mu}\}$ be local coordinates in the Lense-Thirring spacetime and the associated coordinate frame, respectively. In addition, it should be pointed out that some physical quantities in this section and Appendix C are presented in the form of the $1/c$ expansion, where $1/c$ is used as the WFSM parameter~\cite{Blanchet:2013haa}. Since the Lense-Thirring metric is only expanded up to $1/c^3$ order, the framework of the linearized General Relativity is sufficient to analyze gyroscopic precession~\cite{Wu:2021uws,fRtheory}, and in such a case, the coordinates $(x^{\mu})=:(ct,x^{i})$ are treated as though they were the Minkowski coordinates in flat space~\cite{Thorne:1980ru,Blanchet:1985sp}.

Consider a torque-free gyroscope moving in the Lense-Thirring spacetime, and denote $x^{\mu}(\tau)$ as its worldline with $\tau$ as the proper time. Assuming that the four-force acting on the gyroscope is $f$, from Eq.~(\ref{equ3.67}), its four-acceleration $a$ is determined by
\begin{equation}
\label{equ4.1}f=ma
\end{equation}
with $m$ as its rest mass. In fact, Eq.~(\ref{equ4.1}) should be derived from the Mathisson-Papapetrou-Tulczyjew-Dixon (MPTD) equations, where the term related to the curvature tensor has been omitted because the gyroscope scale is very much smaller than the characteristic dimensions of the gravitational field~\cite{Weinberg2014}. In accordance with Refs.~\cite{MTW1973,Ignazio1995}, the spin $s$ of the gyroscope (i.e., its angular momentum vector) is always orthogonal to its four-velocity $u$ and experiences Fermi-Walker transport along its worldline,
\begin{eqnarray}
\label{equ4.2}s\cdot u&=&0,\\
\label{equ4.3}\frac{D_{F}s}{d\tau}&=&\frac{Ds}{d\tau}\pm\frac{1}{c^2}(u\wedge a)\cdot s=0.
\end{eqnarray}
It will be seen that starting from the above three equations, the precessional angular velocity of the gyroscope spin can be derived. Besides, gyroscopic precession can also be discussed based on MPTD equations, and interested readers may consult Refs.~\cite{Ramirez:2017pmp,Deriglazov:2017jub}. Since the four-velocity of the gyroscope satisfies
\begin{eqnarray}
\label{equ4.4}u^2=\pm c^2\Rightarrow u\cdot a=0,
\end{eqnarray}
by use of Eqs.~(\ref{equA6}) and (\ref{equA7}), Eq.~(\ref{equ4.3}) is equivalent to
\begin{eqnarray}
\label{equ4.5}u\cdot\nabla s=\mp\frac{1}{c^2}(a\cdot s)u.
\end{eqnarray}
Thus, Eqs.~(\ref{equ4.2}) and (\ref{equ4.5}) directly result in
\begin{eqnarray}
\label{equ4.6}\frac{ds^2}{d\tau}=u\cdot\nabla s^2=2s\cdot\left(u\cdot\nabla s\right)=0,
\end{eqnarray}
which means that $s^2$ remains fixed along the worldline of the gyroscope.

As shown in Appendix C, the Lense-Thirring metric satisfies Eqs.~(\ref{equ3.37a})---(\ref{equ3.37d}), which implies that we are capable of assuming
that there exists a collection of fiducial observers who are distributed over space and at rest in the coordinate system $x^{\mu}$, and as a consequence, a local orthonormal tetrad $\{\gamma_{\alpha}\}$ in the Lense-Thirring spacetime could be directly defined by means of the corresponding formulas in Sec.~\ref{Sec:3.2}. Based on the detailed calculation in Appendix C, the tetrad $\{\gamma_{\alpha}\}$ determined up to $1/c^3$ order is given by
\begin{equation}\label{equ4.7}
\left\{\begin{array}{l}
\displaystyle \gamma_{0}=\left(1+\frac{1}{c^{2}}U\right)g_{0},\smallskip\\
\displaystyle \gamma_{i}=-\frac{4}{c^3}V_{i}g_{0}+\left(1-\frac{1}{c^{2}}U\right)g_{i},
\end{array}\right.
\end{equation}
where the potentials $U$ and $U_{i}$ are, respectively,
\begin{equation}\label{equ4.8}
\left\{\begin{array}{lll}
\displaystyle U&=\displaystyle \frac{GM}{r},\smallskip\\
\displaystyle V_{i}&=\displaystyle -\frac{GJ\epsilon_{3ij}x^{j}}{2r^3}.
\end{array}\right.
\end{equation}
Here, $G$ is the gravitational constant, $M$ and $J$ are the mass and the conserved angular momentum of the gravitating source, respectively, and $r:=\sqrt{x^{i}x^{i}}$. Before analyzing the motion of the spin $s$ of the gyroscope, its relativistic dynamics needs to be discussed. Let $t_{0}$ be the proper time of the fiducial observer, which is related to the coordinate time $t$ by Eq.~(\ref{equ3.48}), and from Eq.~(\ref{equC1}), the expression of $dt_{0}/dt$ up to $1/c^3$ order is
\begin{eqnarray}
\label{equ4.9}\frac{dt_{0}}{dt}&=&1-\frac{1}{c^2}U.
\end{eqnarray}
As in Eqs.~(\ref{equ3.49}) and (\ref{equ3.50}), the four-velocity $u$ of the gyroscope can be expanded in the tetrad $\{\gamma_{\alpha}\}$,
\begin{eqnarray}
\label{equ4.10}u&=&\gamma_{u}\left(c\gamma_{0}+u^{i}\gamma_{i}\right)\quad \text{with}\quad \gamma_{u}=\frac{dt_{0}}{d\tau},
\end{eqnarray}
and then, Eq.~(\ref{equ3.54}) indicates that its spacetime split with $\gamma_{0}$ yields
\begin{eqnarray}
\label{equ4.11}u\gamma^{0}=\gamma_{u}\left(c+\boldsymbol{u}\right),
\end{eqnarray}
where $\boldsymbol{u}:=u^{i}\boldsymbol{\sigma}_{i}$ is the relative velocity measured in the orthonormal tetrad of the fiducial observer. Due to $u^2=\pm c^2$, the Lorentz factor $\gamma_{u}$ has the expression~(\ref{equ3.55}), and thus, by expanding it up to $1/c^3$ order, one gets
\begin{eqnarray}
\label{equ4.12}\gamma_{u}&=&1+\frac{1}{2c^2}\boldsymbol{u}^2.
\end{eqnarray}
Furthermore, based on Eqs.~(\ref{equ3.63}) and (\ref{equ3.68})---(\ref{equ3.70}), the spacetime splits of the four-acceleration of the gyroscope and the four-force acting on it are able to be given, respectively, and in view of Eq.~(\ref{equ4.1}), we only give the result of the four-force,
\begin{eqnarray}
\label{equ4.13}f\gamma^{0}&=&\gamma_{u}\left(\frac{\boldsymbol{f}\cdot\boldsymbol{u}}{c}+\boldsymbol{f}\right).
\end{eqnarray}
In the Lense-Thirring spacetime, after inserting Eqs.~(\ref{equC14}) and (\ref{equC15}) into Eqs.~(\ref{equ3.68}) and (\ref{equ3.69}), the expressions of the relative force $\boldsymbol{f}$ exerted on the gyroscope and the corresponding power $\boldsymbol{f}\cdot\boldsymbol{u}$ delivered by it up to $1/c^3$ order are derived,
\begin{eqnarray}
\label{equ4.14}\boldsymbol{f}
&=&m\bigg(\boldsymbol{a}-\boldsymbol{\nabla}U-\frac{1}{c^2}\boldsymbol{u}^2\boldsymbol{\nabla}U-\frac{1}{c^{2}}U\boldsymbol{\nabla}U
+\frac{2}{c^2}\left(\boldsymbol{u}\cdot\boldsymbol{\nabla}U\right)\boldsymbol{u}-\frac{4}{c^2}\boldsymbol{u}\times\left(\boldsymbol{\nabla}\times\boldsymbol{V}\right)+\frac{1}{mc^2}(\boldsymbol{u}\cdot\boldsymbol{f})\boldsymbol{u}+\frac{1}{2mc^2}\boldsymbol{u}^2\boldsymbol{f}\bigg)\nonumber\\
\end{eqnarray}
and
\begin{eqnarray}
\label{equ4.15}\boldsymbol{f}\cdot\boldsymbol{u}&=&m\left(\boldsymbol{u}\cdot\left(\boldsymbol{a}-\boldsymbol{\nabla}U\right)+\frac{1}{c^2}\left(\boldsymbol{u}\cdot\boldsymbol{\nabla}U\right)\boldsymbol{u}^2-\frac{1}{c^{2}}\left(\boldsymbol{u}\cdot\boldsymbol{\nabla}U\right)U+\frac{3}{2mc^2}(\boldsymbol{u}\cdot\boldsymbol{f})\boldsymbol{u}^2\right)
\end{eqnarray}
with $\boldsymbol{\nabla}:=\boldsymbol{\sigma}^{k}\partial_{k}$ and $\boldsymbol{V}:=V_{i}\boldsymbol{\sigma}_{i}$. It could be verified that these two equations are compatible. By plugging the potential $U$ into Eq.~(\ref{equ4.14}), one will find that $-m\boldsymbol{\nabla}U$ is the Newtonian gravitational force acting on the gyroscope, and hence, at the leading order, Eqs.~(\ref{equ4.14}) and (\ref{equ4.15}) reduce to the corresponding results in Newtonian gravity, which means that Eq.~(\ref{equ4.14}) is a three-dimensional analogue of Newton's second law for the gyroscope in the Lense-Thirring spacetime. Evidently, the terms at the next-leading order fall into three classes that depend on $U$, $\boldsymbol{V}$, and $\boldsymbol{f}$, respectively, and as implied from Eq.~(\ref{equ4.8}), they should be resulted from gyroscopic motion through the spacetime curved by the mass of the source, rotation of the source, and gyroscopic non-geodesic motion. It will be seen that due to the same reasons, the spin of the gyroscope also experiences three types of precession. In Eqs.~(\ref{equ4.14}) and (\ref{equ4.15}), the corrections to the results in Newtonian gravity are presented in a very elegant way, which intuitively displays the powerful potential of the signature invariant GA framework formulated in Sec.~\ref{Sec:second} for application in spacetime physics.

Next, we begin to review the basis process of evaluating the precessional angular velocity of the gyroscope spin in the language of STA. Let $\{\gamma_{(\alpha)}\}$ be a local orthonormal frame comoving with the gyroscope, and by definition, the timelike vector $\gamma_{(0)}$ is given by $\gamma_{(0)}=u/c$. In order to determine the other three spacelike vectors $\gamma_{(i)}$ of $\{\gamma_{(\alpha)}\}$, the pure Lorentz boost between the gyroscope's four-velocity $u$ and the fiducial observer's four-velocity $c\gamma_{0}$ needs to be presented. According to Eqs.~(\ref{equ3.34}) and (\ref{equ3.35}), under the pure Lorentz boost generated by the rotor
\begin{eqnarray}
\label{equ4.16}\hat{L}:=\frac{c^2\pm u(c\gamma_{0})}{\sqrt{2c^2\left(c^2\pm u\cdot(c\gamma_{0})\right)}},
\end{eqnarray}
the vector $u$ is mapped to $c\gamma_{0}$ by
\begin{eqnarray}
\label{equ4.17}c\gamma_{0}=\tilde{\hat{L}}u\hat{L}.
\end{eqnarray}
After inserting Eq.~(\ref{equ4.11}) into Eq.~(\ref{equ4.16}), the rotor $\hat{L}$ is expressed as
\begin{eqnarray}
\label{equ4.18}\hat{L}=L_{0}+\boldsymbol{L}
\end{eqnarray}
with
\begin{equation}\label{equ4.19}
\left\{\begin{array}{lll}
\displaystyle L_{0}&=\displaystyle \frac{1+\gamma_{u}}{\sqrt{2\left(1+\gamma_{u}\right)}},\smallskip\\
\displaystyle \boldsymbol{L}&=\displaystyle \frac{\gamma_{u}(\boldsymbol{u}/c)}{\sqrt{2\left(1+\gamma_{u}\right)}}.
\end{array}\right.
\end{equation}
In addition, based on Eq.~(\ref{equ4.16}), one is able to directly check that being a rotor, $\hat{L}$ satisfies
\begin{eqnarray}
\label{equ4.20}\hat{L}\tilde{\hat{L}}=\tilde{\hat{L}}\hat{L}=1.
\end{eqnarray}
As indicated in Ref.~\cite{MTW1973}, the comoving orthonormal frame $\{\gamma_{(\alpha)}\}$ of the gyroscope is related to the tetrad $\{\gamma_{\alpha}\}$ by
\begin{eqnarray}
\label{equ4.21}\gamma_{(\alpha)}=\hat{L}\gamma_{\alpha}\tilde{\hat{L}},
\end{eqnarray}
and thus, the other three spacelike vectors $\gamma_{(i)}$ of $\{\gamma_{(\alpha)}\}$ are fully determined. One consequence of Eq.~(\ref{equ4.21}) is that
\begin{eqnarray}
\label{equ4.22}\gamma^{(\alpha)}=\hat{L}\gamma^{\alpha}\tilde{\hat{L}},
\end{eqnarray}
where $\{\gamma^{(\alpha)}\}$ is the reciprocal frame of $\{\gamma_{(\alpha)}\}$. Now, let us expand the spin $s$ of the gyroscope in its comoving frame $\{\gamma_{(\alpha)}\}$,
\begin{eqnarray}
\label{equ4.23}s=s^{(\alpha)}\gamma_{(\alpha)},
\end{eqnarray}
and by virtue of Eq.~(\ref{equ4.2}) and $u=c\gamma_{(0)}$, we have
\begin{eqnarray}
\label{equ4.24}s^{(0)}=s\cdot\gamma^{(0)}=\pm\frac{1}{c}s\cdot u=0.
\end{eqnarray}
In this case, Eq.~(\ref{equ4.6}) states that
\begin{eqnarray}
\label{equ4.25}\frac{d\left(s^{(i)}s_{(i)}\right)}{d\tau}=\mp\frac{d\left(s^{(i)}s^{(j)}\delta_{ij}\right)}{d\tau}=0.
\end{eqnarray}
The above two equations suggest that in the comoving frame $\{\gamma_{(\alpha)}\}$ of the gyroscope, its spin $\left(s^{(1)},s^{(2)},s^{(3)}\right)$ is a purely spatial vector with constant length, and therefore, from the viewpoint of the observer comoving with it, the spin $\left(s^{(1)},s^{(2)},s^{(3)}\right)$ experiences a spatial rotation. That is to say, the spin of the gyroscope always precesses relative to its comoving frame $\{\gamma_{(\alpha)}\}$. The objective of the derivation in this section is to first write down the equation satisfied by $\left(s^{(1)},s^{(2)},s^{(3)}\right)$, and then derive the expression of the precessional angular velocity of the gyroscope spin up to $1/c^3$ order within the signature invariant GA framework formulated in Sec.~\ref{Sec:second}.

From Eqs.~(\ref{equ3.41}), (\ref{equ4.5}), and (\ref{equA14}), the differential equation satisfied by the spin $s$ of the gyroscope is
\begin{eqnarray}
\label{equ4.26}\frac{ds}{d\tau}=\mp\frac{1}{c^2}(a\cdot s)u-\omega(u)\cdot s,
\end{eqnarray}
where from Eqs.~(\ref{equ3.42a}) and (\ref{equ3.42b}),
\begin{eqnarray}
\label{equ4.27}\frac{ds}{d\tau}=u\cdot\partial s=\frac{ds^{\alpha}}{d\tau}\gamma_{\alpha}.
\end{eqnarray}
Motivated by this, we could consider the derivative of $s^{(i)}$ with respect to $\tau$, namely $ds^{(i)}/d\tau$ so as to obtain the equation fulfilled by $\left(s^{(1)},s^{(2)},s^{(3)}\right)$. On the basis of Eqs.~(\ref{equ4.22}) and $\left\langle AB\right\rangle=\left\langle BA\right\rangle$, there is
\begin{eqnarray}
\label{equ4.28}s^{(\alpha)}=s\cdot\gamma^{(\alpha)}=\left\langle s\hat{L}\gamma^{\alpha}\tilde{\hat{L}}\right\rangle=\left\langle\tilde{\hat{L}}s\hat{L}\gamma^{\alpha}\right\rangle=s'\cdot\gamma^{\alpha}
\end{eqnarray}
with
\begin{eqnarray}
\label{equ4.29}s':=\tilde{\hat{L}}s\hat{L},
\end{eqnarray}
and as a result, the effect of the pure Lorentz boost generated by the rotor $\hat{L}$ can be seen by taking $s$ to $s'$. By means of Eqs.~(\ref{equ4.24}), (\ref{equ4.27}), and (\ref{equ4.28}), one will find that
\begin{eqnarray}
\label{equ4.30}s'=s^{(i)}\gamma_{i}\Rightarrow \frac{ds'}{d\tau}=\frac{ds^{(i)}}{d\tau}\gamma_{i},
\end{eqnarray}
from which, the spacetime splits of $s'$ and $ds'/d\tau$ with $\gamma_{0}$ are, respectively,
\begin{eqnarray}
\label{equ4.31}s'\gamma^{0}&=&\boldsymbol{s'}=s^{(i)}\boldsymbol{\sigma}_{i},\\
\label{equ4.32}\frac{ds'}{d\tau}\gamma^{0}&=&\frac{d\boldsymbol{s'}}{d\tau}=\frac{ds^{(i)}}{d\tau}\boldsymbol{\sigma}_{i}.
\end{eqnarray}
Thus, instead of $ds^{(i)}/d\tau$, the expression of $d\boldsymbol{s'}/d\tau$ could be deduced hereafter, and since it is more convenient to be handle $d\boldsymbol{s'}/d\tau$ in GA, working with $\boldsymbol{s'}$ will greatly facilitate the calculations. With the aid of Eqs.~(\ref{equ4.17}), (\ref{equ4.26}), (\ref{equ4.29}), and (\ref{equ4.32}), $d\boldsymbol{s'}/d\tau$ is evaluated as
\begin{eqnarray}
\label{equ4.33}\frac{d\boldsymbol{s'}}{d\tau}&=&\left(\tilde{\hat{L}}\frac{ds}{d\tau}\hat{L}+\frac{d\tilde{\hat{L}}}{d\tau}s\hat{L}+\tilde{\hat{L}}s\frac{d\hat{L}}{d\tau}\right)\gamma^{0}\nonumber\\
&=&\mp\frac{1}{c}a\cdot s-\tilde{\hat{L}}\left(\omega(u)\cdot s\right)\hat{L}\gamma^{0}+\left[\left(\frac{d\tilde{\hat{L}}}{d\tau}\hat{L}\right)s'+s'\left(\tilde{\hat{L}}\frac{d\hat{L}}{d\tau}\right)\right]\gamma^{0}.
\end{eqnarray}
Define
\begin{eqnarray}
\label{equ4.34}a'=\tilde{\hat{L}}a\hat{L},
\end{eqnarray}
and because of Eqs.~(\ref{equ4.4}), (\ref{equ4.20}), (\ref{equ4.22}), and $u=c\gamma_{(0)}$,
\begin{eqnarray}
\label{equ4.35}a'\cdot\gamma^{0}=\Big\langle a'\gamma^{0}\Big\rangle=\left\langle a\gamma^{(0)}\right\rangle=a\cdot\gamma^{(0)}=\pm\frac{1}{c}a\cdot u=0
\end{eqnarray}
holds, which results in that the spacetime split of $a'$ with $\gamma_{0}$ is
\begin{eqnarray}
\label{equ4.36}a'\gamma^{0}&=&\boldsymbol{a'}.
\end{eqnarray}
Via this result, $a\cdot s$ is able to be written as
\begin{eqnarray}
\label{equ4.37}a\cdot s=\left\langle as\right\rangle=\left\langle a's'\right\rangle=\pm\left\langle a'\gamma^{0}\gamma^{0}s'\right\rangle=\mp\boldsymbol{a'}\cdot \boldsymbol{s'},
\end{eqnarray}
where Eqs.~(\ref{equ4.29})---(\ref{equ4.31}) and $\gamma^{0}\gamma_{i}=-\gamma_{i}\gamma^{0}$ are used. By further using Eqs.~(\ref{equA6}) and (\ref{equA19}), one gets
\begin{eqnarray}
\label{equ4.38}-\tilde{\hat{L}}\left(\omega(u)\cdot s\right)\hat{L}&=&\tilde{\hat{L}}\left(s\cdot\omega(u)\right)\hat{L}=s'\cdot\left(\tilde{\hat{L}}\omega(u)\hat{L}\right),
\end{eqnarray}
and here, because $\tilde{\hat{L}}\omega(u)\hat{L}$ is an even multivector satisfying
$$\widetilde{\tilde{\hat{L}}\omega(u)\hat{L}}=-\tilde{\hat{L}}\omega(u)\hat{L},$$
it is a bivector. In addition, Eq.~(\ref{equ4.20}) provides
\begin{eqnarray}
\label{equ4.39}\frac{d\tilde{\hat{L}}}{d\tau}\hat{L}=\widetilde{\tilde{\hat{L}}\frac{d\hat{L}}{d\tau}}=-\tilde{\hat{L}}\frac{d\hat{L}}{d\tau},
\end{eqnarray}
which means that like $\tilde{\hat{L}}\omega(u)\hat{L}$, $\tilde{\hat{L}}(d\hat{L}/d\tau)$ is also a bivector. Thus, by employing Eqs.~(\ref{equA6}) and (\ref{equA14}), the term inside the square brackets in Eq.~(\ref{equ4.33}) is
\begin{eqnarray}
\label{equ4.40}\left(\frac{d\tilde{\hat{L}}}{d\tau}\hat{L}\right)s'+s'\left(\tilde{\hat{L}}\frac{d\hat{L}}{d\tau}\right)=2s'\times\left(\tilde{\hat{L}}\frac{d\hat{L}}{d\tau}\right)=s'\cdot\left(2\tilde{\hat{L}}\frac{d\hat{L}}{d\tau}\right).
\end{eqnarray}
After inserting Eqs.~(\ref{equ4.37}), (\ref{equ4.38}), and (\ref{equ4.40}) into Eq.~(\ref{equ4.33}), $d\boldsymbol{s'}/d\tau$ is rewritten as
\begin{eqnarray}
\label{equ4.41}\frac{d\boldsymbol{s'}}{d\tau}&=&\frac{1}{c}\boldsymbol{a'}\cdot\boldsymbol{s'}+\left(s'\cdot\varOmega(\tau)\right)\gamma^{0}
\end{eqnarray}
with the bivector $\varOmega(\tau)$ defined by
\begin{eqnarray}
\label{equ4.42}\varOmega(\tau)=\tilde{\hat{L}}\omega(u)\hat{L}+2\tilde{\hat{L}}\frac{d\hat{L}}{d\tau}.
\end{eqnarray}
This result indicates that the motion of the spin of the gyroscope relative to the comoving frame $\{\gamma_{(\alpha)}\}$ is completely determined by the bivector field $\varOmega(\tau)$ along its worldline, where $\varOmega(\tau)$ is dependent on the rotor $\hat{L}$ generating the pure Lorentz boost from the gyroscope's four-velocity $u$ to the fiducial observer's four-velocity $c\gamma_{0}$ and the bivector connection $\omega(u)$ associated with the tetrad $\{\gamma_{\alpha}\}$.
Like the bivector connection $\omega(u)$ in Eqs.~(\ref{equ3.60a})---(\ref{equ3.61c}), the bivector $\varOmega(\tau)$ is also able to be decomposed into the electric part $\varOmega^{(E)}(\tau)$ and the magnetic part $\varOmega^{(B)}(\tau)$,
\begin{subequations}
\begin{eqnarray}
\label{equ4.43a}\varOmega^{(E)}(\tau):&=&\Big(\varOmega(\tau)\cdot\big(\gamma^{k}\wedge\gamma^{0}\big)\Big)\gamma_{0}\wedge\gamma_{k},\\
\label{equ4.43b}\varOmega^{(B)}(\tau):&=&\sum_{i<j}\Big(\varOmega(\tau)\cdot\big(\gamma^{j}\wedge\gamma^{i}\big)\Big)\gamma_{i}\wedge\gamma_{j},\\
\label{equ4.43c}\varOmega(\tau)&=&\varOmega^{(E)}(\tau)+\varOmega^{(B)}(\tau),
\end{eqnarray}
\end{subequations}
and they satisfy
\begin{subequations}
\begin{eqnarray}
\label{equ4.44a}\gamma_{0}\varOmega(\tau)\gamma^{0}&=&-\varOmega^{(E)}(\tau)+\varOmega^{(B)}(\tau),\\
\label{equ4.44b}\varOmega^{(E)}(\tau)&=&\frac{1}{2}\left(\varOmega(\tau)-\gamma_{0}\varOmega(\tau)\gamma^{0}\right),\\
\label{equ4.44c}\varOmega^{(B)}(\tau)&=&\frac{1}{2}\left(\varOmega(\tau)+\gamma_{0}\varOmega(\tau)\gamma^{0}\right).
\end{eqnarray}
\end{subequations}
Consequently, based on Eqs.~(\ref{equ4.30}), (\ref{equ4.31}), and the relevant formulas in Appendix A, the second term in Eq.~(\ref{equ4.41}) can be recast as
\begin{eqnarray}
\label{equ4.45}\left(s'\cdot\varOmega(\tau)\right)\gamma^{0}&=&\left(s'\cdot\varOmega(\tau)\right)\cdot\gamma^{0}+\left(s'\cdot\varOmega(\tau)\right)\wedge\gamma^{0}\nonumber\\
&=&\left(\gamma^{0}\wedge s'\right)\cdot\varOmega(\tau)+\left(s^{(i)}\gamma_{i}\cdot\varOmega(\tau)\right)\wedge\gamma^{0}\nonumber\\
&=&-\boldsymbol{s'}\cdot\varOmega^{(E)}(\tau)+\left(s^{(i)}\gamma_{i}\cdot\varOmega^{(B)}(\tau)\right)\wedge\gamma^{0}+\left(s^{(i)}\gamma_{i}\wedge\varOmega^{(B)}(\tau)\right)\cdot\gamma^{0}\nonumber\\
&=&-\boldsymbol{s'}\cdot\varOmega^{(E)}(\tau)+\left\langle s^{(i)}\gamma_{i}\varOmega^{(B)}(\tau)\gamma^{0}\right\rangle_{2}\nonumber\\
&=&-\boldsymbol{s'}\cdot\varOmega^{(E)}(\tau)+\boldsymbol{s'}\times\varOmega^{(B)}(\tau),
\end{eqnarray}
where in the third and fifth steps, $\gamma^{0}\gamma_{i}=-\gamma_{i}\gamma^{0}$ has been used. Finally, by substituting this result back in Eq.~(\ref{equ4.41}), we arrive at
\begin{eqnarray}
\label{equ4.46}\frac{d\boldsymbol{s'}}{d\tau}&=&\boldsymbol{s'}\cdot\left(\frac{\boldsymbol{a'}}{c}-\varOmega^{(E)}(\tau)\right)+\boldsymbol{s'}\times\varOmega^{(B)}(\tau),
\end{eqnarray}
which is the differential equation describing the motion of the spin of the gyroscope relative to its comoving frame $\{\gamma_{(\alpha)}\}$. As analyzed previously, in this frame, the gyroscope spin $\left(s^{(1)},s^{(2)},s^{(3)}\right)$ experiences a spatial rotation, and therefore, Eq.~(\ref{equ4.46}) should depict the precession of
$\boldsymbol{s'}=s^{(i)}\boldsymbol{\sigma}_{i}$. If the condition
\begin{eqnarray}
\label{equ4.47}\frac{\boldsymbol{a'}}{c}=\varOmega^{(E)}(\tau)
\end{eqnarray}
holds, Eq.~(\ref{equ4.46}) reduces to
\begin{eqnarray}
\label{equ4.48}\frac{d\boldsymbol{s'}}{d\tau}&=&\boldsymbol{s'}\times\varOmega^{(B)}(\tau),
\end{eqnarray}
and then,
\begin{eqnarray}
\label{equ4.49}\frac{d\boldsymbol{s'}^2}{d\tau}&=&2\boldsymbol{s'}\cdot\left(\boldsymbol{s'}\times\varOmega^{(B)}(\tau)\right)=2\left\langle\boldsymbol{s'}^2\varOmega^{(B)}(\tau)\right\rangle=2\boldsymbol{s'}^2\left\langle\varOmega^{(B)}(\tau)\right\rangle=0,
\end{eqnarray}
where in terms of the three-dimensional meaning in the relative space, the conservation of $\boldsymbol{s'}^2$ along the worldline of the gyroscope means that Eq.~(\ref{equ4.48}) is the equation depicting the precession of  $\boldsymbol{s'}$. In this case, based on Eqs.~(\ref{equB13}), (\ref{equB14}), and (\ref{equB20}), Eq.~(\ref{equ4.48}) is capable of being transformed into
\begin{eqnarray}
\label{equ4.50}\frac{d\boldsymbol{s'}}{d\tau}&=&-\left\langle\boldsymbol{s'}\varOmega^{(B)}(\tau)II\right\rangle_{2}=-\left\langle\boldsymbol{s'}\varOmega^{(B)}(\tau)I\right\rangle_{2}I
=-\left(\boldsymbol{s'}\times\left(\varOmega^{(B)}(\tau)I\right)\right)I=\left(-\varOmega^{(B)}(\tau)I\right)\times_{3}\boldsymbol{s'},
\end{eqnarray}
which clearly suggests that $-\varOmega^{(B)}(\tau)I$, as a relative vector, is the precessional angular velocity of $\boldsymbol{s'}$ in the conventional sense, and because the cross product (denoted by $\times_{3}$) is rarely employed in GA, the relative bivector $\varOmega^{(B)}(\tau)$ could be regarded as the precessional angular velocity of $\boldsymbol{s'}$. That is to say, in the comoving frame $\{\gamma_{(\alpha)}\}$ of the gyroscope, its spin always precesses with $\varOmega^{(B)}(\tau)$ as the precessional angular velocity. In addition, one should also note that since (\ref{equ4.46}) or (\ref{equ4.48}) has been represented in the ``common'' even subalgebra of the STAs of signatures $(\pm,\mp,\mp,\mp)$, a signature invariant GA derivation of the precessional angular velocity of the gyroscope spin could be found.

In order to make a further analysis on Eq.~(\ref{equ4.46}), we need to derive the expressions of $\varOmega^{(E)}(\tau)$, $\varOmega^{(B)}(\tau)$, and $\boldsymbol{a'}$. Let us first evaluate the corresponding results of $\varOmega^{(E)}(\tau)$ and $\varOmega^{(B)}(\tau)$, and as shown in Eqs.~(\ref{equ4.43a})---(\ref{equ4.43c}), their expressions can directly be read out from that of the bivector $\varOmega(\tau)$. By plugging Eqs.~(\ref{equ4.18}) and (\ref{equ3.60c}) into Eq.~(\ref{equ4.42}), the bivector $\varOmega(\tau)$ is able to be expressed as
\begin{eqnarray}
\label{equ4.51}\varOmega(\tau)&=&\left\langle(L_{0}-\boldsymbol{L})\left(\omega^{(E)}(u)+\omega^{(B)}(u)\right)(L_{0}+\boldsymbol{L})\right\rangle_{2}+\left\langle2(L_{0}-\boldsymbol{L})\left(\frac{dL_{0}}{d\tau}+\frac{d\boldsymbol{L}}{d\tau}\right)\right\rangle_{2}\nonumber\\
&=&L_{0}^2\omega^{(E)}(u)+L_{0}^2\omega^{(B)}(u)+\boldsymbol{L}^2\omega^{(E)}(u)+\boldsymbol{L}^2\omega^{(B)}(u)+2L_{0}\omega^{(E)}(u)\times\boldsymbol{L}+2L_{0}\omega^{(B)}(u)\times\boldsymbol{L}
\nonumber\\
&&-2\left(\omega^{(E)}(u)\cdot\boldsymbol{L}\right)\boldsymbol{L}-2\left(\omega^{(B)}(u)\wedge\boldsymbol{L}\right)\cdot\boldsymbol{L}+2L_{0}\frac{d\boldsymbol{L}}{d\tau}-2\frac{dL_{0}}{d\tau}\boldsymbol{L}-2\boldsymbol{L}\times\frac{d\boldsymbol{L}}{d\tau},
\end{eqnarray}
in which Eqs.~(\ref{equ3.60a}), (\ref{equ3.60b}), (\ref{equ4.19}),  (\ref{equB15}), and
\begin{equation}\label{equ4.52}
\boldsymbol{L}^2\omega^{(B)}(u)=\left\langle\omega^{(B)}(u)\boldsymbol{L}\boldsymbol{L}\right\rangle_{2}=\left(\omega^{(B)}(u)\times\boldsymbol{L}\right)\times\boldsymbol{L}+\left(\omega^{(B)}(u)\wedge\boldsymbol{L}\right)\cdot\boldsymbol{L}
\end{equation}
have been used. Equations~(\ref{equ4.43a})---(\ref{equ4.43c}) suggest that the timelike vector $\gamma_{0}$ or $\gamma^{0}$ only appears in the electric part $\varOmega^{(E)}(\tau)$ of the bivector $\varOmega(\tau)$, and thus, from Eq.~(\ref{equ4.51}),
\begin{subequations}
\begin{eqnarray}
\label{equ4.53a}\varOmega^{(E)}(\tau)&=&L_{0}^2\omega^{(E)}(u)+\boldsymbol{L}^2\omega^{(E)}(u)+2L_{0}\omega^{(B)}(u)\times\boldsymbol{L}
-2\left(\omega^{(E)}(u)\cdot\boldsymbol{L}\right)\boldsymbol{L}+2L_{0}\frac{d\boldsymbol{L}}{d\tau}-2\frac{dL_{0}}{d\tau}\boldsymbol{L},\\
\label{equ4.53b}\varOmega^{(B)}(\tau)&=&L_{0}^2\omega^{(B)}(u)+\boldsymbol{L}^2\omega^{(B)}(u)+2L_{0}\omega^{(E)}(u)\times\boldsymbol{L}
-2\left(\omega^{(B)}(u)\wedge\boldsymbol{L}\right)\cdot\boldsymbol{L}-2\boldsymbol{L}\times\frac{d\boldsymbol{L}}{d\tau}.
\end{eqnarray}
\end{subequations}
After substituting Eq.~(\ref{equ4.19}) in the above results, the expressions of $\varOmega^{(E)}(\tau)$ and $\varOmega^{(B)}(\tau)$ are derived,
\begin{subequations}
\begin{eqnarray}
\label{equ4.54a}\varOmega^{(E)}(\tau)&=&\gamma_{u}\omega^{(E)}(u)+\frac{\gamma_{u}}{c}\omega^{(B)}(u)\times\boldsymbol{u}
-\frac{\gamma_{u}^2}{c^2(1+\gamma_{u})}\left(\omega^{(E)}(u)\cdot\boldsymbol{u}\right)\boldsymbol{u}+\frac{\gamma_{u}^4}{c^3(1+\gamma_{u})}(\boldsymbol{u}\cdot\boldsymbol{a})\boldsymbol{u}+\frac{\gamma_{u}^2}{c}\boldsymbol{a},\\
\label{equ4.54b}\varOmega^{(B)}(\tau)&=&\gamma_{u}\omega^{(B)}(u)+\frac{\gamma_{u}}{c}\omega^{(E)}(u)\times\boldsymbol{u}
-\frac{\gamma_{u}^2}{c^2(1+\gamma_{u})}\left(\omega^{(B)}(u)\wedge\boldsymbol{u}\right)\cdot\boldsymbol{u}-\frac{\gamma_{u}^3}{c^2(1+\gamma_{u})}\boldsymbol{u}\times\boldsymbol{a}.
\end{eqnarray}
\end{subequations}
We turn now to the evaluation of $\boldsymbol{a'}$. Due to $\gamma_{i}\gamma^{0}=-\gamma^{0}\gamma_{i}$ and $\boldsymbol{u}=u^{i}\boldsymbol{\sigma}_{i}=u^{i}\gamma_{i}\gamma^{0}$, one gets $\boldsymbol{u}\gamma^{0}=-\gamma^{0}\boldsymbol{u}$, which leads to $\hat{L}\gamma^{0}=\gamma^{0}\tilde{\hat{L}}$ via Eqs.~(\ref{equ4.18}) and (\ref{equ4.19}). As a consequence, by means of Eqs.~(\ref{equ4.34}), (\ref{equ4.36}), (\ref{equ4.1}), and (\ref{equ4.13}),
\begin{eqnarray}
\label{equ4.55}\boldsymbol{a'}&=&\left\langle\tilde{\hat{L}}a\hat{L}\gamma^{0}\right\rangle_{2}=\left\langle\tilde{\hat{L}}a\gamma^{0}\tilde{\hat{L}}\right\rangle_{2}=\frac{1}{m}\left\langle\tilde{\hat{L}}f\gamma^{0}\tilde{\hat{L}}\right\rangle_{2}\nonumber\\
&=&\frac{\gamma_{u}}{m}\left[\frac{1}{c}(\boldsymbol{f}\cdot\boldsymbol{u})\left\langle\left(\tilde{\hat{L}}\right)^2\right\rangle_{2}+\left\langle\tilde{\hat{L}}\boldsymbol{f}\tilde{\hat{L}}\right\rangle_{2}\right].
\end{eqnarray}
Define the unit relative vector
\begin{eqnarray}
\label{equ4.56}\boldsymbol{e}_{u}&=&\frac{\boldsymbol{u}}{\sqrt{\boldsymbol{u}^2}},
\end{eqnarray}
and the components of the relative force $\boldsymbol{f}$ parallel and perpendicular to it can be determined by following the method presented in Sec.~\ref{Sec:3.1},
\begin{subequations}
\begin{eqnarray}
\label{equ4.57a}\boldsymbol{f}_{\parallel}&=&\big(\boldsymbol{f}\cdot\boldsymbol{e}_{u}\big)\boldsymbol{e}_{u}=\frac{1}{\boldsymbol{u}^2}\big(\boldsymbol{f}\cdot\boldsymbol{u}\big)\boldsymbol{u},\\ \label{equ4.57b}\boldsymbol{f}_{\perp}&=&\big(\boldsymbol{f}\times\boldsymbol{e}_{u}\big)\times\boldsymbol{e}_{u}=\big(\boldsymbol{f}\times\boldsymbol{e}_{u}\big)\boldsymbol{e}_{u}=\frac{1}{\boldsymbol{u}^2}\big(\boldsymbol{f}\times\boldsymbol{u}\big)\boldsymbol{u},
\end{eqnarray}
\end{subequations}
where they satisfy
\begin{eqnarray}
\label{equ4.58}\boldsymbol{f}&=&\boldsymbol{f}_{\parallel}+\boldsymbol{f}_{\perp}
\end{eqnarray}
and
\begin{subequations}
\begin{eqnarray}
\label{equ4.59a}\boldsymbol{f}_{\parallel}\boldsymbol{e}_{u}&=&\boldsymbol{e}_{u}\boldsymbol{f}_{\parallel},\\ \label{equ4.59b}\boldsymbol{f}_{\perp}\boldsymbol{e}_{u}&=&-\boldsymbol{e}_{u}\boldsymbol{f}_{\perp}.
\end{eqnarray}
\end{subequations}
Thus, together with Eqs.~(\ref{equ4.18}), (\ref{equ4.19}), and (\ref{equ4.56}), there are
\begin{subequations}
\begin{eqnarray}
\label{equ4.60a}\boldsymbol{f}_{\parallel}\tilde{\hat{L}}&=&\tilde{\hat{L}}\boldsymbol{f}_{\parallel},\\
\label{equ4.60b}\boldsymbol{f}_{\perp}\tilde{\hat{L}}&=&\hat{L}\boldsymbol{f}_{\perp}.
\end{eqnarray}
\end{subequations}
Based on these two results, (\ref{equ4.57a}), and (\ref{equ4.58}), $\boldsymbol{a'}$ is able to be rewritten as
\begin{eqnarray}
\label{equ4.61}\boldsymbol{a'}
&=&\frac{\gamma_{u}}{m}\left[\frac{1}{c}(\boldsymbol{f}\cdot\boldsymbol{u})\left\langle\left(\tilde{\hat{L}}\right)^2\right\rangle_{2}+\left\langle\left(\tilde{\hat{L}}\right)^2\boldsymbol{f}_{\parallel}\right\rangle_{2}+\left\langle\boldsymbol{f}_{\perp}\right\rangle_{2}\right]\nonumber\\
&=&c\gamma_{u}\omega^{(E)}(u)+\gamma_{u}\omega^{(B)}(u)\times\boldsymbol{u}
-\frac{\gamma_{u}^2}{c(1+\gamma_{u})}\left(\omega^{(E)}(u)\cdot\boldsymbol{u}\right)\boldsymbol{u}+\frac{\gamma_{u}^4}{c^2(1+\gamma_{u})}(\boldsymbol{u}\cdot\boldsymbol{a})\boldsymbol{u}+\gamma_{u}^2\boldsymbol{a},
\end{eqnarray}
where Eqs.~(\ref{equ4.18}) and (\ref{equ4.19}) have been used again.
Comparing Eq.~(\ref{equ4.61}) with Eq.~(\ref{equ4.54a}), it is easy to verify that Eq.~(\ref{equ4.47}) holds, and as noted before, the spin of the gyroscope always precesses relative to its comoving frame $\{\gamma_{(\alpha)}\}$ with $\varOmega^{(B)}(\tau)$ as the precessional angular velocity. Here, by making use of Eqs.~(\ref{equ3.55}) and (\ref{equ3.68}),
the expression of $\varOmega^{(B)}(\tau)$ in Eq.~(\ref{equ4.54b}) can be recast as
\begin{eqnarray}
\label{equ4.62}\varOmega^{(B)}(\tau)
&=&\frac{\gamma_{u}}{c(1+\gamma_{u})}\omega^{(E)}(u)\times\boldsymbol{u}+\omega^{(B)}(u)-\frac{\gamma_{u}}{mc^2(1+\gamma_{u})}\boldsymbol{u}\times\boldsymbol{f}.
\end{eqnarray}
This is the general formula for the precessional angular velocity of the spin of a gyroscope moving in curved spacetime. In the Lense-Thirring spacetime, one only needs to insert Eqs.~(\ref{equC14}), (\ref{equC15}), and (\ref{equ4.12}) into the above result, and then, the expression of $\varOmega^{(B)}(\tau)$ up to $1/c^3$ order is obtained,
\begin{eqnarray}
\label{equ4.63}\varOmega^{(B)}(\tau)
&=&\varOmega^{(B)}_{d}(\tau)+\varOmega^{(B)}_{LT}(\tau)+\varOmega^{(B)}_{T}(\tau)
\end{eqnarray}
with
\begin{equation}\label{equ4.64}
\left\{\begin{array}{lll}
\displaystyle \varOmega^{(B)}_{d}(\tau)&:=\displaystyle \frac{3}{2c^2}\boldsymbol{u}\times\boldsymbol{\nabla}U,\smallskip\\
\displaystyle \varOmega^{(B)}_{LT}(\tau)&:=\displaystyle \frac{2}{c^2}\boldsymbol{\nabla}\times\boldsymbol{V},\smallskip\\
\displaystyle \varOmega^{(B)}_{T}(\tau)&:=\displaystyle -\frac{1}{2mc^2}\boldsymbol{u}\times\boldsymbol{f}.
\end{array}\right.
\end{equation}
In the three-dimensional relative space, $\varOmega^{(B)}_{d}(\tau)$, $\varOmega^{(B)}_{LT}(\tau)$, and $\varOmega^{(B)}_{T}(\tau)$ describe three types of precession of the gyroscope spin in complete generality under the WFSM approximation. If the gyroscope does not experience any force, namely $\boldsymbol{f}=0$, $\varOmega^{(B)}(\tau)
=\varOmega^{(B)}_{d}(\tau)+\varOmega^{(B)}_{LT}(\tau)$ is the precessional angular velocity of its spin brought about by the curved spacetime in General Relativity.
As implied from Eqs.~(\ref{equ4.8}) and (\ref{equ4.64}), $\varOmega^{(B)}_{d}(\tau)$ and $\varOmega^{(B)}_{LT}(\tau)$
are resulted from gyroscopic motion through the spacetime curved by the mass of the source and rotation of the source, respectively, and hence, they should be the de Sitter precession and the Lense-Thirring precession~\cite{Ignazio1995}. Besides, $\varOmega^{(B)}_{T}(\tau)$, associated with the relative force $\boldsymbol{f}$ acting on the gyroscope, explicitly represents the Thomas precession of its spin, which is caused by gyroscopic non-geodesic motion. In the fine structure of atomic spectra, Thomas precession plays a significant role~\cite{MTW1973}.

Recall that the three-dimensional operator $\boldsymbol{\nabla}$ appearing in Eq.~(\ref{equ4.64}) is defined by $\boldsymbol{\nabla}=\boldsymbol{\sigma}^{k}\partial_{k}$ (cf.~(\ref{equ2.41})), and on the basis of it, the expression of $\varOmega^{(B)}_{d}(\tau)$ can be readily derived by the potential $U$ in Eq.~(\ref{equ4.8}),
\begin{eqnarray}
\label{equ4.65}\varOmega^{(B)}_{d}(\tau)=\displaystyle \frac{3GM}{2c^2r^3}\boldsymbol{r}\times\boldsymbol{u},
\end{eqnarray}
where $\boldsymbol{r}:=x^{i}\boldsymbol{\sigma}_{i}$ is the relative position vector of the gyroscope, and due to $r=\sqrt{x^{i}x^{i}}$, there is
$r=\sqrt{\boldsymbol{r}^2}$. In order to deduce the expression of $\varOmega^{(B)}_{LT}(\tau)$, some tricks need to be applied. In the language of GA, the relative angular momentum bivector $\boldsymbol{J}$, is more convenient to describe the rotation of the source. Eqs.~(\ref{equ4.8}) and (\ref{equC1}) indicate that the source is rotating around the $x^{3}$ axis, so its relative angular momentum vector is
\begin{eqnarray}
\label{equ4.66}\boldsymbol{J}_{pseu}&:=&J\boldsymbol{\sigma}_{3},
\end{eqnarray}
and then, from Eqs.~(\ref{equB14}), (\ref{equB20}), (\ref{equB2}), and (\ref{equB6}), its relative angular momentum bivector should be
\begin{eqnarray}
\label{equ4.67}\boldsymbol{J}&=&\boldsymbol{J}_{pseu}I=J\boldsymbol{\sigma}_{1}\times\boldsymbol{\sigma}_{2}=\frac{1}{2}J\epsilon_{3ij}\boldsymbol{\sigma}_{i}\times\boldsymbol{\sigma}_{j}.
\end{eqnarray}
Thus, via Eqs.~(\ref{equ4.8}) and (\ref{equB15}), one is able to express $\boldsymbol{V}$ as
\begin{eqnarray}
\label{equ4.68}\boldsymbol{V}&=&V_{i}\boldsymbol{\sigma}_{i}=-\frac{GJ\epsilon_{3ij}x^{j}}{2r^3}\boldsymbol{\sigma}_{i}
=\frac{GJ\epsilon_{3ij}x^{k}}{4r^3}\boldsymbol{\sigma}_{k}\times(\boldsymbol{\sigma}_{i}\times\boldsymbol{\sigma}_{j})
=\frac{G}{2r^3}\boldsymbol{r}\times\boldsymbol{J}.
\end{eqnarray}
Keeping in mind that the relative angular momentum bivector $\boldsymbol{J}$ of the source is conserved, the following identity holds,
\begin{eqnarray}
\label{equ4.69}\boldsymbol{\nabla}\times\left(\frac{1}{r^3}\boldsymbol{r}\times\boldsymbol{J}\right)&=&\boldsymbol{\nabla}\left(\frac{1}{r^3}\right)\times\left(\boldsymbol{r}\times\boldsymbol{J}\right)+\frac{1}{r^3}\boldsymbol{\sigma}_{i}\times\left(\partial_{i}\boldsymbol{r}\times\boldsymbol{J}\right)\nonumber\\
&=&-\frac{3}{r^5}\boldsymbol{r}\times\left(\boldsymbol{r}\times\boldsymbol{J}\right)+\frac{1}{r^3}\boldsymbol{\sigma}_{i}\times\left(\boldsymbol{\sigma}_{i}\times\boldsymbol{J}\right)\nonumber\\
&=&\left[-\frac{3}{r^5}\boldsymbol{r}\times\left(\boldsymbol{r}\times\boldsymbol{J}_{pseu}\right)+\frac{1}{r^3}\boldsymbol{\sigma}_{i}\times\left(\boldsymbol{\sigma}_{i}\times\boldsymbol{J}_{pseu}\right)\right]I\nonumber\\
&=&\left[-\frac{1}{r^3}\boldsymbol{J}_{pseu}+\frac{3}{r^5}\left(\boldsymbol{r}\cdot\boldsymbol{J}_{pseu}\right)\boldsymbol{r}\right]I\nonumber\\
&=&-\frac{1}{r^3}\boldsymbol{J}+\frac{3}{r^5}(\boldsymbol{r}\wedge\boldsymbol{J})\boldsymbol{r},
\end{eqnarray}
where in the third and fifth steps, $\boldsymbol{J}=\boldsymbol{J}_{pseu}I$ and (\ref{equA5}) have been used, and as a result, by use of Eqs.~(\ref{equ4.68}) and (\ref{equ4.69}), the expression of $\varOmega^{(B)}_{LT}(\tau)$ is obtained,
\begin{eqnarray}
\label{equ4.70}\varOmega^{(B)}_{LT}(\tau)=-\frac{G}{c^2r^5}\Big(r^2\boldsymbol{J}-3(\boldsymbol{r}\wedge\boldsymbol{J})\boldsymbol{r}\Big).
\end{eqnarray}
As discussed earlier, $\varOmega^{(B)}_{d}(\tau)$, $\varOmega^{(B)}_{LT}(\tau)$, and $\varOmega^{(B)}_{T}(\tau)$ are capable of being directly transformed into their corresponding expressions in the conventional sense by multiplying $-I$, namely,
\begin{equation}\label{equ4.71}
\left\{\begin{array}{lll}
\displaystyle -\varOmega^{(B)}_{d}(\tau)I&=\displaystyle \frac{3GM}{2c^2r^3}\boldsymbol{r}\times_{3}\boldsymbol{u},\smallskip\\
\displaystyle -\varOmega^{(B)}_{LT}(\tau)I&=\displaystyle -\frac{G}{c^2r^5}\Big(r^2\boldsymbol{J}_{pseu}-3(\boldsymbol{r}\cdot\boldsymbol{J}_{pseu})\boldsymbol{r}\Big),\smallskip\\
\displaystyle -\varOmega^{(B)}_{T}(\tau)I&=\displaystyle -\frac{1}{2mc^2}\boldsymbol{u}\times_{3}\boldsymbol{f}.
\end{array}\right.
\end{equation}
Although these expressions presented here seem to be identical to those in Refs.~\cite{Eric2014,Wu:2021uws}, one still needs to note that since the relative velocity $\boldsymbol{u}$ in $-\varOmega^{(B)}_{d}(\tau)I$ and $-\varOmega^{(B)}_{T}(\tau)I$ is measured in the orthonormal tetrad $\{\gamma_{\alpha}\}$ of the fiducial observer instead of in the coordinate frame $\{g_{\mu}\}$, their above expressions are slightly different from those obtained in tensor language. In despite of this, a straightforward calculation~\cite{Ignazio1995,Wu:2021uws} shows that the difference between the gyroscope's velocities measured in $\{\gamma_{\alpha}\}$ and in $\{g_{\mu}\}$ is at least at $1/c^2$ order, so the above $-\varOmega^{(B)}_{d}(\tau)I$ and $-\varOmega^{(B)}_{T}(\tau)I$ are essentially equivalent to their conventional expressions. These computations in the final part of this section display in detail how to give a signature invariant GA derivation of the precessional angular velocity of the gyroscope spin within the framework provided by the ``common'' even subalgebra of the STAs of signatures $(\pm,\mp,\mp,\mp)$, which could stand as a successful paradigm of the application of this framework in spacetime physics.

In this section, based on the STAs of signatures $(\pm,\mp,\mp,\mp)$ formulated in Sec.~\ref{Sec:second} and the GA techniques constructed in Sec.~\ref{Sec:third}, an efficient treatment of gyroscopic precession is achieved. One significant advantage of GA approach is that only geometric objects are involved during calculation, and thus, many equations are given a degree of clarity which is lost in tensor language. A typical example is that the relationship between the gyroscope spin $s$ and its components $s^{(i)}$ in the comoving frame $\{\gamma_{(\alpha)}\}$ is clearly shown by the equation $s^{(i)}=s\cdot\gamma^{(i)}$, which could help readers understand that instead of $s$, it is the spin $\left(s^{(1)},s^{(2)},s^{(3)}\right)$ in the frame $\{\gamma_{(\alpha)}\}$ that experiences a spatial rotation. However, in the classical derivation with tensor, since one always needs to work with the components of some tensor, the role of $s$ is usually played by its components in the coordinate frame $\{g_{\mu}\}$, and thus, the above equation is replaced by the corresponding component equations~\cite{Ignazio1995,Wu:2021uws}, from which, the relationship between $s$ and $s^{(i)}$ can not be explicitly reflected.

It should be noted that the application of the rotor techniques is also very crucial in simplifying the derivation. In the beginning, Eqs.~(\ref{equ4.24})---(\ref{equ4.26}) imply that in order to obtain the precessional angular velocity of the gyroscope spin $\left(s^{(1)},s^{(2)},s^{(3)}\right)$ in the frame $\{\gamma_{(\alpha)}\}$, the expression of $ds^{(i)}/d\tau$ needs to be given. Then as in Eq.~(\ref{equ4.28}), by employing the rotor techniques, the effect of the pure Lorentz boost generated by the rotor $\hat{L}$ is transformed from $\gamma^{(i)}$ to $s'$, and as a result, one can deal with the geometric object $ds'/d\tau=\left(ds^{(i)}/d\tau\right)\gamma_{i}$ rather than $ds^{(i)}/d\tau$.
Being a common trick in STA, such an approach is extremely useful for computations. The STAs of signatures $(\pm,\mp,\mp,\mp)$ and the GA techniques for General Relativity formulated in Ref.~\cite{Francis:2003xi} are organically integrated in Sec.~\ref{Sec:3.2}, so that physics in curved spacetime is able to be discussed within the signature invariant framework provided in Sec.~\ref{Sec:second}, which is perhaps the most easily overlooked contribution of the present paper. It is based on the results presented in Sec.~\ref{Sec:3.2} that relativistic dynamics of the gyroscope and the precession of its spin can be studied in the two STAs. In particular, within the framework provided by the ``common'' even subalgebra of the two STAs, the three-dimensional generalized equation of motion for the gyroscope and the precessional angular velocity of its spin are able to be derived in a signature invariant manner. The treatment of gyroscopic precession in this section intuitively displays the basic method of dealing with specific problems in curved spacetime within the signature invariant framework. In the future, if the applications of this method could be extended to a wider range, the study of spacetime physics in the language of GA will be greatly promoted.
\section{Summary and discussions~\label{Sec:fifth}}
Since the establishment of STA by David Hestenes, the signature $(+,-,-,-)$ has been widely used~\cite{Hestenes1966,Doran2003}, which may cause inconvenience to the application of STA in relativistic physics because plenty of literatures on relativity adopt the opposite signature $(-,+,+,+)$. Although the STA of signature $(-,+,+,+)$ was also used~\cite{Oppositesignature}, a lack of long-term attention to it results in that its applications are quite limited. In this paper, by following the original idea of Hestenes, the techniques related to relative vector and spacetime split are built up in the STA of signature $(-,+,+,+)$, so that a more convenient approach to relativistic
physics could be given in the language of GA. The further research suggests that the two even subalgebras of the STAs of signatures $(\pm,\mp,\mp,\mp)$ share the same operation rules, so that they could be treated as one algebraic formalism. Consequently, many calculations between vectors involved in a large number of specific problems can be transformed into those in this ``common'' even subalgebra of the two STAs through the techniques on spacetime split, and then be solved efficiently in a signature invariant manner with the help of various operations provided in Appendix B. Thus, the ``common'' even subalgebra of the two STAs provides a signature invariant GA framework for spacetime physics.

When orthogonal transformations in spaces of arbitrary signature are performed, calculations with rotors are demonstrably more efficient than calculations with matrices, which is a remarkable advantage of GA. Therefore, the topic of rotor techniques on Lorentz transformation should be specifically addressed in the STAs of signatures $(\pm,\mp,\mp,\mp)$, and  what needs to be pointed out is that since rotor techniques have not been fully developed in the STA of signature $(-,+,+,+)$, it is significant to explicitly elaborate how to construct the rotors inducing Lorentz boost and spatial rotation in this algebraic formalism. In the present paper, by constructing the rotors on the basis of the exponential function defined on the ``common'' even subalgebra of the two STAs, the general Lorentz boost with velocity in an arbitrary direction and the general spatial rotation in an arbitrary plane are
handled in a signature invariant manner.

Relativistic dynamics of a massive particle in curved spacetime is also studied so as to describe the motion of a gyroscope moving around a gravitating source~\cite{Weinberg2014}. To this end, the two STAs and their ``common'' even subalgebra are first generated by a local orthonormal tetrad, and thus, the corresponding signature invariant GA framework can be set up. Then, after organically integrating the STAs of signatures $(\pm,\mp,\mp,\mp)$ and the GA techniques for General Relativity formulated in Ref.~\cite{Francis:2003xi}, physics in curved spacetime is able to be discussed within the signature invariant framework provided in Sec.~\ref{Sec:second}, which lays the foundation for dealing with gyroscope precession hereafter. With these preparations, for a massive particle, the spacetime splits of the velocity, acceleration, momentum, and force four-vectors with the normalized four-velocity
of the fiducial observer are derived, and as a consequence, a three-dimensional analogue of Newton's second law for this particle in curved spacetime is achieved. Since the result is derived in a comoving orthonormal tetrad of the fiducial observer and is presented in the form of geometric objects, it is an elegant generalization of the classical one in flat spacetime.

As a comprehensive application of the GA techniques constructed before, the last task of this paper is to provide an efficient treatment of gyroscopic precession in the STAs of signatures $(\pm,\mp,\mp,\mp)$. For a gyroscope moving in the Lense-Thirring spacetime, its relativistic dynamics is first discussed, and
some significant results like the three-dimensional generalized equation of motion for the gyroscope are given. Then, by applying the rotor techniques, the geometric object $ds'/d\tau=\left(ds^{(i)}/d\tau\right)\gamma_{i}$ is able to be directly dealt with instead of $ds^{(i)}/d\tau$, which greatly simplifies the following derivation. The result suggests that if Eq.~(\ref{equ4.47}) holds, the spin of the gyroscope always precesses relative to its comoving frame $\{\gamma_{(\alpha)}\}$ with $\varOmega^{(B)}(\tau)$ as the precessional angular velocity. Within the framework provided by the ``common'' even subalgebra of the two STAs, signature invariant expressions of the relevant physical quantities involved in Eq.~(\ref{equ4.46}) are deduced, which clearly indicates that Eq.~(\ref{equ4.47}) holds, and therefore, the gyroscope spin indeed precesses in the frame $\{\gamma_{(\alpha)}\}$.  After expanding $\varOmega^{(B)}(\tau)$ up to $1/c^3$ order, the gyroscope spin's angular velocities of the de Sitter precession, the Lense-Thirring precession, and the Thomas precession are all directly read out, and their expressions, in the form of geometric objects, are equivalent to their conventional ones in component form, respectively.

All physical laws should be independent of the choice of signature, which implies that many significant techniques constructed in the STA of signature $(+,-,-,-)$ can also be introduced to the STA of signature $(-,+,+,+)$, and starting from this motivation, we find that the ``common'' even subalgebra of the two STAs provides a signature invariant GA framework for spacetime physics. In order to pave the way for the applications of these two STAs and their ``common'' even subalgebra, we elaborate in detail the rotor techniques on Lorentz transformation and the method of handling physics in curved spacetime within the signature invariant framework, and they are of theoretical significance and of practical worth. As two successful paradigms, the treatment of relativistic dynamics of a massive particle and gyroscopic precession clearly shows that the GA techniques constructed in this paper are efficient and reliable. Being straightforward generalizations, these techniques could also be applied to gyroscopic precession in alternative theories of gravity, such as $f(R)$ gravity~\cite{fRtheory,Wu:2021uws}, $f(R,\mathcal{G})$ gravity~\cite{fRGtheory}, and $f(X,Y,Z)$ gravity~\cite{Stabile:2010mz}. However, since these topics are usually explored by making use of some complicated mathematical tools (e.g., the symmetric and trace-free formalism in terms of the irreducible Cartesian tensors~\cite{Wu:2021uws}), it is crucial to develop new techniques to apply these tools in STA. In fact, by generalizing various GA techniques in STA of signature $(+,-,-,-)$~\cite{Hestenes1966,Hestenes1984,Hestenes1986,Doran2003}, the approach in this paper could also be applied to other fields, and it has been verified that some topics in classical mechanics and electrodynamics can be described in such a manner. We expect that the applications of this approach will be extended to a wider range in the future, so that the study of spacetime physics in the language of GA could be greatly promoted.
\begin{acknowledgments}
This work was supported by China Postdoctoral Science Foundation (Grant No. 2021M690569), the National Natural Science Foundation of China (Grant No. 12105039).
\end{acknowledgments}
\appendix\label{appendix}
\section{List of operation rules of blades in the STAs of signatures $(\pm,\mp,\mp,\mp)$}
As mentioned in Sec.~\ref{Sec:second}, once operation rules of blades of different grades in both the STAs are given,
one can perform operations between any two multivectors. Based on the general formulas in finite dimensional GA~\cite{Hestenes1984,Doran2003},
a detail list of operation rules of blades in the two STAs is provided as follows, where since all of the following equations are presented in a signature independent form, the signs ``$\pm$'' associated with multivectors have been omitted. In the list below, $a,b,c,d,e,f,g$, and $h$ are vectors, $B,B_{1}$, and $B_{2}$ are 2-blades, $T,T_{1}$, and $T_{2}$ are 3-blades, $F$ is 4-blade, and $A_{r}$ is $r$-blade $(r=1,2,3,4)$.\\
\begin{eqnarray}
\label{equA1}aA_{r}&=&a\cdot A_{r}+a\wedge A_{r},\ A_{r}a=A_{r}\cdot a+A_{r}\wedge a,\quad (r=1,2,3),\\
\label{equA2}B_{1}B_{2}&=&B_{1}\cdot B_{2}+B_{1}\times B_{2}+B_{1}\wedge B_{2},\\
\label{equA3}BT&=&B\cdot T+B\times T,\quad TB=T\cdot B+T\times B,\\
\label{equA4}T_{1}T_{2}&=&T_{1}\cdot T_{2}+\langle T_{1}T_{2}\rangle_{2},\\
\label{equA5}A_{r}F&=&A_{r}\cdot F=(-1)^{3r}F\cdot A_{r}=(-1)^{3r}FA_{r},\quad (r=1,2,3,4),\\
\label{equA6}a\cdot A_{r}&=&(-1)^{r-1}A_{r}\cdot a,\quad a\wedge A_{r}=(-1)^{r}A_{r}\wedge a,\\
a\cdot(b\wedge c\wedge d\wedge \cdots)&=&(a\cdot b)c\wedge d\wedge \cdots-(a\cdot c)b\wedge d\wedge \cdots+(a\cdot d)b\wedge c\wedge \cdots-\cdots\nonumber,\\
\label{equA7}&&
\end{eqnarray}
\begin{eqnarray}
\label{equA8}(a\wedge b)\cdot(c\wedge d)&=&(c\wedge d)\cdot(a\wedge b)=\text{det}\left(\begin{array}{cc}
b\cdot c,&\ b\cdot d\\
a\cdot c,&\ a\cdot d
\end{array}\right),\\
(a\wedge b)\cdot(c\wedge d\wedge e)&=&(c\wedge d\wedge e)\cdot(a\wedge b)\nonumber\\
&=&\big((a\wedge b)\cdot(c\wedge d)\big)e-\big((a\wedge b)\cdot(c\wedge e)\big)d+\big((a\wedge b)\cdot(d\wedge e)\big)c,\nonumber\\
\label{equA9}&&\\
(a\wedge b)\cdot(c\wedge d\wedge e\wedge f)&=&(c\wedge d\wedge e\wedge f)\cdot(a\wedge b)\nonumber\\
&=&\big((a\wedge b)\cdot(c\wedge d)\big)e\wedge f-\big((a\wedge b)\cdot(c\wedge e)\big)d\wedge f\nonumber\\
&+&\big((a\wedge b)\cdot(c\wedge f)\big)d\wedge e+\big((a\wedge b)\cdot(d\wedge e)\big)c\wedge f\nonumber\\
\label{equA10}&-&\big((a\wedge b)\cdot(d\wedge f)\big)c\wedge e+\big((a\wedge b)\cdot(e\wedge f)\big)c\wedge d,\\
\label{equA11}(a\wedge b\wedge c)\cdot(d\wedge e\wedge f)&=&(d\wedge e\wedge f)\cdot(a\wedge b\wedge c)=\text{det}\left(\begin{array}{ccc}
c\cdot d,&\ c\cdot e,&\ c\cdot f\\
b\cdot d,&\ b\cdot e,&\ b\cdot f\\
a\cdot d,&\ a\cdot e,&\ a\cdot f
\end{array}\right),\\
\label{equA12}(a\wedge b\wedge c)\cdot(d\wedge e\wedge f\wedge g)&=&-(d\wedge e\wedge f\wedge g)\cdot(a\wedge b\wedge c)\nonumber\\
&=&\big((a\wedge b\wedge c)\cdot(d\wedge e\wedge f)\big)g-\big((a\wedge b\wedge c)\cdot(d\wedge e\wedge g)\big)f\nonumber\\
&+&\big((a\wedge b\wedge c)\cdot(d\wedge f\wedge g)\big)e-\big((a\wedge b\wedge c)\cdot(e\wedge f\wedge g)\big)d,\\
\label{equA13}(a\wedge b\wedge c\wedge d)\cdot(e\wedge f\wedge g\wedge h)&=&(e\wedge f\wedge g\wedge h)\cdot(a\wedge b\wedge c\wedge d)\nonumber\\
&=&\text{det}\left(\begin{array}{cccc}
d\cdot e,&\ d\cdot f,&\ d\cdot g,&\ d\cdot h,\\
c\cdot e,&\ c\cdot f,&\ c\cdot g,&\ c\cdot h,\\
b\cdot e,&\ b\cdot f,&\ b\cdot g,&\ b\cdot h,\\
a\cdot e,&\ a\cdot f,&\ a\cdot g,&\ a\cdot h,
\end{array}\right),\\
\label{equA14}B\times A_{r}&=&\langle BA_{r}\rangle_{r},\quad A_{r}\times B=\langle A_{r}B\rangle_{r},\quad (r=1,2,3),\\
(a\wedge b)\times(c\wedge d)&=&-(c\wedge d)\times(a\wedge b)\nonumber\\
\label{equA15}&=&(b\cdot c)a\wedge d-(b\cdot d)a\wedge c+(a\cdot d)b\wedge c-(a\cdot c)b\wedge d,\\
(a\wedge b)\times(c\wedge d\wedge e)&=&-(c\wedge d\wedge e)\times(a\wedge b)\nonumber\\
&=&(b\cdot c)a\wedge d\wedge e-(b\cdot d)a\wedge c\wedge e+(b\cdot e)a\wedge c\wedge d\nonumber\\
\label{equA16}&-&(a\cdot c)b\wedge d\wedge e+(a\cdot d)b\wedge c\wedge e-(a\cdot e)b\wedge c\wedge d,\\
\big\langle(a\wedge b\wedge c)(d\wedge e\wedge f)\big\rangle_{2}&=&-\big\langle(d\wedge e\wedge f)(a\wedge b\wedge c)\big\rangle_{2}\nonumber\\
&=&\big((b\wedge c)\cdot(d\wedge e)\big)a\wedge f-\big((b\wedge c)\cdot(d\wedge f)\big)a\wedge e\nonumber\\
&+&\big((b\wedge c)\cdot(e\wedge f)\big)a\wedge d-\big((a\wedge c)\cdot(d\wedge e)\big)b\wedge f\nonumber\\
&+&\big((a\wedge c)\cdot(d\wedge f)\big)b\wedge e-\big((a\wedge c)\cdot(e\wedge f)\big)b\wedge d\nonumber\\
&+&\big((a\wedge b)\cdot(d\wedge e)\big)c\wedge f-\big((a\wedge b)\cdot(d\wedge f)\big)c\wedge e\nonumber\\
\label{equA18}&+&\big((a\wedge b)\cdot(e\wedge f)\big)c\wedge d,
\end{eqnarray}
where Eqs.~(\ref{equA1})---(\ref{equA7}) and (\ref{equA14})---(\ref{equA16}) can be directly obtained according to the corresponding formulas in Refs.~\cite{Hestenes1984,Doran2003}, and  the derivations of Eqs.~(\ref{equA8})---(\ref{equA13}) and (\ref{equA18}) are able to be greatly simplified by making use of the bases (\ref{equ2.2}) (cf.~(\ref{equ2.3})) and the anticommutation of the vector generators. In addition,
as two typical formulas in GA,
\begin{eqnarray}
\label{equA19}a\cdot A_{r}&=&\frac{1}{2}\big(aA_{r}-(-1)^{r}A_{r}a\big),\\
\label{equA20}a\wedge A_{r}&=&\frac{1}{2}\big(aA_{r}+(-1)^{r}A_{r}a\big)
\end{eqnarray}
are also often used in application.
\section{The ``common'' even subalgebra of the STAs of signatures $(\pm,\mp,\mp,\mp)$}
In the ``common'' even subalgebra of the two STAs, all the operation rules are independent of signatures $(\pm,\mp,\mp,\mp)$, and therefore, in this section, the signs ``$\pm$'' associated with multivectors have been omitted for brevity. According to~(\ref{equ2.25}), a basis for this algebraic formalism is
\begin{equation}\label{equB.1}
\{1,\ \ \boldsymbol{\sigma}_{k},\ \ \boldsymbol{\sigma}_{i}\boldsymbol{\sigma}_{j}\ (i<j),\quad \boldsymbol{\sigma}_{1}\boldsymbol{\sigma}_{2}\boldsymbol{\sigma}_{3}\},
\end{equation}
where $\{\boldsymbol{\sigma}_{k}\}$, as the vector generators, provide a representation-free version of the Pauli matrices, and from
the relevant formulas in Sec.~\ref{Sec:second}, they satisfy the following fundamental properties,
\begin{eqnarray}
\label{equB2}I&=&\boldsymbol{\sigma}_{1}\boldsymbol{\sigma}_{2}\boldsymbol{\sigma}_{3},\\
\label{equB3}\boldsymbol{\sigma}_{i}\boldsymbol{\sigma}_{j}+\boldsymbol{\sigma}_{j}\boldsymbol{\sigma}_{i}&=&2\boldsymbol{\sigma}_{i}\cdot\boldsymbol{\sigma}_{j}=2\delta_{ij},\\
\label{equB4}\boldsymbol{\sigma}_{i}\boldsymbol{\sigma}_{j}-\boldsymbol{\sigma}_{j}\boldsymbol{\sigma}_{i}&=&2\boldsymbol{\sigma}_{i}\times\boldsymbol{\sigma}_{j}=2\epsilon_{ijk}\boldsymbol{\sigma}_{k}I,\\
\label{equB5}\boldsymbol{\sigma}_{i}\boldsymbol{\sigma}_{j}&=&\delta_{ij}+\epsilon_{ijk}\boldsymbol{\sigma}_{k}I,\\
\label{equB6}\boldsymbol{\sigma}_{i}\boldsymbol{\sigma}_{j}&=&\boldsymbol{\sigma}_{i}\times\boldsymbol{\sigma}_{j}=-\boldsymbol{\sigma}_{j}\times\boldsymbol{\sigma}_{i}=-\boldsymbol{\sigma}_{i}\boldsymbol{\sigma}_{j},\quad (i\neq j).
\end{eqnarray}
Let $\boldsymbol{a}=a_{i}\boldsymbol{\sigma}_{i}$, $\boldsymbol{b}=b_{j}\boldsymbol{\sigma}_{j}$, and $\boldsymbol{c}=c_{k}\boldsymbol{\sigma}_{k}$, and as shown in Eqs.~(\ref{equ2.28})---(\ref{equ2.30}), three types of basic homogeneous multivectors in this algebraic formalism are, respectively, $\boldsymbol{a}$, $\boldsymbol{a}\times\boldsymbol{b}$, and
\begin{eqnarray}
\label{equB7}\boldsymbol{a}\wedge\left(\boldsymbol{b}\times\boldsymbol{c}\right)&=&\det\left(\begin{array}{ccc}
a_{1},&\ b_{1},&\ c_{1}\\
a_{2},&\ b_{2},&\ c_{2}\\
a_{3},&\ b_{3},&\ c_{3}
\end{array}\right)I.
\end{eqnarray}
By using Eqs.~(\ref{equ2.31})---(\ref{equ2.39}), (\ref{equ2.21}), (\ref{equA5}), and (\ref{equA9}), one gets
\begin{eqnarray}
\label{equB8}\boldsymbol{a}\boldsymbol{b}&=&\boldsymbol{a}\cdot\boldsymbol{b}+\boldsymbol{a}\times\boldsymbol{b},\\
\label{equB9}\boldsymbol{a}\left(\boldsymbol{b}\times\boldsymbol{c}\right)&=&\boldsymbol{a}\times\left(\boldsymbol{b}\times\boldsymbol{c}\right)+\boldsymbol{a}\wedge\left(\boldsymbol{b}\times\boldsymbol{c}\right),\\
\label{equB10}\left(\boldsymbol{b}\times\boldsymbol{c}\right)\boldsymbol{a}&=&\left(\boldsymbol{b}\times\boldsymbol{c}\right)\times\boldsymbol{a}+\left(\boldsymbol{b}\times\boldsymbol{c}\right)\wedge\boldsymbol{a},\\
\label{equB11}\boldsymbol{a}I&=&I\boldsymbol{a}=\boldsymbol{a}\cdot I=I\cdot \boldsymbol{a}=\frac{1}{2}\epsilon_{kij}a_{k}\left(\boldsymbol{\sigma}_{i}\times\boldsymbol{\sigma}_{j}\right),\\
\label{equB12}\left(\boldsymbol{a}\times\boldsymbol{b}\right)\left(\boldsymbol{c}\times\boldsymbol{d}\right)&=&\left(\boldsymbol{a}\times\boldsymbol{b}\right)\cdot\left(\boldsymbol{c}\times\boldsymbol{d}\right)+\left(\boldsymbol{a}\times\boldsymbol{b}\right)\times\left(\boldsymbol{c}\times\boldsymbol{d}\right),\\
\label{equB13}\left(\boldsymbol{a}\times\boldsymbol{b}\right)I&=&I\left(\boldsymbol{a}\times\boldsymbol{b}\right)=\left(\boldsymbol{a}\times\boldsymbol{b}\right)\cdot I=I\cdot \left(\boldsymbol{a}\times\boldsymbol{b}\right)=-\epsilon_{kij}a_{i}b_{j}\boldsymbol{\sigma}_{k},\\
\label{equB14}I^2&=&II=I\cdot I=-1
\end{eqnarray}
with $\boldsymbol{d}=d_{p}\boldsymbol{\sigma}_{p}$,
\begin{eqnarray}
\label{equB15}\boldsymbol{a}\times\left(\boldsymbol{b}\times\boldsymbol{c}\right)&=&-\left(\boldsymbol{b}\times\boldsymbol{c}\right)\times\boldsymbol{a}=\left(\boldsymbol{a}\cdot\boldsymbol{b}\right)\boldsymbol{c}-\left(\boldsymbol{a}\cdot\boldsymbol{c}\right)\boldsymbol{b},\\
\label{equB16}\boldsymbol{a}\wedge\left(\boldsymbol{b}\times\boldsymbol{c}\right)&=&\left(\boldsymbol{b}\times\boldsymbol{c}\right)\wedge\boldsymbol{a}=\boldsymbol{b}\wedge\left(\boldsymbol{c}\times\boldsymbol{a}\right)=\boldsymbol{c}\wedge\left(\boldsymbol{a}\times\boldsymbol{b}\right),\\
\label{equB17}\left(\boldsymbol{a}\times\boldsymbol{b}\right)\cdot\left(\boldsymbol{c}\times\boldsymbol{d}\right)&=&\left(\boldsymbol{b}\cdot\boldsymbol{c}\right)\left(\boldsymbol{a}\cdot\boldsymbol{d}\right)-\left(\boldsymbol{b}\cdot\boldsymbol{d}\right)\left(\boldsymbol{a}\cdot\boldsymbol{c}\right),\\
\label{equB18}\left(\boldsymbol{a}\times\boldsymbol{b}\right)\times\left(\boldsymbol{c}\times\boldsymbol{d}\right)&=&\left(\boldsymbol{b}\cdot\boldsymbol{c}\right)\left(\boldsymbol{a}\times\boldsymbol{d}\right)+\left(\boldsymbol{a}\cdot\boldsymbol{d}\right)\left(\boldsymbol{b}\times\boldsymbol{c}\right)\nonumber\\
&-&\left(\boldsymbol{b}\cdot\boldsymbol{d}\right)\left(\boldsymbol{a}\times\boldsymbol{c}\right)-\left(\boldsymbol{a}\cdot\boldsymbol{c}\right)\left(\boldsymbol{b}\times\boldsymbol{d}\right).
\end{eqnarray}
These formulas constitute the main operation rules of geometric product, inner product, wedge product, and commutator product in the ``common'' even subalgebra of the two STAs.

As noted in Sec.~\ref{Sec:second}, the relative space spanned by $\{\boldsymbol{\sigma}_{k}\}$ is an Euclidean space of dimension 3 with $I$ as a pseudoscalar, and in this space, the inner product and the cross product between two relative vectors are well-defined. From Eqs.~(\ref{equ2.18}) and (\ref{equ2.19}),
the inner product and the cross product between relative vectors $\boldsymbol{a}$ and $\boldsymbol{b}$  are
\begin{eqnarray}
\label{equB19}\boldsymbol{a}\cdot\boldsymbol{b}&=&\left\langle\boldsymbol{a}\boldsymbol{b}\right\rangle=a_{k}b_{k},\\
\label{equB20}\boldsymbol{a}\times_{3}\boldsymbol{b}&=&-I\left(\boldsymbol{a}\times\boldsymbol{b}\right)=\epsilon_{ijk}a_{i}b_{j}\boldsymbol{\sigma}_{k},
\end{eqnarray}
which are identical to their conventional ones, respectively. Eq.~(\ref{equB20}) suggests that in the relative space, $\boldsymbol{a}\times_{3}\boldsymbol{b}$ is actually the dual of $\boldsymbol{a}\times\boldsymbol{b}$,
and with this result, the commutator products in
Eqs.~(\ref{equB15})---(\ref{equB18}) can be replaced by the corresponding cross products. By applying Eqs.~(\ref{equB20}), (\ref{equB13}), and (\ref{equB14}), the equalities
\begin{eqnarray}
\label{equB21}\boldsymbol{a}\times_{3}\left(\boldsymbol{b}\times_{3}\boldsymbol{c}\right)&=&\{\boldsymbol{a}\times\left[\left(\boldsymbol{b}\times\boldsymbol{c}\right)I\right]\}I=\langle\boldsymbol{a}\left(\boldsymbol{b}\times\boldsymbol{c}\right)I\rangle_{2}I=\langle\boldsymbol{a}\left(\boldsymbol{b}\times\boldsymbol{c}\right)\rangle_{2}I^2\nonumber\\
&=&-\boldsymbol{a}\times\left(\boldsymbol{b}\times\boldsymbol{c}\right),\\
\label{equB22}\boldsymbol{a}\cdot\left(\boldsymbol{b}\times_{3}\boldsymbol{c}\right)I&=&-\{\boldsymbol{a}\cdot\left[\left(\boldsymbol{b}\times\boldsymbol{c}\right)I\right]\}I=-\langle\boldsymbol{a}\left(\boldsymbol{b}\times\boldsymbol{c}\right)I\rangle I=-\langle\boldsymbol{a}\left(\boldsymbol{b}\times\boldsymbol{c}\right)\rangle_{4}I^2\nonumber\\
&=&\boldsymbol{a}\wedge\left(\boldsymbol{b}\times\boldsymbol{c}\right),\\
\label{equB23}\left(\boldsymbol{a}\times_{3}\boldsymbol{b}\right)\cdot\left(\boldsymbol{c}\times_{3}\boldsymbol{d}\right)&=&\langle\left(\boldsymbol{a}\times\boldsymbol{b}\right)I\left(\boldsymbol{c}\times\boldsymbol{d}\right)I\rangle=\langle\left(\boldsymbol{a}\times\boldsymbol{b}\right)\left(\boldsymbol{c}\times\boldsymbol{d}\right)I^2\rangle\nonumber\\
&=&-\langle\left(\boldsymbol{a}\times\boldsymbol{b}\right)\left(\boldsymbol{c}\times\boldsymbol{d}\right)\rangle=-\left(\boldsymbol{a}\times\boldsymbol{b}\right)\cdot\left(\boldsymbol{c}\times\boldsymbol{d}\right),\\
\label{equB24}\left(\boldsymbol{a}\times_{3}\boldsymbol{b}\right)\times_{3}\left(\boldsymbol{c}\times_{3}\boldsymbol{d}\right)&=&-\langle\left(\boldsymbol{a}\times\boldsymbol{b}\right)I\left(\boldsymbol{c}\times\boldsymbol{d}\right)I\rangle_{2}I=-\langle\left(\boldsymbol{a}\times\boldsymbol{b}\right)\left(\boldsymbol{c}\times\boldsymbol{d}\right)I^2\rangle_{2}I\nonumber\\
&=&\langle\left(\boldsymbol{a}\times\boldsymbol{b}\right)\left(\boldsymbol{c}\times\boldsymbol{d}\right)\rangle_{2}I=\left(\left(\boldsymbol{a}\times\boldsymbol{b}\right)\times\left(\boldsymbol{c}\times\boldsymbol{d}\right)\right)I
\end{eqnarray}
are deduced, and then, plugging them into Eqs.~(\ref{equB15})---(\ref{equB18}) gives
\begin{eqnarray}
\label{equB25}\boldsymbol{a}\times_{3}\left(\boldsymbol{b}\times_{3}\boldsymbol{c}\right)&=&\left(\boldsymbol{a}\cdot\boldsymbol{c}\right)\boldsymbol{b}-\left(\boldsymbol{a}\cdot\boldsymbol{b}\right)\boldsymbol{c},\\
\label{equB26}\boldsymbol{a}\cdot\left(\boldsymbol{b}\times_{3}\boldsymbol{c}\right)&=&\boldsymbol{b}\cdot\left(\boldsymbol{c}\times_{3}\boldsymbol{a}\right)=\boldsymbol{c}\cdot\left(\boldsymbol{a}\times_{3}\boldsymbol{b}\right)=\det\left(\begin{array}{ccc}
a_{1},&\ b_{1},&\ c_{1}\\
a_{2},&\ b_{2},&\ c_{2}\\
a_{3},&\ b_{3},&\ c_{3}
\end{array}\right),
\end{eqnarray}
\begin{eqnarray}
\label{equB27}\left(\boldsymbol{a}\times_{3}\boldsymbol{b}\right)\cdot\left(\boldsymbol{c}\times_{3}\boldsymbol{d}\right)&=&\left(\boldsymbol{a}\cdot\boldsymbol{c}\right)\left(\boldsymbol{b}\cdot\boldsymbol{d}\right)-\left(\boldsymbol{a}\cdot\boldsymbol{d}\right)\left(\boldsymbol{b}\cdot\boldsymbol{c}\right),\\
\label{equB28}\left(\boldsymbol{a}\times_{3}\boldsymbol{b}\right)\times_{3}\left(\boldsymbol{c}\times_{3}\boldsymbol{d}\right)&=&\left(\boldsymbol{a}\cdot\boldsymbol{c}\right)\left(\boldsymbol{b}\times_{3}\boldsymbol{d}\right)+\left(\boldsymbol{b}\cdot\boldsymbol{d}\right)\left(\boldsymbol{a}\times_{3}\boldsymbol{c}\right)\nonumber\\
&-&\left(\boldsymbol{a}\cdot\boldsymbol{d}\right)\left(\boldsymbol{b}\times_{3}\boldsymbol{c}\right)-\left(\boldsymbol{b}\cdot\boldsymbol{c}\right)\left(\boldsymbol{a}\times_{3}\boldsymbol{d}\right).
\end{eqnarray}
Eqs.~(\ref{equB25})---(\ref{equB28}) are exactly those formulas involving cross product in vector analysis, which implies that the relative space, as an Euclidean space of dimension 3, can be treated as an arena for classical physics.

Functions defined on finite dimensional GA have a wide range of applications in physics, and readers interested in the details on this topics are invited to consult Refs.~\cite{Hestenes1986,Application.electrodynamics,Doran2003}. Here, we only focus on those defined on the ``common'' even subalgebra of the STAs of signatures $(\pm,\mp,\mp,\mp)$. The exponential function is the most common one, and for a multivector $A$, its exponential function is defined by
\begin{eqnarray}
\label{equB29}\text{e}^{A}=\sum_{m=0}^{\infty}\frac{A^{\langle m\rangle}}{m!}
\end{eqnarray}
with
\begin{equation}\label{equB30}
A^{\langle m\rangle}:=\left\{\begin{array}{lll}
\displaystyle (A\cdot A)^{\frac{m}{2}},  &\displaystyle\quad\text{for even}\ m, \smallskip\\
\displaystyle (A\cdot A)^{\frac{m-1}{2}}A,   &\displaystyle\quad\text{for odd}\ m,
\end{array}\right.
\end{equation}
where the series is absolutely convergent~\cite{Application.electrodynamics}. Starting from the equality $A\cdot A=\tilde{A}\cdot\tilde{A}$~\cite{Hestenes1984}, one property of $\text{e}^{A}$
can be directly obtained,
\begin{eqnarray}
\label{equB31}\widetilde{\text{e}^{A}}&=&\text{e}^{\tilde{A}}.
\end{eqnarray}
Note that $A^{\langle m\rangle}$ is not equivalent to $A^{m}:=AA\cdots A$ in the general case. The odd and even parts of the exponential function are referred to as the hyperbolic sine and cosine functions, respectively,
\begin{equation}\label{equB32}
\left\{\begin{array}{l}
\displaystyle \sinh{A}:=\sum_{n=0}^{\infty}\frac{A^{\langle2n+1\rangle}}{(2n+1)!}=\frac{\text{e}^{A}-\text{e}^{-A}}{2},\smallskip\\
\displaystyle \cosh{A}:=\sum_{n=0}^{\infty}\frac{A^{\langle2n\rangle}}{(2n)!}=\frac{\text{e}^{A}+\text{e}^{-A}}{2},
\end{array}\right.
\end{equation}
and thus,
\begin{eqnarray}
\label{equB33}\text{e}^{A}&=&\cosh{A}+\sinh{A}.
\end{eqnarray}
Similarly, the trigonometric functions sine and cosine are also defined by power series in the normal way,
\begin{equation}\label{equB34}
\left\{\begin{array}{l}
\displaystyle \sin{A}=\sum_{n=0}^{\infty}(-1)^{n}\frac{A^{\langle2n+1\rangle}}{(2n+1)!},\smallskip\\
\displaystyle \cos{A}=\sum_{n=0}^{\infty}(-1)^{n}\frac{A^{\langle2n\rangle}}{(2n)!}.
\end{array}\right.
\end{equation}
We are now in a position to find the relationship between hyperbolic and trigonometric functions. Let $J$ be a multivector satisfying $|J^{\langle2\rangle}|=1$ and $JA=AJ$, where $|J^{\langle2\rangle}|$ denotes the absolute value of $J^{\langle2\rangle}$. By inserting $JA$ into Eqs.~(\ref{equB32}) and (\ref{equB34}), the relationships of the functions $\sin,\ \cos,\ \sinh$, and $\cosh$ are derived,
\begin{subequations}
\begin{eqnarray}
\label{equB35a}\sinh{\left(JA\right)}=J\sinh A,\quad&\text{for}&\quad J^{\langle2\rangle}=1,\\
\label{equB35b}\sinh{\left(JA\right)}=J\sin A,\ \, \quad&\text{for}&\quad J^{\langle2\rangle}=-1,\\
\label{equB35c}\cosh{\left(JA\right)}=\cosh A,\ \,\, \quad&\text{for}&\quad J^{\langle2\rangle}=1,\\
\label{equB35d}\cosh{\left(JA\right)}=\cos A,\,\quad\quad&\text{for}&\quad J^{\langle2\rangle}=-1
\end{eqnarray}
\end{subequations}
and
\begin{subequations}
\begin{eqnarray}
\label{equB36a}\sin{\left(JA\right)}=J\sin A,\ \; \quad&\text{for}&\quad J^{\langle2\rangle}=1,\\
\label{equB36b}\sin{\left(JA\right)}=J\sinh A,\quad&\text{for}&\quad J^{\langle2\rangle}=-1,\\
\label{equB36c}\cos{\left(JA\right)}=\cos A,\; \quad\quad&\text{for}&\quad J^{\langle2\rangle}=1,\\
\label{equB36d}\cos{\left(JA\right)}=\cosh A,\ \ \quad&\text{for}&\quad J^{\langle2\rangle}=-1.
\end{eqnarray}
\end{subequations}
Based on these equalities, with Eq.~(\ref{equB33}), we arrive at
\begin{equation}\label{equB37}
\text{e}^{JA}=\left\{\begin{array}{lll}
\displaystyle \cosh{A}+J\sinh{A},  &\displaystyle\quad\text{for} &\displaystyle\quad J^{\langle2\rangle}=1,\smallskip\\
\displaystyle \cos{A}+J\sin{A},   &\displaystyle\quad\text{for} &\displaystyle\quad J^{\langle2\rangle}=-1.
\end{array}\right.
\end{equation}

Next, starting from Eq.~(\ref{equB37}), a simple method to construct rotors will be provided. Let $\alpha$ and $B$ be a real number and a unit 2-blade, respectively, and we shall prove that $\text{e}^{\alpha B}$ is a rotor. Here, as a unit 2-blade, $B$ is defined by
\begin{eqnarray}
\label{equB38}B^2=1\quad \text{or}\quad B^2=-1,
\end{eqnarray}
and one needs to note that in such a case, $B^2=B^{\langle2\rangle}=B\cdot B$ holds. Thus, by substitution of Eq.~(\ref{equB37}), $\text{e}^{\alpha B}$ can be written as
\begin{equation}\label{equB39}
\text{e}^{\alpha B}=\left\{\begin{array}{lll}
\displaystyle \cosh{\alpha}+B\sinh{\alpha},  &\displaystyle\quad\text{for} &\displaystyle\quad B^2=1,\smallskip\\
\displaystyle \cos{\alpha}+B\sin{\alpha},   &\displaystyle\quad\text{for} &\displaystyle\quad B^2=-1,
\end{array}\right.
\end{equation}
and applying this result repeatedly, we find that
\begin{eqnarray}
\label{equB40}\text{e}^{\alpha_{1} B}\text{e}^{\alpha_{2} B}=\text{e}^{(\alpha_{1}+\alpha_{2}) B}
\end{eqnarray}
with $\alpha_{1}$ and $\alpha_{2}$ as arbitrary real numbers, where the particular case,
\begin{eqnarray}
\label{equB41}\text{e}^{\alpha B}\text{e}^{-\alpha B}=1,
\end{eqnarray}
is also worth noting. Besides, with the help of Eq.~(\ref{equB31}), the above equation is equivalent to
\begin{eqnarray}
\label{equB42}\text{e}^{\alpha B}\widetilde{\text{e}^{\alpha B}}=1.
\end{eqnarray}
Eqs.~(\ref{equB39}) and (\ref{equB42}) clearly suggest that $\text{e}^{\alpha B}$ is an even multivector satisfying $\text{e}^{\alpha B}\widetilde{\text{e}^{\alpha B}}=1$. Then, according to the definition of rotor (cf.~Sec.~\ref{Sec:3.1}), one only needs to prove that in spacetime, the map defined by $b\mapsto \text{e}^{\alpha B}b\,\widetilde{\text{e}^{\alpha B}}$ transforms any vector into another one. For an arbitrary vector $b$, Eq.~(\ref{equA1}) provides the decomposition,
\begin{eqnarray}
\label{equB43}b=(b\cdot B) B^{-1}+(b\wedge B) B^{-1}.
\end{eqnarray}
Since Eq.~(\ref{equB38}) indicates that
\begin{eqnarray}
\label{equB44}B^{-1}=B\quad \text{or}\quad B^{-1}=-B,
\end{eqnarray}
one can directly verify that
\begin{subequations}
\begin{eqnarray}
\label{equB45a}(b\cdot B)\wedge B^{-1}&=&0,\\
\label{equB45b}(b\wedge B)\times B^{-1}&=&\langle b B B^{-1}\rangle_{3}-(b\cdot B)\wedge B^{-1}=0.
\end{eqnarray}
\end{subequations}
Thus, by using Eqs.~(\ref{equA1}) and (\ref{equA3}), the following results hold:
\begin{subequations}
\begin{eqnarray}
\label{equB46a}(b\cdot B) B^{-1}&=&(b\cdot B)\cdot B^{-1}=:b_{\parallel},\\
\label{equB46b}(b\wedge B) B^{-1}&=&(b\wedge B)\cdot B^{-1}=:b_{\perp}
\end{eqnarray}
\end{subequations}
with
\begin{eqnarray}
\label{equB47}b&=&b_{\parallel}+b_{\perp}
\end{eqnarray}
and
\begin{subequations}
\begin{eqnarray}
\label{equB48a}b_{\parallel}\wedge B&=&\langle(b\cdot B)B^{-1}B\rangle_{3}=0,\\
\label{equB48b}b_{\perp}\cdot B&=&\langle(b\wedge B)B^{-1}B\rangle_{1}=0,
\end{eqnarray}
\end{subequations}
where $b_{\parallel}$ and $b_{\perp}$ are obviously the components of $b$ parallel and perpendicular to $B$, respectively.
Furthermore, by virtue of Eqs.~(\ref{equA1}) and (\ref{equA6}), two important equalites
\begin{subequations}
\begin{eqnarray}
\label{equB49a}b_{\parallel}B&=&b_{\parallel}\cdot B=-B\cdot b_{\parallel}=-Bb_{\parallel},\\
\label{equB49b}b_{\perp}B&=&b_{\perp}\wedge B=B\wedge b_{\perp}=Bb_{\perp}
\end{eqnarray}
\end{subequations}
are obtained, and then, together with Eqs.~(\ref{equB47}), (\ref{equB31}), (\ref{equB39}), and (\ref{equB40}),
we finally get
\begin{eqnarray}
\label{equB50}\text{e}^{\alpha B}b\,\widetilde{\text{e}^{\alpha B}}=\text{e}^{\alpha B}\left(b_{\parallel}+b_{\perp}\right)\text{e}^{-\alpha B}=\text{e}^{2\alpha B}b_{\parallel}+b_{\perp}=\cos{(2\alpha)}b_{\parallel}-\sin{(2\alpha)}b_{\parallel}\cdot B+b_{\perp}.
\end{eqnarray}
Evidently, $\text{e}^{\alpha B}b\,\widetilde{\text{e}^{\alpha B}}$ is a vector, and therefore, the map defined by $b\mapsto \text{e}^{\alpha B}b\,\widetilde{\text{e}^{\alpha B}}$ is indeed a transformation in spacetime. In the STAs of signatures $(\pm,\mp,\mp,\mp)$, the rotor $\text{e}^{\alpha B}$ constructed above can be employed to handle Lorentz boost and
spatial rotation, and we will discuss this topic in Sec.~\ref{Sec:3.1}.
\section{A local orthonormal tetrad $\{\gamma_{\alpha}\}$ and the bivector connection $\omega(u)$ associated with it in the Lense-Thirring spacetime}
The Lense-Thirring metric in isotropic coordinates has the form of the $1/c$ expansion~\cite{Wu:2021uws},
\begin{equation}\label{equC1}
\left\{\begin{array}{ll}
\displaystyle g_{00}&=\displaystyle \pm\left(1-\frac{2}{c^{2}}U\right),\smallskip\\
\displaystyle g_{0i}&=\displaystyle \pm\frac{4}{c^{3}}V_{i},\smallskip\\
\displaystyle g_{ij}&=\displaystyle \mp\delta_{ij}\left(1+\frac{2}{c^{2}}U\right),
\end{array}\right.
\end{equation}
where the potentials $U$ and $V_{i}$ are defined in Eq.~(\ref{equ4.8}), and by applying Eqs.~(\ref{equA8}), (\ref{equA11}), and (\ref{equA13}), the following quantities expanded up to $1/c^3$ order are given,
\begin{subequations}
\begin{eqnarray}
\label{equC2a}(g_{1}\wedge g_{0})\cdot(g_{0}\wedge g_{1})&=&-1,\\
\label{equC2b}(g_{1}\wedge g_{0})\cdot(g_{0}\wedge g_{2})&=&0,\\
\label{equC2c}(g_{1}\wedge g_{0})\cdot(g_{0}\wedge g_{3})&=&0,\\
\label{equC2d}(g_{1}\wedge g_{0})\cdot(g_{1}\wedge g_{2})&=&\frac{4}{c^{3}}V_{2},
\end{eqnarray}
\end{subequations}
\begin{subequations}
\begin{eqnarray}
\label{equC3a}(g_{2}\wedge g_{1}\wedge g_{0})\cdot(g_{0}\wedge g_{1}\wedge g_{2})&=&\pm\left(1+\frac{2}{c^{2}}U\right),\\
\label{equC3b}(g_{2}\wedge g_{1}\wedge g_{0})\cdot(g_{0}\wedge g_{1}\wedge g_{3})&=&0,\\
\label{equC3c}(g_{2}\wedge g_{1}\wedge g_{0})\cdot(g_{0}\wedge g_{2}\wedge g_{3})&=&0,\\
\label{equC3d}(g_{2}\wedge g_{1}\wedge g_{0})\cdot(g_{1}\wedge g_{2}\wedge g_{3})&=&\pm\frac{4}{c^{3}}V_{3},
\end{eqnarray}
\end{subequations}
and
\begin{eqnarray}
\label{equC4}(g_{3}\wedge g_{2}\wedge g_{1}\wedge g_{0})\cdot(g_{0}\wedge g_{1}\wedge g_{2}\wedge g_{3})&=&-\left(1+\frac{4}{c^{2}}U\right).
\end{eqnarray}
Eqs.~(\ref{equC1}), (\ref{equC2a}), (\ref{equC3a}), and (\ref{equC4}) show that Eqs.~(\ref{equ3.37a})---(\ref{equ3.37d}) hold, which implies that we are capable of assuming that there exists a collection of fiducial observers who are distributed over space and at rest in the coordinate system of $g_{\mu\nu}$. As a result, with the help of the relevant formulas in Appendix A, by inserting Eqs.~(\ref{equ3.36}) and (\ref{equC2a})---(\ref{equC4}) into Eq.~(\ref{equ3.38}), a local orthonormal tetrad $\{\gamma_{\alpha}\}$ determined up to $1/c^3$ order in the Lense-Thirring spacetime is acquired,
\begin{equation*}
\left\{\begin{array}{l}
\displaystyle \gamma_{0}=\left(1+\frac{1}{c^{2}}U\right)g_{0},\smallskip\\
\displaystyle \gamma_{i}=-\frac{4}{c^3}V_{i}g_{0}+\left(1-\frac{1}{c^{2}}U\right)g_{i},
\end{array}\right.
\end{equation*}
namely Eq.~(\ref{equ4.7}).

Next, the bivector connection $\omega(u)$ associated with $\{\gamma_{\alpha}\}$ will be derived, and the relevant computations are greatly simplified by the condition ``up to $1/c^3$ order''. Plugging Eqs.~(\ref{equ3.36}) and (\ref{equC1})---(\ref{equC4}) into Eq.~(\ref{equ3.47}) yields
\begin{equation}\label{equC5}
\left\{\begin{array}{l}
\displaystyle g_{0}=\left(1-\frac{1}{c^{2}}U\right)\gamma_{0},\smallskip\\
\displaystyle g_{i}=\frac{4}{c^3}V_{i}\gamma_{0}+\left(1+\frac{1}{c^{2}}U\right)\gamma_{i},
\end{array}\right.
\end{equation}
and with them, one is able to deduce
\begin{equation}\label{equC6}
\left\{\begin{array}{l}
\displaystyle g_{0}\wedge g_{1}\wedge g_{2}=\left(1+\frac{1}{c^{2}}U\right)\gamma_{0}\gamma_{1}\gamma_{2},\smallskip\\
\displaystyle g_{0}\wedge g_{1}\wedge g_{3}=\left(1+\frac{1}{c^{2}}U\right)\gamma_{0}\gamma_{1}\gamma_{3},\smallskip\\
\displaystyle g_{0}\wedge g_{2}\wedge g_{3}=\left(1+\frac{1}{c^{2}}U\right)\gamma_{0}\gamma_{2}\gamma_{3},\smallskip\\
\displaystyle g_{1}\wedge g_{2}\wedge g_{3}=\frac{4}{c^{3}}V_{1}\gamma_{0}\gamma_{2}\gamma_{3}-\frac{4}{c^{3}}V_{2}\gamma_{0}\gamma_{1}\gamma_{3}+\frac{4}{c^{3}}V_{3}\gamma_{0}\gamma_{1}\gamma_{2}+\left(1+\frac{3}{c^{2}}U\right)\gamma_{1}\gamma_{2}\gamma_{3},\smallskip\\
\displaystyle g_{0}\wedge g_{1}\wedge g_{2}\wedge g_{3}=\left(1+\frac{2}{c^{2}}U\right)\gamma_{0}\gamma_{1}\gamma_{2}\gamma_{3},
\end{array}\right.
\end{equation}
where the orthogonality and the anticommutation of $\{\gamma_{\alpha}\}$ are used. Thus, let $\{\gamma^{\beta}\}$ be the reciprocal tetrad of $\{\gamma_{\alpha}\}$, and then, the reciprocal frame $\{g^{\mu}\}$ of the coordinate frame can be constructed by Eq.~(\ref{equ3.46}),
\begin{equation}\label{equC7}
\left\{\begin{array}{l}
\displaystyle g^{0}=\left(1+\frac{1}{c^{2}}U\right)\gamma^{0}-\frac{4}{c^3}V_{j}\gamma^{j},\smallskip\\
\displaystyle g^{i}=\left(1-\frac{1}{c^{2}}U\right)\gamma^{i},
\end{array}\right.
\end{equation}
with which, one further obtains
\begin{equation}\label{equC8}
\left\{\begin{array}{l}
\displaystyle g^{0}\wedge g^{j}=\gamma^{0}\wedge\gamma^{j}-\frac{4}{c^3}V_{k}\gamma^{k}\wedge\gamma^{j},\smallskip\\
\displaystyle g^{i}\wedge g^{j}=\left(1-\frac{2}{c^{2}}U\right)\gamma^{i}\wedge\gamma^{j}.
\end{array}\right.
\end{equation}
Besides, Eqs.~(\ref{equC1}) and (\ref{equC5}) provide
\begin{equation}\label{equC9}
\left\{\begin{array}{l}
\displaystyle \partial_{0}g_{00}=\partial_{0}g_{0i}=0,\smallskip\\
\displaystyle \partial_{j}g_{00}=\mp\frac{2}{c^2}\partial_{j}U,\smallskip\\
\displaystyle \partial_{j}g_{0i}=\pm\frac{4}{c^3}\partial_{j}V_{i},\smallskip\\
\displaystyle \partial_{0}g_{k0}=\partial_{0}g_{ki}=0,\smallskip\\
\displaystyle \partial_{j}g_{k0}=\pm\frac{4}{c^3}\partial_{j}V_{k},\smallskip\\
\displaystyle \partial_{j}g_{ki}=\mp\frac{2}{c^2}\partial_{j}U\delta_{ki}
\end{array}\right.
\end{equation}
and
\begin{equation}\label{equC10}
\left\{\begin{array}{l}
\displaystyle g_{0}\cdot\partial g_{0}=g_{0}\cdot\partial g_{i}=0,\smallskip\\
\displaystyle g_{j}\cdot\partial g_{0}=-\frac{1}{c^2}\partial_{j}U\gamma_{0}\smallskip\\
\displaystyle g_{j}\cdot\partial g_{i}=\frac{4}{c^3}\partial_{j}V_{i}\gamma_{0}+\frac{1}{c^2}\partial_{j}U\gamma_{i},
\end{array}\right.
\end{equation}
respectively, where in the derivation of Eq.~(\ref{equC10}), Eq.~(\ref{equ3.42a}) has been employed. The substitution of Eqs.~(\ref{equC8})---(\ref{equC10}) in Eq.~(\ref{equ3.45}) gives rise to the connection bivectors $\omega(g_{0})$ and $\omega(g_{k})$ expanded up to $1/c^3$ order,
\begin{equation}\label{equC11}
\left\{\begin{array}{l}
\displaystyle \omega(g_{0})=-\frac{1}{c^2}\partial_{j}U\boldsymbol{\sigma}^{j}-\frac{2}{c^3}\partial_{j}V_{k}\boldsymbol{\sigma}^{k}\times \boldsymbol{\sigma}^{j},\smallskip\\
\displaystyle \omega(g_{i})=\frac{1}{c^2}\partial_{j}U\boldsymbol{\sigma}^{i}\times \boldsymbol{\sigma}^{j}+\frac{2}{c^3}\partial_{j}V_{i}\boldsymbol{\sigma}^{j}-\frac{2}{c^3}\partial_{i}V_{j}\boldsymbol{\sigma}^{j},
\end{array}\right.
\end{equation}
in which, $\{\boldsymbol{\sigma}^{k}:=\gamma_{0}\gamma^{k}=\pm\gamma^{0}\gamma^{k}\}$ is the reciprocal frame of $\{\boldsymbol{\sigma}_{k}\}$, and as in Eq.~(\ref{equ2.8}), there is $\boldsymbol{\sigma}^{i}\times \boldsymbol{\sigma}^{j}=\mp\gamma^{i}\wedge\gamma^{j}$. Together with Eqs.~(\ref{equ4.7}), (\ref{equ4.10}), and (\ref{equ4.12}), the four-velocity $u$ of the gyroscope can also be expanded in the coordinate frame $\{g_{\mu}\}$,
\begin{eqnarray}
\label{equC12}u&=&\left(1+\frac{1}{2c^2}\boldsymbol{u}^2\right)\left[c\left(1+\frac{1}{c^{2}}U\right)g_{0}+u^{i}\left(-\frac{4}{c^3}V_{i}g_{0}+\left(1-\frac{1}{c^{2}}U\right)g_{i}\right)\right]\nonumber\\
&=&\left[c\left(1+\frac{1}{2c^2}\boldsymbol{u}^2+\frac{1}{c^{2}}U\right)-\frac{4}{c^3}u^{i}V_{i}\right]g_{0}+\left(1+\frac{1}{2c^2}\boldsymbol{u}^2-\frac{1}{c^{2}}U\right)u^{i}g_{i},
\end{eqnarray}
and then, by applying Eq.~(\ref{equ3.44}), the expression of the bivector connection $\omega(u)$ associated with $\{\gamma_{\alpha}\}$ up to $1/c^3$ order is achieved,
\begin{eqnarray}
\label{equC13}\omega(u)&=&c\left(1+\frac{1}{2c^2}\boldsymbol{u}^2+\frac{1}{c^{2}}U\right)\omega(g_{0})+\left(1+\frac{1}{2c^2}\boldsymbol{u}^2-\frac{1}{c^{2}}U\right)u^{i}\omega(g_{i})\nonumber\\
&=&-\frac{1}{c}\partial_{j}U\boldsymbol{\sigma}^{j}-\frac{2}{c^2}\partial_{j}V_{k}\boldsymbol{\sigma}^{k}\times \boldsymbol{\sigma}^{j}-\frac{1}{2c^3}\boldsymbol{u}^2\partial_{j}U\boldsymbol{\sigma}^{j}-\frac{1}{c^{3}}U\partial_{j}U\boldsymbol{\sigma}^{j}\nonumber\\
&&+\frac{1}{c^2}u^{i}\partial_{j}U\boldsymbol{\sigma}^{i}\times \boldsymbol{\sigma}^{j}+\frac{2}{c^3}u^{i}\left(\partial_{j}V_{i}-\partial_{i}V_{j}\right)\boldsymbol{\sigma}^{j}\nonumber\\
&=&-\frac{1}{c}\boldsymbol{\nabla}U+\frac{2}{c^2}\boldsymbol{\nabla}\times\boldsymbol{V}-\frac{1}{2c^3}\boldsymbol{u}^2\boldsymbol{\nabla}U-\frac{1}{c^{3}}U\boldsymbol{\nabla}U+\frac{1}{c^2}\boldsymbol{u}\times\boldsymbol{\nabla}U-\frac{2}{c^3}\boldsymbol{u}\times\left(\boldsymbol{\nabla}\times\boldsymbol{V}\right)
\end{eqnarray}
with $\boldsymbol{\nabla}:=\boldsymbol{\sigma}^{k}\partial_{k}$ and $\boldsymbol{V}:=V_{i}\boldsymbol{\sigma}_{i}$. Finally, according to Eqs.~(\ref{equ3.60a})---(\ref{equ3.60c}), the corresponding expressions of the electric part $\omega^{(E)}(u)$ and the magnetic part $\omega^{(B)}(u)$ of $\omega(u)$ are, respectively, evaluated as
\begin{eqnarray}
\label{equC14}\omega^{(E)}(u)&=&-\frac{1}{c}\boldsymbol{\nabla}U-\frac{1}{2c^3}\boldsymbol{u}^2\boldsymbol{\nabla}U-\frac{1}{c^{3}}U\boldsymbol{\nabla}U-\frac{2}{c^3}\boldsymbol{u}\times\left(\boldsymbol{\nabla}\times\boldsymbol{V}\right),\\
\label{equC15}\omega^{(B)}(u)&=&\frac{2}{c^2}\boldsymbol{\nabla}\times\boldsymbol{V}+\frac{1}{c^2}\boldsymbol{u}\times\boldsymbol{\nabla}U.
\end{eqnarray}


\begin{thebibliography}{99}
\centerline{\textbf{REFERENCES}}
\bigskip
\bibitem{Clifford1882} W. K. Clifford, \emph{Mathematical Papers} (Macmillan, London, 1882).
\bibitem{Hestenes1966} D. Hestenes, \emph{Space-Time Algebra} (Gordon and Breach, New York, 1966).
\bibitem{Hestenes1984} D. Hestenes and G. Sobczyk, \emph{Clifford Algebra to Geometric Calculus} (Reidel, Dordrecht, 1984).
\bibitem{Hestenes1986} D. Hestenes, \emph{New Foundations for Classical Mechanics} (Kluwer Academic Publishers, Dordrecht, 1999).
\bibitem{Application.electrodynamics}
B. Jancewicz, \emph{Multivectors and Clifford Algebra in Eelectrodynamics} (World Scientific, Singapore, 1989);\\
D. Hestenes, Primer on Geometric Algebra for Introductory Mathematics and Physics, http://geocalc.clas.asu.edu/pdf/Pri\\
merGeometricAlgebra.pdf;\\
J. Dressel, K. Y. Bliokh, and F. Nori, \emph{Spacetime Algebra as a Powerful Tool for Electromagnetism}, Phys. Rep. \textbf{589}, 1 (2015).
\bibitem{Application.gravity}
D. Hestenes, \emph{Curvature Calculations with Spacetime Algebra}, Int. J. Theor. Phys. \textbf{25}, 581 (1986);\\
A. N. Lasenby, C. J. L. Doran, and S. F. Gull, \emph{Gravity, Gauge Theories and Geomeric Algebra}, Phil. Trans. R. Soc. Lond. A \textbf{356}, 487 (1998);\\
A. M. Lewis, C. J. L. Doran, and A. N. Lasenby, \emph{Quadratic Lagrangians and Topology in Gauge Theory Gravity}, Gen. Rel. Grav. \textbf{32}, 161 (2000);\\
M. Pav\v{s}i\v{c}, \emph{Towards the Unification of Gravity and Other Interactions: What Has Been Missed?}, J. Phys. Conf. Ser. \textbf{222}, 012017 (2010);\\
A. N. Lasenby, \emph{Geometric Algebra, Gravity and Gravitational Waves}, Adv. Appl. Clifford Algebras \textbf{29}, 79 (2019).
\bibitem{Application.quantum}
C. J. L. Doran, A. N. Lasenby, and S. F. Gull, \emph{States and Operators in the Spacetime Algebra}, Found. Phys. \textbf{23}, 1239 (1993);\\
C. J. L. Doran, A. N. Lasenby, S. F. Gull, S. Somaroo, and A. D. Challinor, \emph{Spacetime Algebra and Electron Physics}, Adv. Imag. Elect. Phys. \textbf{95}, 271 (1996);\\
A. N. Lasenby, C. J. L. Doran, and S. F. Gull, \emph{A Multivector Derivative Approach to Lagrangian Field Theory}, Found. Phys. \textbf{23}, 1295 (1993);\\
A. M. Lewis, C. J. L. Doran, and  A. N. Lasenby, \emph{Electron Scattering without Spin Sums}, Int. J. Theor. Phys. \textbf{40}, 363 (2001).
\bibitem{Doran2003} C. J. L. Doran and A. N. Lasenby, \emph{Geometric Algebra for Physicists} (Cambridge University Press, Cambridge, 2003).
\bibitem{Rotortechnology}
D. Hestenes, \emph{Proper Particle Mechanics}, J. Math. Phys. \textbf{15}, 1768 (1974);\\
D. Hestenes, \emph{Proper Dynamics of a Rigid Point Particle}, J. Math. Phys. \textbf{15}, 1778 (1974).
\bibitem{Sabbata2006} V. de Sabbata and B. K. Datta, \emph{Geometric Algebra and Applications to Physics} (Taylor $\&$ Francis Group, New York, 2007).
\bibitem{MTW1973} C. W. Misner, K. S. Thorne, and J. A. Wheeler, \emph{Gravitation} (W. H. Freeman and Company, San Francisco, 1973).
\bibitem{Ignazio1995} I. Ciufolini and J. A. Wheeler, \emph{Gravitation and Inertia} (Princeton University Press, Princeton, 1995).
\bibitem{Eric2010} \'{E}. Gourgoulhon, \emph{Special Relativity in General Frames From Particles to Astrophysics} (Springer, Berlin, 2013).
\bibitem{Oppositesignature}
K. Greider, \emph{Relativistic Quantum Theory with Correct Consevation Laws}, Phys. Rev. Lett. \textbf{44}, 1718 (1980);\\
K. R. Greider, \emph{A Unifying Clifford Algebra Formalism for Relativistic Fields}, Found. Phys. \textbf{14}, 467 (1984);\\
W. Pezzaglia, \emph{Clifford Algebra Geometric-Multispinor Particles and Multivector-Current Gauge Fields}, Found. Phys. Lett. \textbf{5}, 57 (1992);\\
W. M. Pezzaglia, Jr. and J. J. Adams, \emph{Should Metric Signature Matter in Clifford Algebra Formulations of Physical Theories?}, e-Print Archive: gr-qc/9704048;\\
M. Pav\v{s}i\v{c}, \emph{The Landscape of Theoretical Physics: A Global View from Point Particles to the Brane World and beyond, in Search of a Unifying Principle} (Kluwer Academic, Dordrecht, 2001);\\
J. Vaz, \emph{The Clifford Algebra of Physical Space and Elko Spinors}, Int. J. Theor. Phys. \textbf{57}, 582 (2018).
\bibitem{Wu:2021uws} B. Wu and X. Zhang, \emph{Multipole Analysis on Gyroscopic Precession in $f(R)$ Gravity with Irreducible Cartesian Tensors}, Phys. Rev. D \textbf{104}, 024052 (2021).
\bibitem{Everitt:2011hp} C. W. F. Everitt \emph{et al.}, \emph{Gravity Probe B: Final Results of a Space Experiment to Test General Relativity}, Phys. Rev. Lett. \textbf{106}, 221101 (2011).
\bibitem{Lasenby:2016lfl} A. N. Lasenby, \emph{Geometric Algebra as a Unifying Language for Physics and Engineering and Its Use in the Study of Gravity}, Adv. Appl. Clifford Algebras \textbf{27}, 733 (2017).
\bibitem{Francis:2003xi} M. R. Francis and A. Kosowsky, \emph{Geometric Algebra Techniques for General Relativity}, Annals Phys. \textbf{311}, 459 (2004).
\bibitem{Weinberg2014} S. Weinberg, \emph{Gravitation and Cosmology: Principles and Applications of the General Theory of Relativity} (Wiley, New York, 2014).
\bibitem{Snygg1997} J. Snygg, \emph{Clifford Algebra} (Oxford University Press, New York, 1997).
\bibitem{Blanchet:2013haa} L. Blanchet, \emph{Gravitational Radiation from Post-Newtonian Sources and Inspiralling Compact Binaries}, Living Rev. Relativity \textbf{17}, 2 (2014).
\bibitem{fRtheory}
J. N\"{a}f and P. Jetzer, \emph{On the $1/c$ Expansion of f(R) Gravity}, Phys. Rev. D \textbf{81}, 104003 (2010);\\
N. Castel-Branco, J. P\'aramos, and R. March, \emph{Perturbation of the Metric around a Spherical Body from a Nonminimal Coupling between Matter and Curvature}, Phys. Lett. B \textbf{735}, 25 (2014);\\
A. Dass and S. Liberati, \emph{The Gyroscopic Frequency of Metric $f(R)$ and Generalised Brans-Dicke Theories: Constraints from Gravity Probe-B}, Gen. Relativ. Gravit. \textbf{51}, 108 (2019).
\bibitem{fRGtheory}
M. F. Shamir and A. Komal, \emph{Energy Bounds in $f(R,G)$ Gravity with Anisotropic Background}, Int. J. Geom. Meth. Mod. Phys. \textbf{14}, 1750169 (2017);\\
S. D. Odintsov, V. K. Oikonomou, and S. Banerjee, \emph{Dynamics of Inflation and Dark Energy from $F(R,G)$ Gravity}, Nucl. Phys. B \textbf{938}, 935 (2019).
\bibitem{Stabile:2010mz} A. Stabile, \emph{The Most General Fourth Order Theory of Gravity at Low Energy}, Phys. Rev. D \textbf{82}, 124026 (2010).
\bibitem{Jackson1998} J. D. Jackson, \emph{Classical Electrodynamics} (John Wiley \& Sons, New York, 1999).
\bibitem{Landau1971} L. D. Landau and E. M. Lifshitz, \emph{The Classical Theory of Fields} (Butterworth-Heinemann, Oxford, 1980).
\bibitem{Wald1984} R. M. Wald, \emph{General Relativity} (The University of Chicago Press, Chicago, 1984).
\bibitem{Michael2019} M. Tsamparlis, \emph{Special Relativity: An Introduction with 200 Problems and Solutions} (Springer, New York, 2010).
\bibitem{Yepez:2011bw} J. Yepez, \emph{Einstein's Vierbein Field Theory of Curved Space}, e-Print Archive: gr-qc/1106.2037.
\bibitem{Maurizio2017} M. Gasperini, \emph{Theory of Gravitational Interactions} (Springer, Roma, 2013).
\bibitem{Peter2016} P. Hoyng, \emph{Relativistic Astrophysics and Cosmology: a Pirmer} (Springer, Berlin, 2006).
\bibitem{Hawking1973} S. W. Hawking and G. F. R. Ellis, \emph{The Large Scale Structure of Space-Time} (Cambridge University Press, Cambridge, 1973).
\bibitem{Thorne:1980ru} K. S. Thorne, \emph{Multipole Expansions of Gravitational Radiation}, Rev. Mod. Phys. \textbf{52}, 299 (1980).
\bibitem{Blanchet:1985sp} L. Blanchet and T. Damour, \emph{Radiative Gravitational Fields in General Relativity I. General Structure of the Field Outside the Source}, Phil. Trans. R. Soc. A \textbf{320}, 379 (1986).
    \bibitem{Ramirez:2017pmp} W. G. Ram\'{i}rez and A. A. Deriglazov, \emph{Relativistic Effects due to Gravimagnetic Moment of a Rotating Body}, Phys. Rev. D \textbf{96}, 124013 (2017).
\bibitem{Deriglazov:2017jub} A. A. Deriglazov and W. G. Ram\'{i}rez, \emph{Recent Progress on the Description of Relativistic Spin: Vector Model of Spinning Particle and Rotating Body with Gravimagnetic Moment in General Relativity}, Adv. Math. Phys. \textbf{2017}, 7397159 (2017).
\bibitem{Eric2014} E. Poisson and C. M. Will, \emph{Gravity: Newtonian, Post-Newtonian, Relativistic} (Cambridge University Press, Cambridge, 2014).
\end{thebibliography}
\end{document}